\documentclass[12pt]{article}
\usepackage[doublespacing]{setspace}
\usepackage{mathtools}
\usepackage{booktabs}
\usepackage[english]{babel} 
\usepackage[protrusion=true,expansion=true]{microtype} 
\usepackage{amsmath,amsfonts,amsthm}
\usepackage{ amssymb }
\usepackage{enumitem}
\usepackage{graphicx}
\usepackage{array}
\usepackage[toc,page]{appendix}
\usepackage{xcolor}
\usepackage{tikz}
\usetikzlibrary{arrows.meta,positioning}

\usepackage{booktabs} 
\usepackage{natbib}
\usepackage{setspace}
\usepackage{hyperref}
\hypersetup{hidelinks}
\usepackage{microtype} 
\usepackage{tabularx}
\usepackage{subcaption}
\usepackage[font = small,labelfont=bf,textfont=it]{caption} 
\usepackage{footnote}
\usepackage{algorithm}
\usepackage[noend]{algpseudocode}
\usepackage{etoolbox}
\usepackage{multirow}
\algblock{ParFor}{EndParFor}
\algnewcommand\algorithmicparfor{\textbf{for}}
\algnewcommand\algorithmicpardo{}
\algnewcommand\algorithmicendparfor{}
\algrenewtext{ParFor}[1]{\algorithmicparfor\ #1\ }
\algrenewtext{EndParFor}{\algorithmicendparfor}

\makeatletter
\AtBeginDocument{%

  \newenvironment{proof}[1][\proofname]{\par
    \pushQED{\qed}%
    \normalfont \topsep6\p@\@plus6\p@\relax
    \trivlist
    \item[\hskip\labelsep
          \itshape
      #1\@addpunct{.}\hspace{1ex}]
  }{%
    \popQED\endtrivlist\@endpefalse
  }
}

\theoremstyle{plain} 
\newtheorem{theorem}{Theorem}
\newtheorem{corollary}{Corollary}
\newtheorem{proposition}[theorem]{Proposition}
\newtheorem{lemma}{Lemma}
\theoremstyle{remark} 

\newtheorem{remark}{Remark}
\theoremstyle{definition} 

\newtheorem{assumption}{Assumption}

\newcommand{\distras}[1]{%
  \savebox{\mybox}{\hbox{\kern3pt$\scriptstyle#1$\kern3pt}}%
  \savebox{\mysim}{\hbox{$\sim$}}%
  \mathbin{\overset{#1}{\kern\z@\resizebox{\wd\mybox}{\ht\mysim}{$\sim$}}}%
}
\newcolumntype{C}[1]{>{\centering\let\newline\\\arraybackslash\hspace{0pt}}m{#1}}

\newcommand{\bl}{\textcolor{blue}}

\newcommand{\ww}[1]{\textcolor{blue}{\bf [WW:~#1]}}

\graphicspath{{Fig/}}

\newcommand{\blind}{1}

\usepackage[top=1in, bottom=1in, left=0.8in, right=0.8in]{geometry}

\begin{document}

\if1\blind
{
\setcounter{footnote}{0}
 \renewcommand{\thefootnote}{\fnsymbol{footnote}}
 \centering{\bf\Large Diffusion Non-Additive Model for Multi-Fidelity Simulations with Tunable Precision
 }\\
  \vspace{0.3in}
  \centering{Junoh Heo$^{1}$, Romain Boutelet$^{1}$, Wenjia Wang$^{2}$, and Chih-Li Sung$^{1,}$\footnote{Corresponding author. Address for correspondence: Chih-Li Sung, Department of Statistics and Probability, Michigan State University, East Lansing, MI 48824, USA. Email: \texttt{sungchih@msu.edu}}\vspace{0in}\\
        $^{1}$Department of Statistics and Probability, Michigan State University\\
        $^{2}$Department of Industrial Systems Engineering and Management, National University of Singapore\\
        }
    \date{\vspace{-7ex}}
} \fi

\if0\blind
{
 \centering{\bf\Large Diffusion Non-Additive Model for Multi-Fidelity Simulations with Tunable Precision}\\
    \begin{center}
    {\LARGE\bf }
\end{center}
} \fi

\begin{abstract}
Computer simulations are indispensable for analyzing complex systems, yet high‑fidelity models often incur prohibitive computational costs. Multi‑fidelity frameworks address this challenge by combining inexpensive low‑fidelity simulations with costly high‑fidelity simulations to improve both accuracy and efficiency. However, certain scientific problems demand even more accurate results than the highest‑fidelity simulations available, particularly when a tuning parameter controlling simulation accuracy is available, but the exact solution corresponding to a zero-valued parameter remains out of reach. In this paper, we introduce the Diffusion Non-Additive (DNA) model, inspired by generative diffusion models, which captures nonlinear dependencies across fidelity levels using Gaussian process priors and extrapolates to the exact solution. The DNA model:
(i) accommodates complex, non-additive relationships across fidelity levels; (ii) employs a nonseparable covariance kernel to model interactions between the tuning parameter and input variables, improving predictive performance;
(iii) provides closed‑form expressions for the posterior predictive mean and variance, allowing efficient inference and uncertainty quantification; and (iv) establishes rigorous theoretical bounds on the prediction error, leading to an optimal experimental design strategy. The methodology is validated on a suite of numerical studies and real-world case studies. An \textsf{R} package implementing the proposed methodology is available to support practical applications.
\end{abstract}
\noindent%
{\it Keywords}: Surrogate Model, Experimental Design, Gaussian Process,  Finite Element Method, Uncertainty Quantification

\section{Introduction}\label{sec:intro}

Computer simulations based on mathematical models have become indispensable for understanding and solving complex engineering and physical problems, particularly when real-world experimentation is costly, hazardous, or infeasible. These simulations often involve models governed by differential equations, which describe various physical phenomena such as heat transfer, structural deformation, electromagnetic fields, and turbulence flows \citep{mak2018efficient,shi2024prediction, jin2025multi, sendrea2024review}. 

Many such simulations incorporate a tuning parameter that controls the trade-off between computational cost and simulation accuracy. A lower value of the tuning parameter corresponds to a higher-fidelity approximation, but typically demands significantly greater computational resources. This setting naturally gives rise to \textit{multi-fidelity modeling}, where data from both inexpensive, low-fidelity simulations and expensive, high-fidelity simulations are used jointly to improve efficiency and predictive performance.

\begin{figure}[h]
\begin{center}
\includegraphics[width=1\textwidth]{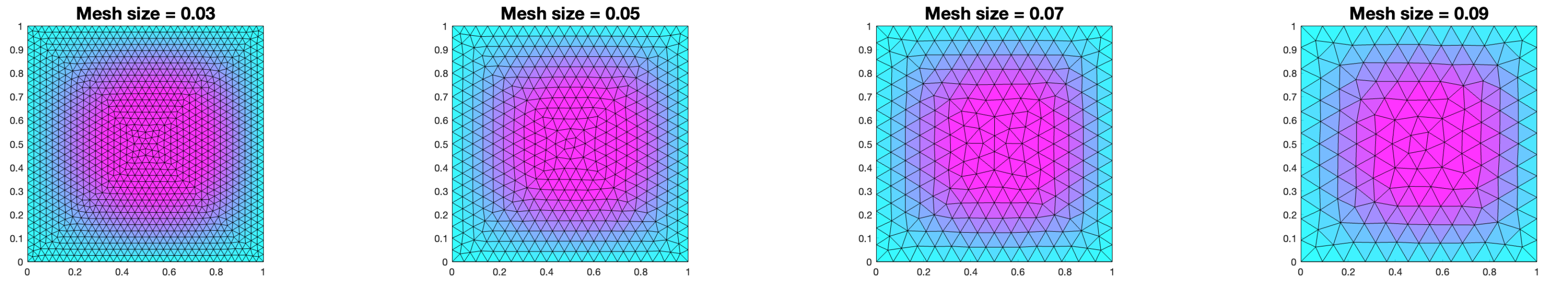} 
\end{center}
\caption{The visualizations of FEM solutions of Poisson's equation at $x=0.5$ for mesh sizes $t=0.03$, $0.05$, $0.07$, and $0.09$.}
\label{fig:Poisson_visualization}
\end{figure}

As a concrete example, consider the finite element method (FEM), one of the most widely used numerical methods for solving differential equations in science and engineering \citep{dhatt2012finite}. FEM discretizes a complex domain into smaller elements and approximates the governing equations over these elements. The mesh size in FEM serves as the tuning parameter: a finer mesh (i.e., smaller mesh size) improves simulation accuracy by more closely approximating the “true” solution—the numerical result obtained in the limit of zero mesh size \citep{tuo2014surrogate}. Figure~\ref{fig:Poisson_visualization} illustrates FEM solutions at different mesh sizes, highlighting the impact of discretization on simulation accuracy.

Beyond general FEM solutions, multi-fidelity problems commonly arise in engineering, statistics, mathematics, and the sciences, where balancing accuracy against computational cost is a key concern. In signal processing \citep{clark2020multi}, for instance, higher sampling rates enhance the precision of signal reconstruction by capturing finer details, yet they require greater data storage and processing resources. Optimization problems \citep{hallmann2020tolerance} exhibit a similar trade-off: tighter convergence tolerances provide more accurate solutions but demand additional iterations and computation time. Bayesian approximation techniques \citep{li2014stochastic}, such as Markov Chain Monte Carlo (MCMC) or Quasi-Monte Carlo, yield more reliable estimates with larger sample sizes, though at an increased computational cost. In time-dependent partial differential equations  \citep{de2016numerical}, achieving accuracy at a target time $T$ requires simulating from an initial time up to $T$, with longer simulations incurring proportionally greater expense. Similarly, in hyperparameter optimization for deep learning \citep{yang2026bayesian}, different stopping criteria lead to varying fidelity levels: early stopping results in higher validation loss and lower-fidelity approximations of model performance, while training to full convergence provides high-fidelity estimates at a substantially greater cost.

A wide range of multi-fidelity modeling approaches has been developed in the literature. Notable examples include \cite{kennedy2000predicting}, \cite{qian2006building}, \cite{qian2008bayesian}, \cite{le2014recursive}, \cite{perdikaris2017nonlinear}, \cite{kerleguer2024bayesian}, and \cite{heo2023active}, along with work on experimental design \citep{yuchi2023design, sung2022stacking} and optimization \citep{picheny2013quantile,he2017optimization}. However, most of these methods focus on emulating the highest fidelity simulator available, often without explicitly modeling the tuning parameter that controls fidelity. Exceptions include the work by \cite{tuo2014surrogate}, which targets the exact solution in the limit of zero mesh size, as well as its adaptive kernel extension by \cite{boutelet2025active} and multi-dimensional tuning parameter generalization by \cite{ji2024conglomerate}. More recently, \cite{oates2025probabilistic} extended Richardson extrapolation within a probabilistic framework, focusing on extrapolating the tuning parameters. 

Despite these advancements, a fundamental limitation of many existing approaches lies in their underlying model structure. By assuming an additive structure of the models, these methods often lack the flexibility to capture complex, non-additive relationships across fidelities. Furthermore, it is even more challenging to ensure rigorous theoretical guarantees regarding prediction error and convergence beyond simple linear or additive formulations, particularly when extrapolating to the exact solution. This absence of a theoretical foundation exacerbates the challenge of experimental design, specifically how to optimally allocate a limited computational budget across these levels. While several experimental design strategies have been recently proposed for multi-fidelity simulations \citep{ehara2023adaptive,sung2022stacking,chen2025fixed}, these frameworks are built upon the auto-regressive model by \cite{kennedy2000predicting}. 
Consequently, there is a need for a flexible modeling framework that not only captures complex dependencies but also provides rigorous multi-fidelity sample allocation strategies.

In this paper, we introduce a novel and flexible model, the Diffusion Non-Additive (DNA) model, designed for multi-fidelity data with an associated tuning parameter. The DNA model offers \textit{tunable precision}, meaning it can predict outputs at any specified fidelity level, including extrapolation to the exact solution in the limit as the tuning parameter approaches zero. Inspired by generative diffusion models in deep learning \citep{ho2020denoising}, the DNA model adopts a recursive structure: lower-fidelity outputs serve as inputs to higher-fidelity levels, and the relationships across fidelities are governed by a Gaussian process (GP) prior. This nonparametric prior enables the model to capture complex, non-additive, and nonlinear dependencies across fidelity levels.

The DNA model offers several key innovations:
(i) To the best of our knowledge, it is the first to integrate diffusion-inspired modeling into the multi-fidelity framework.
(ii) It employs a nonseparable covariance kernel, drawing on ideas from spatial statistics for spatial-temporal modeling, to model interactions between input variables and the tuning parameter. This not only improves predictive accuracy but also offers interpretable representations of physical relationships.
(iii) Although recursive GP structures typically involve computationally intensive procedures (e.g., MCMC) to estimate posterior distributions, we derive closed-form expressions for the posterior mean and variance under nested designs, enabling efficient inference and rigorous uncertainty quantification.
(iv) It provides convergence error bounds for the prediction and leverages these theoretical insights to derive an optimal multi-fidelity experimental design strategy, along with explicit cost–accuracy complexity guarantees.

The organization of this article is as follows. Section~\ref{sec:DNA} introduces the DNA model. Section~\ref{sec:DNAposterior} derives an analytical posterior of the DNA model. Section~\ref{sec:theoretical} establishes the theoretical guarantees of the framework, detailing the convergence error bounds, the cost complexity, the optimal sample allocation, and the construction of the experimental design. In Sections~\ref{sec:studies} and \ref{sec:real}, we conduct a comprehensive analysis of various methods through extensive numerical experiments and real-world case studies. Section~\ref{sec:nonnested} extends the DNA model to non-nested designs. Finally, Section~\ref{sec:conclusion} concludes with a discussion and future directions.

\section{DNA Model for Multi-Fidelity Data}\label{sec:DNA}

\subsection{Problem Setup}

Let $f(t_l, \mathbf{x})$ represent the scalar simulation output of the computer code with input parameter $\mathbf{x} \in \Omega \subseteq \mathbb{R}^d$ and tuning parameter $t_l$ at fidelity level $l$. We assume that simulations are conducted at $L$ distinct fidelity levels to train an \textit{emulator} that approximates the true simulator, where a higher fidelity level corresponds to a simulator with more accurate outputs but also higher computational costs per run. The smaller the tuning parameter, the higher the fidelity, such that $t_1 > t_2 > \cdots > t_L > 0$.

The primary objective is to construct an efficient emulator for the \emph{exact solution} $f(0, \mathbf{x})$, which corresponds to the limiting case as the tuning parameter approaches zero. Directly evaluating $f(0, \mathbf{x})$ is typically infeasible due to prohibitive computational costs or numerical constraints. For example, in FEM, the tuning parameter corresponds to the mesh size, and achieving the zero limit is often restricted by software resolution capabilities.

For each fidelity level $l$, simulations are performed at $n_l$ design points, denoted by $\mathcal{X}_l = \{\mathbf{x}_i^{[l]}\}_{i=1}^{n_l}\subset \Omega$, at tuning parameter $t_l$. These simulations yield corresponding outputs $\mathbf{y}_l := \{f(t_l, \mathbf{x})\}_{\mathbf{x} \in \mathcal{X}_l}$, representing the vector of outputs for $f(t_l, \mathbf{x})$ at design points $\mathbf{x} \in \mathcal{X}_l$. Each element of $\mathbf{y}_l$ is denoted by $y_i^{[l]} = f(t_l, \mathbf{x}_i^{[l]})$. 

Assume that the designs $\mathcal{X}_l$ are sequentially nested, i.e.,
\begin{equation}
\mathcal{X}_L \subseteq \mathcal{X}_{L-1} \subseteq \cdots \subseteq \mathcal{X}_1 \subseteq \Omega,
\label{eq:nested}
\end{equation}
and $\mathbf{x}_i^{[l]} = \mathbf{x}_i^{[l-1]}$ for $i = 1, \ldots, n_l$.  In other words, design points for a higher-fidelity simulator are a subset of the design points for a lower-fidelity simulator. This nested design is common in the multi-fidelity modeling literature and has been shown to enable more efficient inference in various approaches \citep{qian2009nested, qian2009construction, haaland2010approach,huang2026nested}. We extend our approach to accommodate non-nested designs in Section~\ref{sec:nonnested}.

\subsection{Diffusion Non-Additive (DNA) Model}

For notational clarity, let $f_l(\mathbf{x}) := f(t_l, \mathbf{x})$. To model the relationship between different fidelity levels, we assume a functional dependence through an unknown function $W$:
\begin{align}\label{eq:DNAmodel}
\begin{cases}
&f_1(\mathbf{x}) = W_1(\mathbf{x}),\\
&f_l(\mathbf{x}) = W(t_l, \mathbf{x}, f_{l-1}(\mathbf{x})), \quad\text{for}\quad l = 2, \ldots, L,
\end{cases}  
\end{align}
where $t_1 > t_2 > \cdots > t_L > 0$. This recursive formulation satisfies the Markov property, as each fidelity level $f_l(\mathbf{x})$ depends only on the immediately preceding level $f_{l-1}(\mathbf{x})$, without direct dependence on earlier levels.

The structure suggests a \textit{diffusion}-like process, where fidelity levels evolve through a recursive relationship. The term ``diffusion'' is inspired by its analogy in deep learning models \citep{ho2020denoising}, where similar recursive structures propagate information across layers to enable progressive refinement. Each fidelity level $f_l(\mathbf{x})$ represents the simulator output under a coarser or less accurate configuration, such as a relaxed solver tolerance or fewer iterations. As the tuning parameter $t_l$ increases, these choices accumulate approximation error, gradually biasing the output away from the exact solution $f(0,\mathbf{x})$ as illustrated in Figure \ref{fig:Poisson_visualization}. This degradation during the data generation procedure via FEM can be viewed as a \textit{forward diffusion} across fidelity levels. On the other hand, the recursion $f_l(\mathbf{x}) = W(t_l, \mathbf{x}, f_{l-1}(\mathbf{x}))$ exploits this one-step Markov dependence to refine predictions toward $f(0,\mathbf{x})$, resembling a \textit{backward diffusion} process. The hierarchical structure of the model is visualized in the top panel of Figure \ref{fig:model}.

\begin{figure}[!t]
    \centering
\resizebox{0.97\linewidth}{!}{
\begin{tikzpicture}[
    node distance=1.5cm and 0.8cm,
    >=Stealth,
    every node/.style={draw, circle, minimum size=1.5cm, inner sep=0pt, font=\bfseries},
    every path/.style={->}
]

\node (x) {\( f(t,\mathbf{x}) \)};
\node[right=1.5cm of x] (xL) {\( f_L(\mathbf{x}) \)};
\node[draw=none, minimum size=0.7cm, right=0.7cm of xL] (rightdotsT) {\( \dots \)};
\node[right=0.7cm of rightdotsT] (xt) {\( f_l(\mathbf{x}) \)};
\node[right=1.5cm of xt] (xtm1) {\( f_{l-1}(\mathbf{x}) \)};
\node[draw=none, minimum size=0.7cm, right=0.7cm of xtm1] (leftdotsT) {\( \dots \)};
\node[right=0.7cm of leftdotsT] (x1) {\( f_1(\mathbf{x}) \)};

\draw[->, shorten >=2pt] (x1) -- (leftdotsT);
\draw[->, shorten >=2pt] (leftdotsT) -- (xtm1);
\draw[->] (xtm1) -- (xt) node[midway, above=-0.3cm, draw=none]
  {\(W(t_l, \mathbf{x}, f_{l-1}(\mathbf{x}))\)};
\draw[->, shorten >=2pt] (xt) -- (rightdotsT);
\draw[->, shorten >=2pt] (rightdotsT) -- (xL);
\draw[->] (xL) -- (x) node[midway, above=-0.2cm, draw=none]
  {\(W(t, \mathbf{x}, f_L(\mathbf{x}))\)};

\draw[dashed, ->, bend left=30] (xtm1) to[out=-30, in=200] (xt);
\draw[dashed, ->, bend left=30] (xL) to[out=-30, in=200] (x);

\node[below=1.5cm of x] (y) {\( \mu^*(t,\mathbf{x}) \)};
\node[right=1.5cm of y] (yL) {\( \mu^*_L(\mathbf{x}) \)};
\node[draw=none, minimum size=0.7cm, right=0.7cm of yL] (bottomrightdotsM) {\( \dots \)};
\node[right=0.7cm of bottomrightdotsM] (yt) {\( \mu^*_l(\mathbf{x}) \)};
\node[right=1.5cm of yt] (ytm1) {\( \mu^*_{l-1}(\mathbf{x}) \)};
\node[draw=none, minimum size=0.7cm, right=0.7cm of ytm1] (bottomdotsM) {\( \dots \)};
\node[right=0.7cm of bottomdotsM] (mu1) {\( \mu^*_1(\mathbf{x}) \)};

\draw[->, shorten >=2pt] (mu1) -- (bottomdotsM);
\draw[->, shorten >=2pt] (bottomdotsM) -- (ytm1);
\draw[->] (ytm1) -- (yt) node[midway, above=-0.9cm, draw=none]
  {\(h_1(\mathbf{x}|t_l,\mu^*_{l-1}(\mathbf{x}),\sigma^{*2}_{l-1}(\mathbf{x}))\)};
\draw[->, shorten >=2pt] (yt) -- (bottomrightdotsM);
\draw[->, shorten >=2pt] (bottomrightdotsM) -- (yL);
\draw[->] (yL) -- (y) node[midway, above=-0.7cm, draw=none]
  {\(h_1(\mathbf{x}|t,\mu^*_L(\mathbf{x}),\sigma^{*2}_L(\mathbf{x}))\)};

\draw[dashed, ->, bend left=30] (ytm1) to[out=-30, in=200] (yt);
\draw[dashed, ->, bend left=30] (yL) to[out=-30, in=200] (y);

\node[below=0.5cm of y] (z) {\( \sigma^{*2}(t,\mathbf{x}) \)};
\node[right=1.5cm of z] (zL) {\( \sigma^{*2}_L(\mathbf{x}) \)};
\node[draw=none, minimum size=0.7cm, right=0.7cm of zL] (bottomrightdotsV) {\( \dots \)};
\node[right=0.7cm of bottomrightdotsV] (zt) {\( \sigma^{*2}_l(\mathbf{x}) \)};
\node[right=1.5cm of zt] (ztm1) {\( \sigma^{*2}_{l-1}(\mathbf{x}) \)};
\node[draw=none, minimum size=0.7cm, right=0.7cm of ztm1] (bottomdotsV) {\( \dots \)};
\node[right=0.7cm of bottomdotsV] (sigma1) {\( \sigma^{*2}_1(\mathbf{x}) \)};

\draw[->, shorten >=2pt] (sigma1) -- (bottomdotsV);
\draw[->, shorten >=2pt] (bottomdotsV) -- (ztm1);
\draw[->] (ztm1) -- (zt) node[midway, below=-0.9cm, draw=none]
  {\(h_2(\mathbf{x}|t_l,\mu^*_{l-1}(\mathbf{x}),\sigma^{*2}_{l-1}(\mathbf{x}))\)};
\draw[->, shorten >=2pt] (zt) -- (bottomrightdotsV);
\draw[->, shorten >=2pt] (bottomrightdotsV) -- (zL);
\draw[->] (zL) -- (z) node[midway, below=-0.7cm, draw=none]
  {\(h_2(\mathbf{x}|t,\mu^*_L(\mathbf{x}),\sigma^{*2}_L(\mathbf{x}))\)};

\draw[dashed, ->, bend left=-30] (ztm1) to[out=30, in=150] (zt);
\draw[dashed, ->, bend left=-30] (zL) to[out=30, in=150] (z);

\node[left=0.2cm of x, draw=none] (model) {\textbf{Model}};
\node[below=0.8cm of model, draw=none] (posteriormean) {\textbf{Posterior}};
\node[below=-1cm of posteriormean, draw=none] (mean) {\textbf{Mean}};
\node[below=-0.3cm of mean, draw=none] (posteriorvar) {\textbf{Posterior}};
\node[below=-1.2cm of posteriorvar, draw=none] (var) {\textbf{Variance}};

\end{tikzpicture}
}
    \vspace{-1.25cm} \caption{The Markov chain representation of the hierarchical modeling structure, illustrating the relationships between fidelity levels and their corresponding uncertainties. The first row represents the model formulation, where \( W \) is assumed to follow a GP prior. The middle and bottom rows present the inference process, showing the posterior mean calculations and posterior variance evaluations, respectively, with \( h_1 \) and \( h_2 \) having closed-form expressions.}
\label{fig:model}
\end{figure}

Notably, the model proposed in \cite{tuo2014surrogate} assumes $f_l(\mathbf{x}) = f(0, \mathbf{x}) + \delta_l(\mathbf{x})$, which can be interpreted as a special case of the proposed model \eqref{eq:DNAmodel}. Specifically, the relationship in \eqref{eq:DNAmodel} can be expressed as:
\[
W(t_l, \mathbf{x}, f_{l-1}(\mathbf{x})) = f_{l-1}(\mathbf{x}) + (\delta_l(\mathbf{x})-\delta_{l-1}(\mathbf{x})),
\]
where $\delta_l(\mathbf{x})$ is unknown discrepancy function that converges to $0$ as $t_l \to 0$. 

Unlike the additive structure in \cite{tuo2014surrogate}, the proposed model \eqref{eq:DNAmodel} is more flexible as it directly incorporates the lower-fidelity output $f_{l-1}(\mathbf{x})$ as an input, allowing for non-additive relationships. This flexibility enables the model to capture complex, non-linear dependencies across fidelity levels. Combining these properties, we name the model the \textit{Diffusion Non-Additive (DNA)} model.

Instead of explicitly specifying $W$, we model it as a random process using a Gaussian Process (GP) prior \citep{rasmussen2006gaussian, gramacy2020surrogates}. This GP assumption allows us to flexibly capture complex relationships without overly constraining the model structure. Specifically, we assume:
\begin{equation}\label{eq:firstlayerGP}
    W_1(\mathbf{x}) \sim \mathcal{GP}(\alpha_1, \tau_1^2 K_1(\mathbf{x}, \mathbf{x}')),\quad\text{and}
\end{equation}
\begin{equation}\label{eq:DNA}
W(t, \mathbf{x}, y) \sim \mathcal{GP}\left(\alpha, \tau^2 K\left((t, \mathbf{x}, y), (t', \mathbf{x}', y')\right)\right),
\end{equation}
where $\alpha_1$ and $\alpha$ are a constant mean, $\tau_1^2$ and $\tau^2$ are positive scale parameters, and $K_1$ and $K$ are positive-definite kernel functions. 

The kernel function $K_1$ can be chosen from commonly used options such as the squared exponential kernel or the Matérn kernel. For instance, the squared exponential kernel is defined as:
\begin{equation}\label{eq:sqrdexp}
K_1(\mathbf{x}, \mathbf{x}') = \prod_{j=1}^d \exp \left( -\frac{(x_j - x_j')^2}{\theta_{1j}} \right),
\end{equation}
where  $(\theta_{11}, \ldots, \theta_{1d})$ are the \textit{lengthscale} hyperparameters. For the kernel $K((t, \mathbf{x}, y), (t', \mathbf{x}', y'))$, we propose the following \textit{nonseparable} formulation:
\begin{align}\label{eq:Gneiting4}
K((t, &\mathbf{x}, y), (t', \mathbf{x}', y'))= \\&\left(\frac{(t-t')^2}{\theta_t} + 1\right)^{ - \left(\frac{\beta(d+1)}{2}+\delta \right) } \exp \left( -\left(\frac{(t-t')^2}{\theta_t} + 1\right)^{-\beta}\left[\frac{ (y-y')^{2} }{\theta_y} +\sum_{j=1}^d\frac{   (x_j-x_j')^2 }{\theta_j }\right] \right) ,\nonumber
\end{align}
where $\theta_t,\theta_y,$ and $(\theta_1,\ldots,\theta_d)$ are the lengthscale parameters. This kernel is inspired by nonseparable spatio-temporal models \citep{cressie1999classes,gneiting2002nonseparable,porcu2006nonseparable,rodrigues2010class} and is proven to be a valid positive-definite function by \cite{gneiting2002nonseparable}. 
In this context, the input variables are analogous to spatial coordinates, while the tuning parameter serves as a temporal component. 

When $t=t'$, the kernel simplifies to:
\begin{equation}\label{eq:spatiokernel}
K((t, \mathbf{x}, y), (t, \mathbf{x}', y'))=\exp \left( -\left(\frac{ (y-y')^{2} }{\theta_y} +\sum_{j=1}^d\frac{   (x_j-x_j')^2 }{\theta_j }\right) \right),
\end{equation}
which resembles conventional kernels based only on input variables and the output $y$, similar to \cite{heo2023active}. Conversely, for $(\mathbf{x}, y)=(\mathbf{x}', y')$, the kernel reduces to:
\begin{equation}\label{eq:temporalkernel}
K((t, \mathbf{x}, y), (t', \mathbf{x}, y))=\left(\frac{(t-t')^2}{\theta_t} + 1\right)^{ - \left(\frac{\beta(d+1)}{2}+\delta \right) }.
\end{equation}
This expression  belongs to the family of \textit{inverse multiquadric} kernels in $t$ and $t'$ \citep{song2025efficient}, which are generally written as
$K(t, t')=\left(\frac{(t-t')^2}{\theta} + 1\right)^{ -\nu }$,
where $\nu$ is the smoothness parameter.

The covariance kernel $K((t, \mathbf{x}, y), (t', \mathbf{x}', y'))$ is nonseparable, meaning that it cannot be decomposed into the sum or the product of a purely input kernel \eqref{eq:spatiokernel} and a purely tuning parameter kernel \eqref{eq:temporalkernel}. This coupling allows the model to modulate the interaction between $(\mathbf{x},y)$ and the tuning parameter $t$, with the parameter $0\leq\beta\leq 1$ playing a key role in governing this interaction. As $\beta$ increases, the coupling becomes stronger, and the correlation between $(\mathbf{x},y)$ and $(\mathbf{x}',y')$ decays more quickly when the tuning parameters $t$ and $t'$ differ. Importantly, when $\beta=0$, meaning there is no interaction between the tuning parameter and the inputs, the kernel simplifies to a separable form:
\begin{equation*}
K((t, \mathbf{x}, y), (t, \mathbf{x}', y'))=\left(\frac{(t-t')^2}{\theta_t} + 1\right)^{ - \delta } \exp \left( -\left(\frac{ (y-y')^{2} }{\theta_y} +\sum_{j=1}^d\frac{   (x_j-x_j')^2 }{\theta_j }\right) \right),
\end{equation*}
making the separable kernel a special case of the proposed kernel function.

\begin{remark}\label{thm:conditionnumber}
The requirement $\delta > 0$ not only controls the baseline decay in $t$, ensuring that the effects of the tuning parameters remain non-negligible, but also improves numerical stability by enhancing the conditioning of the kernel matrix. To illustrate this, consider the condition number $\kappa(\mathbf{K})$, defined as the ratio of the largest to smallest eigenvalue of the kernel matrix $\mathbf{K}$. 
Consider the special case of a common input design $\mathcal{X} := \mathcal{X}_1 = \cdots = \mathcal{X}_L$. Suppose $\theta_y \to \infty$ and $\beta = 0$. For the kernel matrix $\mathbf{K}$ constructed from equation~\eqref{eq:Gneiting4}, it can be shown that, when $\delta = 0$, the condition number $\kappa(\mathbf{K}) = \infty$, indicating that the matrix is numerically ill-conditioned. In contrast, when $\delta > 0$, the condition number $\kappa(\mathbf{K})<\infty$. A formal proof of this result is provided in Supplementary Materials~\ref{proofcondition}. 
Based on this, we impose a lower bound of 0.5 on $\delta$ during parameter estimation. 
\end{remark}

It is also worth noting that, unlike the recent Recursive Non‑Additive (RNA) model \citep{heo2023active} and the nonlinear auto-regressive GP model \citep{perdikaris2017nonlinear}, which construct a separate GP at each fidelity level $l$ with its own hyperparameters, the DNA model incorporates the tuning parameter $t_l$ as an additional input and embeds all non-lowest‑fidelity outputs into a single GP. By augmenting the input space to $(t,\mathbf{x},y)$, DNA employs a shared lengthscale and scale parameter across levels $2,\dots,L$, reducing the total number of lengthscale hyperparameters from $(dL+L-1)$ in RNA to $(2d+1)$. 
Moreover, the explicit inclusion of $t$ facilitates DNA to perform direct inference for higher-fidelity outputs (i.e., extrapolation to smaller tuning parameters and even the exact solution, i.e., $t=0$).

\subsection{Parameter Estimation}\label{eq:estimation}

The proposed model involves several unknown parameters, including the parameters $\alpha_1$, $\alpha$,  \(\left\{\theta_{1j}\right\}_{j=1}^d\), \(\left\{\theta_{j}\right\}_{j=1}^d\), \(\theta_y\), \(\theta_{t}\), \(\tau_1^2\), \(\tau^2\), \(\beta\), and $\delta$, which are estimated using the maximum likelihood estimation (MLE) method. Specifically, under the GP assumption for \(W_1(\mathbf{x})\) in \eqref{eq:firstlayerGP}, it follows that:
\[
\mathbf{y}_1 \sim \mathcal{N}(\alpha_1\mathbf{1}_{n_1}, \tau_1^2 \mathbf{K}_1),
\]
where \(\mathbf{K}_1\) is an \(n_1 \times n_1\) matrix with elements \((\mathbf{K}_1)_{ij} = K_1(\mathbf{x}_i^{[1]}, \mathbf{x}_j^{[1]})\). Similarly, under the DNA model \eqref{eq:DNA}, the observed simulations \((\mathbf{y}_2^T, \ldots, \mathbf{y}_L^T)^T\) follow a multivariate normal distribution:
\[
Y_{-1} := \begin{bmatrix}
    \mathbf{y}_2 \\
    \vdots \\
    \mathbf{y}_L
\end{bmatrix}
\sim \mathcal{N}_{N_{-1}}\left(\alpha \mathbf{1}_{N_{-1}}, \tau^2 \mathbf{K}\right),
\]
where \(N_{-1} = \sum_{l=2}^L n_l\), \(\mathbf{K}\) is an \(N_{-1} \times N_{-1}\) matrix with elements
\[
\mathbf{K}_{ij} = K\left(((\mathbf{t}_{-1})_i, (\mathbf{X}_{-1})_i, (Y_{-L})_i), ((\mathbf{t}_{-1})_j, (\mathbf{X}_{-1})_j, (Y_{-L})_j)\right),
\]
and \(\mathbf{X}_{-1} = (\mathcal{X}_2^T, \ldots, \mathcal{X}_L^T)^T\), \(\mathbf{t}_{-1} = (t_2\mathbf{1}^T_{n_2}, \ldots, t_L\mathbf{1}^T_{n_L})^T\), and \(Y_{-L} = (\mathbf{y}_1[1:n_2]^T, \mathbf{y}_2[1:n_3]^T, \ldots, \mathbf{y}_{L-1}[1:n_L]^T)^T\), where $\mathbf{y}_l[1:n_k]$ denotes the first $n_k$ samples of $\mathbf{y}_l$. The MLE estimates for  $\alpha_1,\alpha,\tau^2_1$ and $\tau^2$ are given by 
$$
\hat{\alpha}_1=\frac{\mathbf{y}_1^T\mathbf{K}^{-1}_1\mathbf{y}_1}{\mathbf{1}^T_{n_1}\mathbf{K}^{-1}_1\mathbf{1}_{n_1}},\quad\hat{\alpha}=\frac{Y^T_{-1}\mathbf{K}^{-1}Y_{-1}}{\mathbf{1}^T_{N_{-1}}\mathbf{K}^{-1}\mathbf{1}_{N_{-1}}},
$$
and 
$$
\hat{\tau}_1^2=\frac{(\mathbf{y}_1-\hat{\alpha}_1\mathbf{1}_{n_1})^T\mathbf{K}^{-1}_1(\mathbf{y}_1-\hat{\alpha}_1\mathbf{1}_{n_1})}{n_1},\quad\hat{\tau}^2=\frac{(Y_{-1}-\hat{\alpha}\mathbf{1}_{N_{-1}})^T\mathbf{K}^{-1}(Y_{-1}-\hat{\alpha}\mathbf{1}_{N_{-1}})}{N_{-1}}.
$$
The remaining kernel parameters associated with  \(K\) are estimated by maximizing the profile log-likelihood (up to an additive constant):
\begin{equation}\label{eq:likelihood}
-\frac{1}{2} \log(\det(\mathbf{K})) + \frac{N_{-1}}{2}\log\left((Y_{-1} -  \hat{\alpha}\mathbf{1}_{N_{-1}})^T \mathbf{K}^{-1} (Y_{-1} - \hat{\alpha} \mathbf{1}_{N_{-1}})\right).
\end{equation}
These parameters can be efficiently estimated using optimization algorithms such as the quasi-Newton method \citep{byrd1995limited}. A similar MLE approach is used to estimate the parameters associated with \(K_1\). The gradient of the log-likelihood, provided in the Supplementary Materials~\ref{supp:gradient}, 
has a closed-form expression, which facilitates faster and more stable convergence.

\section{Posterior of DNA Model}\label{sec:DNAposterior}
In this section, we derive the posterior distribution of \( f(t, \mathbf{x}) \) for \( 0 \leq t < t_L \) and \( \mathbf{x} \in \Omega \), given the observations \( Y_N := \{\mathbf{y}_l\}_{l=1}^L \), where \( N = \sum_{l=1}^L n_l \). 
Due to the complex non-additive relationship in \eqref{eq:DNAmodel} and the GP prior, computing the posterior can be computationally demanding, often requiring numerical techniques such as Monte Carlo integration. To address this, we derive the closed-form expressions for the posterior mean and variance in a recursive manner, which significantly improves computational efficiency.

By the GP assumptions, the posterior distribution of \( f_l \) given \( Y_N \) and \( f_{l-1} \) at a new input location \( \mathbf{x} \) is normally distributed. Specifically, for \( l = 1 \), we have:
\begin{align}
    W_1(\mathbf{x}) &| \mathbf{y}_1 \sim \mathcal{N}(\mu_1(\mathbf{x}), \sigma_1^2(\mathbf{x})),\quad\text{with}\nonumber\\
\mu_1(\mathbf{x}) &= \alpha_1 +\mathbf{k}_1(\mathbf{x})^T \mathbf{K}_1^{-1} (\mathbf{y}_1-\alpha_1 \mathbf{1}_{n_1}) , \quad \text{and} \label{eq:gppostmean} \\
\sigma^2_1(\mathbf{x}) &= \tau^2_1 \left( 1 - \mathbf{k}_1(\mathbf{x})^T \mathbf{K}_1^{-1} \mathbf{k}_1(\mathbf{x}) \right), \label{eq:gppostvar}
\end{align}
where \( \mathbf{k}_1(\mathbf{x}) \) is an \( n_1 \times 1 \) matrix with elements \( (\mathbf{k}_1(\mathbf{x}))_{i,1} = K_1(\mathbf{x}, \mathbf{x}^{[1]}_i) \). For \( l \geq 2 \) and \( 0 \leq t < t_L \), the posterior distributions of \( f_l(\mathbf{x}) \) and $f(t, \mathbf{x})$ are:
\begin{align*}
f_l(\mathbf{x}) | Y_N, f_{l-1}(\mathbf{x}) &\sim \mathcal{N}(\mu_l(\mathbf{x}, f_{l-1}(\mathbf{x})), \sigma_l^2(\mathbf{x}, f_{l-1}(\mathbf{x}))),\\
    f(t, \mathbf{x}) | Y_N, f_L(\mathbf{x}) &\sim \mathcal{N}(\mu_{L+1}(t, \mathbf{x}, f_L(\mathbf{x})), \sigma_{L+1}^2(t, \mathbf{x}, f_L(\mathbf{x}))),
\end{align*}
with for $l=2,\ldots,L$,
\begin{align*}
\mu_l(\mathbf{x}, f_{l-1}(\mathbf{x})) &= \alpha + \mathbf{k}(t_l, \mathbf{x}, f_{l-1}(\mathbf{x}))^T \mathbf{K}^{-1} (Y_{-1} - \alpha \mathbf{1}_{N_{-1}} ), \\
\sigma^2_l(\mathbf{x}, f_{l-1}(\mathbf{x})) &= \tau^2 \left( 1 - \mathbf{k}(t_l, \mathbf{x}, f_{l-1}(\mathbf{x}))^T \mathbf{K}^{-1} \mathbf{k}(t_l, \mathbf{x}, f_{l-1}(\mathbf{x})) \right), \quad\text{and}\\
\mu_{L+1}(t, \mathbf{x}, f_L(\mathbf{x})) &= \alpha + \mathbf{k}(t, \mathbf{x}, f_L(\mathbf{x}))^T \mathbf{K}^{-1} (Y_{-1} - \alpha \mathbf{1}_{N_{-1}})\nonumber \\
\sigma^2_{L+1}(t, \mathbf{x}, f_L(\mathbf{x})) &= \tau^2 \left( 1 - \mathbf{k}(t, \mathbf{x}, f_L(\mathbf{x}))^T \mathbf{K}^{-1} \mathbf{k}(t, \mathbf{x}, f_L(\mathbf{x})) \right),\nonumber
\end{align*}
where \( \mathbf{k}(t, \mathbf{x}, y) \) is an \( N_{-1} \times 1 \) matrix with 
$
\left(\mathbf{k}(t, \mathbf{x}, y)\right)_{i,1} = K\left((t, \mathbf{x}, y), ((\mathbf{t}_{-1})_i, (\mathbf{X}_{-1})_i, (Y_{-L})_i)\right).$

The posterior distribution of \( f(t, \mathbf{x}) \) for \( 0 \leq t < t_L \) can then be expressed as:
\begin{align*}
p(f(t, \mathbf{x}) | Y_N) = \int  \cdots \int p(f(t, \mathbf{x}) | Y_N, f_L(\mathbf{x})) &p(f_L(\mathbf{x}) | Y_N, f_{L-1}(\mathbf{x}))p(f_{L-1}(\mathbf{x}) | Y_N, f_{L-2}(\mathbf{x}))\cdots \\
& p(f_2(\mathbf{x}) | Y_N, f_1(\mathbf{x})) p(f_1(\mathbf{x}) | Y_N) \, \mathrm{d}f_L(\mathbf{x}) \cdots \mathrm{d}f_1(\mathbf{x}).
\end{align*}
While each component \( p \) is a normal distribution, evaluating this integral using Monte Carlo simulations directly can be computationally expensive, especially when \( L \) is large. To address this, we extend the approach of \cite{heo2023active} and derive the posterior mean and variance in a closed-form recursive fashion, as stated in the following proposition.

\begin{proposition}\label{prop:closedform}
Under the kernel functions \eqref{eq:sqrdexp} and \eqref{eq:Gneiting4}, the posterior mean and variance of $f(t,\mathbf{x})$ given the data $Y_N$ for $0\leq t<t_L$ can be expressed in a recursive fashion: 
\begin{align*}\label{eq:posteriormean}
\mu^*&(t,\mathbf{x})=\mathbb{E}[f(t,\mathbf{x})|Y_N]=h_1(\mathbf{x}|t,\mu^*_L(\mathbf{x}),\sigma^{*2}_L(\mathbf{x})),\\
\sigma^{*2}&(t,\mathbf{x})=\mathbb{V}[f(t,\mathbf{x})|Y_N]=h_2(\mathbf{x}|t,\mu^*_L(\mathbf{x}),\sigma^{*2}_L(\mathbf{x})),\nonumber
\end{align*}
and for $l=2,\ldots,L$,
\begin{align*}
\mu_l^*&(\mathbf{x})=\mathbb{E}[f_l(\mathbf{x})|Y_N]=h_1(\mathbf{x}|t_l,\mu^*_{l-1}(\mathbf{x}),\sigma^{*2}_{l-1}(\mathbf{x})),\\
\sigma^{*2}_l&(\mathbf{x})=\mathbb{V}[f_l(\mathbf{x})|Y_N]=h_2(\mathbf{x}|t_l,\mu^*_{l-1}(\mathbf{x}),\sigma^{*2}_{l-1}(\mathbf{x})),
\end{align*}
where
$$
h_1(\mathbf{x}|t,\mu,\sigma^2)=
\alpha + \sqrt{\frac{\theta_{y}}{\theta_{y}+2
\sigma^2 }}\sum^{N_{-1}}_{i=1} r_i c_i^{\frac{d+1}{2}+\frac{\delta}{\beta}} \exp\left( -c_i \left\{\sum_{j=1}^d\frac{(x_{j}-(\mathbf{X}_{-1})_{ij})^2}{\theta_{j}} + \frac{((Y_{-L})_i-\mu)^2}{\theta_{y}+2\sigma^2} \right\}\right)
$$
and 
\begin{align*}
h_2(\mathbf{x}|t,\mu,\sigma^2)=\tau^2 - &(\mu-\alpha)^2 + \left( \sum_{i,k=1}^{N_{-1}} \zeta_{ik}(t,\mu,\sigma^2)\left(r_i r_k - \tau^2 (\mathbf{K}^{-1})_{ik} \right) (c_ic_k)^{\frac{d+1}{2}+\frac{\delta}{\beta}} \right. \\
&\times \left. \exp{ \left(-\sum_{j=1}^d\frac{c_i(x_{j}-(\mathbf{X}_{-1})_{ij})^2+c_k(x_{j}-(\mathbf{X}_{-1})_{kj})^2}{\theta_{j}}  \right)} \right),    
\end{align*}
where $r_i = (\mathbf{K}^{-1} (Y_{-1}-\alpha \mathbf{1}_{N_{-1}}))_i$, $c_i:=c(t,\mathbf{t}_i)=\left(\frac{(t-\mathbf{t}_i)^2}{\theta_t} + 1\right)^{-\beta}$ and 
\begin{align*}
\zeta_{ik}(t,\mu,\sigma^2) &= \sqrt{\frac{\theta_{y}}{\theta_{y}+2(c_i+c_k)\sigma^2}} \\
&\times \exp{\left( - \frac{c_i\left((Y_{-L})_i-\mu\right)^2+c_k\left((Y_{-L})_k-\mu\right)^2+\frac{2}{\theta_{y}}c_ic_k\sigma^2\left((Y_{-L})_i-(Y_{-L})_k\right)^2}{\theta_{y}+2(c_i+c_k)\sigma^2} \right)}.
\end{align*}
For $l=1$, it follows that $\mu^*_1(\mathbf{x})=\mu_1(\mathbf{x})$ and $\sigma^{*2}_1(\mathbf{x})=\sigma^{2}_1(\mathbf{x})$ as in \eqref{eq:gppostmean} and \eqref{eq:gppostvar}, respectively.
\end{proposition}

The full derivations for Proposition \ref{prop:closedform} are provided in Supplementary Materials~\ref{supp:sqexposterior}. 
While the nonseparable kernel in \eqref{eq:Gneiting4} is based on the squared exponential kernel, we also develop a nonseparable variant based on the Matérn kernel \citep{stein1999interpolation}, another widely used choice. Supplementary Materials~\ref{app:matern} 
includes this alternative kernel and its closed-form expressions for the posterior mean and variance are provided in Supplementary Materials~\ref{supp:maternposterior}. 
These derivations build on the work of \cite{kyzyurova2018coupling}, \cite{ming2021linked}, and \cite{heo2023active}.

Proposition \ref{prop:closedform} enables efficient computation of the posterior mean and variance through a recursive formulation. The hierarchical structure of the posterior inference process is illustrated in Figure \ref{fig:model}. To approximate the posterior distribution, we adopt the \textit{moment-matching} method, leveraging the Gaussian approximation based on the derived posterior mean and variance. This Gaussian approximation to the DNA model minimizes the Kullback–Leibler divergence between the true distribution and its Gaussian representation \citep{minka2001expectation}. The unknown parameters in the posterior distribution are replaced by their corresponding estimates. An \textsf{R} package, \textsf{DNAmf}, implementing the proposed methods is available on \textsf{R} CRAN repository.

\section{Convergence Error Bound and Experimental Design}\label{sec:theoretical}
We now investigate the theoretical error bounds of the DNA model, which will be used to derive insights for constructing multi-fidelity experiments in Section \ref{sec:doe}. The theoretical proofs are given in Supplementary Materials~\ref{appendix:proofSec4}.

\subsection{Convergence Error Bound of the DNA model}
We first define the necessary functional spaces and graph sets, followed by a set of regularity assumptions. For $l=2,\dots,L$, we define the slice functions $g_l(\mathbf{x},y):=W(t_l,\mathbf{x},y)$, and for prediction at the target precision, we define $g_{L+1}(t,\mathbf{x},y):=W(t,\mathbf{x},y).$ Next, we define the domains and the graph sets over which these functions operate. Let $\Omega \subset \mathbb{R}^{d}$ be a compact input space, which implies the joint domain $[0,t_L] \times \Omega$ is also compact. Within this space, we represent the manifold of the input variables and their corresponding lower-fidelity outputs through the graph sets: $\mathcal{M}_{l-1} = \{(\mathbf{x},f_{l-1}(\mathbf{x})) : \mathbf{x} \in \Omega\} \subset \Omega \times \mathbb{R}$ for $l=2,\dots,L$,
and similarly for the target precision space: $\mathcal{M}_{L} = \{(t,\mathbf{x},f_{L}(\mathbf{x})) : (t,\mathbf{x}) \in [0,t_L] \times \Omega\} \subset [0,t_L] \times \Omega \times \mathbb{R}$.
Because $\Omega$ is compact and the functions $f_l$ are continuous, these graph sets $M_{l-1}$ and $\mathcal{M}_{L}$ are inherently compact. 

To evaluate the model within the previously defined graph sets, we map these input points into the embedded graph design, pairing the input coordinates directly with their lower-fidelity outputs:
$\mathcal{Z}_l^{xy}=\{(\mathbf{x}_i^{[l]},f_{l-1}(\mathbf{x}_i^{[l]}))\}_{i=1}^{n_l}\subset \mathcal{M}_{l-1}.$
Finally, to quantify the density and coverage of these mapped designs across their respective spaces, we employ the concept of fill distance \citep{wendland2004scattered}. For a general compact set $D\subset\mathbb{R}^m$ and a finite $S\subset D$, the fill distance is defined as
\[
h_{S,D}=\sup_{z\in D}\min_{s\in S}\|z-s\|_2.
\]

To formally bound the convergence error, we impose the following standard regularity assumptions regarding the underlying functions and the design space.

\begin{assumption}[RKHS regularity]\label{assum_RKHSregularity}
Assume that $f_l$ for $l=2,\dots,L$ are deterministic functions, $g_l|_{M_{l-1}}\in\mathcal{N}_{K_l}(M_{l-1})$, $g_{L+1}|_{\mathcal{M}_L} \in \mathcal{N}_{K}(\mathcal{M}_L)$, where $\mathcal N_K(D)$ denotes the reproducing kernel Hilbert space (RKHS) induced by kernel $K$ on domain $D$, and $K_l((\mathbf{x},y),(\mathbf{x}',y')):=K((t_l,\mathbf{x},y),(t_l,\mathbf{x}',y')).$
\end{assumption}

\begin{assumption}[Lipschitz fidelity maps]\label{assum_Lipsf}
For $l=1,\dots,L$, there exists $B_l<\infty$ such that
\[
|f_l(\mathbf{x})-f_l(\mathbf{x}')|\le B_l\|\mathbf{x}-\mathbf{x}'\|_2,\qquad\forall \mathbf{x},\mathbf{x}'\in \Omega.
\]
In addition, for any $t,t'\in [0,t_1]$, $|f(t,\mathbf{x})-f(t',\mathbf{x})|\le B_t|t-t'|$.
\end{assumption}

\begin{assumption}[Lipschitz posterior means in $y$]\label{assum_Lipsfl}
For $l=2,\dots,L$, the posterior mean $\mu_l(\mathbf{x},y)$ is differentiable in $y$ and
\[\Lambda_l=\sup_{(\mathbf{x},y)\in \Omega\times\mathbb{R}}\left|\partial_y\mu_l(\mathbf{x},y)\right|<\infty.\]
Similarly, $\Lambda_{L+1}=\sup_{(t,\mathbf{x},y)\in [0,t_L] \times \Omega\times\mathbb{R}}\left|\partial_y\mu_{L+1}(t,\mathbf{x},y)\right| + \left|\partial_t\mu_{L+1}(t,\mathbf{x},y)\right|<\infty$.
\end{assumption}

\begin{assumption}[Quasi-uniformity]\label{assum_Quasi_uniform}
For $l=1,\dots,L$, there exists $c_\Omega>0$ such that $h_{\mathcal{X}_l,\Omega}\le c_\Omega n_l^{-1/d}.$
\end{assumption}
The fill distance $h_{\mathcal{X}_l,\Omega}$ represents the radius of the largest ball in the domain which does not contain a design point \citep{wendland2004scattered}. Bounding this distance guarantees that the absence of large sampling gaps will reduce the prediction errors. 


With these conditions established, our first step is to bound the interpolation error at any given fidelity level $l$ using a standard kernel interpolation result. 

\begin{lemma}[\citealp{wendland2004scattered}]\label{lemma_inter_bound}
Let $\eta>0$ be the approximation order associated with the smoothness of the kernel $K_l$. Then, there exists a constant $C$ such that 
\[
\sup_{z\in M_{l-1}}|g_l(z)-\mu_l(z)| \le C h_{\mathcal{Z}_l^{xy},M_{l-1}}^{\eta}\|g_l\|_{\mathcal N_{K_l}(M_{l-1})}.
\]
\end{lemma}
The convergence rate is strictly dictated by the smoothness of the chosen kernel. For example, a kernel from the Mat\'ern class defined in Supplementary Materials~\ref{app:matern} yields a polynomial approximation order equal to its smoothness parameter $\eta=\nu$.  On the other hand, the nonseparable kernel in \eqref{eq:Gneiting4}, which is based on the squared exponential kernel, exhibits spectral convergence, meaning the polynomial error bound holds for any $\eta>0$.


\begin{lemma}\label{graph_fill_x}
Under Assumption \ref{assum_Lipsf}, $h_{\mathcal{Z}_l^{xy},M_{l-1}} \le \sqrt{1+B_{l-1}^2}\, h_{\mathcal{X}_l,\Omega}.$
\end{lemma}

We define the levelwise residuals as
\[
\varepsilon_l(\mathbf{x}) = |f_l(\mathbf{x})-\mu_l(\mathbf{x},f_{l-1}(\mathbf{x}))|, \quad l=1,\dots,L+1.
\]
Let $\hat f_1(\mathbf{x})=\mu_1(\mathbf{x})$, $\hat f_l(\mathbf{x})=\mu_l(\mathbf{x},\hat f_{l-1}(\mathbf{x}))$ for $l=2,\dots,L$, and $\hat f(t,\mathbf{x})=\mu_{L+1}(t,\mathbf{x},\hat f_L(\mathbf{x}))$. The following lemma explicitly tracks how these errors propagate from the lowest fidelity up to the target precision.

\begin{lemma}\label{lemma:recursive-error}
Under Assumption~\ref{assum_Lipsfl}, the following holds for all \((t,\mathbf{x})\in[0,t_L] \times \Omega\):
\[
|f(t,\mathbf{x})-\hat f(t,\mathbf{x})| \le \varepsilon_{L+1}(t,\mathbf{x}) + \sum_{l=2}^{L} \Big(\prod_{s=l}^{L} \Lambda_{s+1}\Big)\,\varepsilon_l(\mathbf{x}) + \Big(\prod_{s=1}^{L} \Lambda_{s+1}\Big)\,\varepsilon_1(\mathbf{x}).
\]
\end{lemma}


The following corollary follows from Assumption~\ref{assum_Quasi_uniform} and Lemma ~\ref{lemma_inter_bound}.

\begin{corollary}\label{cor:rates}
For each $l=2,\ldots,L$, the levelwise residual satisfies
\[
\sup_{\mathbf{x}\in\Omega}\varepsilon_l(\mathbf{x})
\le
C\big(\sqrt{1+B_{l-1}^2}\,c_\Omega \big)^{\eta}
n_l^{-\eta/d}
\|g_l\|_{\mathcal{N}_{K_l}(M_{l-1})}.
\]
\end{corollary}

Combining these explicit levelwise convergence rates with the recursive propagation dynamics of Lemma~\ref{lemma:recursive-error} yields the pointwise error bound for the DNA model in the following theorem.

\begin{theorem}
\label{thm:graph-error-bound}
For all \((t,\mathbf{x})\in[0,t_L] \times\Omega\), the pointwise prediction error is bounded by:
\[
\begin{aligned}
|f(t,\mathbf{x})-& \hat f(t,\mathbf{x})| \le\ 
(B_t+\Lambda_{L+1})(t_L-t) +  \\
&+\sum_{l=2}^L \Big(\prod_{s=l}^{L} \Lambda_{s+1}\Big)\, C\big(\sqrt{1+B_{l-1}^2}\,c_\Omega \big)^{\eta} n_l^{-\eta/d}\|g_l\|_{\mathcal N_{K_l}(M_{l-1})} + \Big(\prod_{s=1}^{L} \Lambda_{s+1}\Big)\,C c_\Omega^\eta n_1^{-\eta/d}\|g_1\|_{\mathcal N_{K_1}(\Omega)}.
\end{aligned}
\]
\end{theorem}

Theorem~\ref{thm:graph-error-bound} explicitly decouples the prediction error into a target-precision extrapolation error (i.e., the first term) and accumulated interpolation errors propagating across fidelity levels. This decomposition highlights a fundamental trade-off in multi-fidelity modeling: achieving high accuracy requires a sufficiently small terminal fidelity parameter $t_L$ to reduce the extrapolation error, while sufficiently large sample sizes $\{n_l\}_{l=1}^L$ are needed to control the interpolation errors at each level. The explicit dependence on $\{n_l\}_{l=1}^L$ further motivates a principled sample allocation strategy to control the prediction error under a finite computational budget.

\subsection{Optimal Sample Allocation}

Building on the explicit error bound established in Theorem~\ref{thm:graph-error-bound}, we can strategically optimize the allocation of computational resources across fidelity levels. Assume $\lambda_{\max} = \max_{l}\{\Lambda_{l}\},c_B = \max_{l}\{B_{l}\}$, and $c_g = \max_{l}\{\|g_l\|_{\mathcal N_{K_l}(M_{l-1})}\}$. To minimize the error bound subject to a total computational budget $C_{\rm total}$, we consider the Lagrangian
\[
\mathcal{L}(n_1, \dots, n_L, \lambda) 
= \sum_{l=1}^L \lambda_{\max}^{L-l+1} n_l^{-\eta/d} 
+ \lambda \left( \sum_{l=1}^L C_l n_l - C_{\rm total} \right).
\]
Solving the first-order optimality conditions yields the optimal sample allocation
$n_l \propto \left(\lambda_{\max}^{L-l+1}C^{-1}_l\right)^{\frac{\eta}{\eta + d}}$.
Normalizing to satisfy the budget constraint gives
\begin{align}\label{optimal_sample_size}
n_l 
= \frac{C_{\rm total}}{\sum_{k=1}^L \lambda_{\max}^{\frac{(L-k+1)d}{\eta+d}} C_k^{\frac{\eta}{\eta+d}}}
\left(\frac{\lambda_{\max}^{L-l+1}}{C_l}\right)^{\frac{d}{\eta+d}}.
\end{align}

This expression reveals how the optimal allocation depends on the cost $C_l$, the Lipschitz constant $\lambda_{\max}$, the input dimension $d$, and the smoothness $\eta$. In particular, for fixed $\lambda_{\max}$, $d$, and $\eta$, the sample size $n_l$ decreases with the cost $C_l$ at the rate $C_l^{-\frac{d}{\eta+d}}$. The effect of $\lambda_{\max}$ is more subtle, since it enters both the numerator and denominator of \eqref{optimal_sample_size}. In general, increasing $\lambda_{\max}$ changes the relative allocation across fidelity levels rather than uniformly increasing or decreasing all $n_l$. When the dependence across successive fidelity levels is stronger, the allocation tends to place relatively more weight on lower-cost levels, as information propagates more effectively across fidelities.

This optimal allocation strategy allows us to address a fundamental question: what total computational cost is required to guarantee a prescribed target accuracy $\epsilon$? We establish this cost–complexity trade-off in the following theorem.

\begin{theorem}\label{thm:complexity}
Suppose $\Omega$ is compact and convex. Suppose the tuning parameters $\{t_1, t_2, \cdots\}$ follow a geometric sequence for increasing fidelity levels, i.e., $t_l=t_0T^{-l}$ for $l\in\mathbb{N}^+$, where  $t_0 > 0$ and $T>1$.
Assume there exist positive constants $\gamma,c_\Lambda,c_B,c_g$ and $c_1$ such that, for $l=1,\ldots,L$:
\begin{enumerate}
\item $\Lambda_l\le c_\Lambda,\Lambda_{L+1}\le c_\Lambda, B_l\le c_B,B_t\le c_B$, and $\|g_l\|_{\mathcal N_{K_l}(M_{l-1})}\le c_g$.
\item The computational cost $C_l$ is bounded as $C_l\leq c_1t_l^{-\gamma}$.
\end{enumerate}
Assuming an error tolerance of $0 < \epsilon<e^{-1}$, there then exist choices of $L$ and $n_1, \cdots, n_L$ for which $\hat{f}(0,\mathbf{x})$ achieves the desired prediction bound
$|f(0,\mathbf{x})-\hat{f}(0,\mathbf{x})|\leq \epsilon, \quad \mathbf{x} \in \Omega$,
with a total computational cost $C_{\rm total}$ bounded by
\begin{equation*}
C_{\rm total} \leq c_2\epsilon^{-\frac{d}{\eta}-\gamma},
\label{eq:compcost}
\end{equation*}
with a  positive constant $c_2$.
\end{theorem}



Theorem~\ref{thm:complexity} shows that the proposed multi-fidelity DNA model achieves a prescribed accuracy with a computational cost scaling as $\mathcal{O}(\epsilon^{-\frac{d}{\eta}-\gamma})$. Importantly, this rate depends only on intrinsic problem difficulty (through $d$ and $\eta$) and simulator cost (through $\gamma$), indicating that the recursive multi-fidelity structure does not introduce additional computational overhead.

\subsection{Experimental Design Construction for DNA Model}\label{sec:doe}

With the optimal sample allocation developed in the previous subsection, we propose a systematic procedure for constructing an experimental design for the DNA model. Our goal is to specify both the tuning parameters and the number of fidelity levels $L$ using  the optimal sample allocation.

In many realistic settings, the coarsest and finest tuning parameters, $t_1$ and $t_L$, are constrained by computational feasibility and are therefore prespecified. Given these endpoints, we assume that the intermediate tuning parameters follow a geometric sequence,
$t_l = T^{-1}t_{l-1}$ with $T>1$, as is commonly adopted in the multilevel literature \citep{giles2008multilevel,giles2015multilevel,oberkampf2025verification}. This implies
$t_l=t_1T^{-(l-1)}$ and $T=\left(t_1/t_L\right)^{1/(L-1)}$.  Suppose that the computational cost at level $l$ satisfies
$C_l=c_1t_l^{-\gamma}$ for some constants $c_1>0$ and $\gamma>0$. Let $a=d/(\eta+d)$ and $b=1-a$.
Then, the optimal allocation in \eqref{optimal_sample_size} simplifies to
\begin{equation}\label{eq:new_allocation}
n_l
=
\frac{C_{\rm total} t_1^\gamma}{c_1}
\frac{
\lambda_{\max}^{a(L-l+1)}
\left(t_1/t_L\right)^{-\frac{\gamma a(l-1)}{L-1}}
}{
\sum_{k=1}^L
\lambda_{\max}^{a(L-k+1)}
\left(t_1/t_L\right)^{\frac{\gamma b(k-1)}{L-1}}
}.
\end{equation}
To determine the number of fidelity levels, we impose a minimum sample size requirement $n_{\min}$ at each level and choose the largest $L$ such that $\min_{1\le l\le L} n_l \ge n_{\min}$.

The resulting procedure is summarized as follows:
\begin{itemize}
    \item Step 1: Specify $t_1$, $t_L$, $n_{\min}$, $C_{\rm total}$, $\lambda_{\max}$, $\eta$, $c_1$, and $\gamma$.
    \item Step 2: For each candidate $L\geq 2$, compute $T=\left(t_1/t_L\right)^{1/(L-1)}$
    and obtain the sample sizes $n_1,\ldots,n_L$ using \eqref{eq:new_allocation}.
    \item Step 3: Select the largest $L$ such that $\min_{1\le l\le L} n_l \ge n_{\min}$.
    \item Step 4: For each level $l=1,\ldots,L$, generate space-filling design points (e.g., Latin hypercube samples) with sample size $n_l$ over $\Omega$, using either a nested or non-nested design.
\end{itemize}

In practice, the quantities required for the design can be specified using a combination of prior knowledge, computational constraints, and simple pilot studies. The cost parameters $c_1$ and $\gamma$ can be estimated empirically by running the simulator at a few different fidelity levels and fitting a log--log regression of computational cost against the tuning parameter $t_l$ \citep{boutelet2025active}. The parameters $\lambda_{\max}$ and $\eta$ characterize the dependence across fidelity levels and the smoothness of the underlying function, respectively. When prior knowledge is unavailable, they can be set to reasonable default values (e.g., $\lambda_{\max}\approx 1$ and $\eta=4$) and refined if necessary through sensitivity analysis. Alternatively, preliminary runs can be used to estimate these quantities from data.

\begin{figure}[ht]
    \centering
    \includegraphics[width=0.9\linewidth]{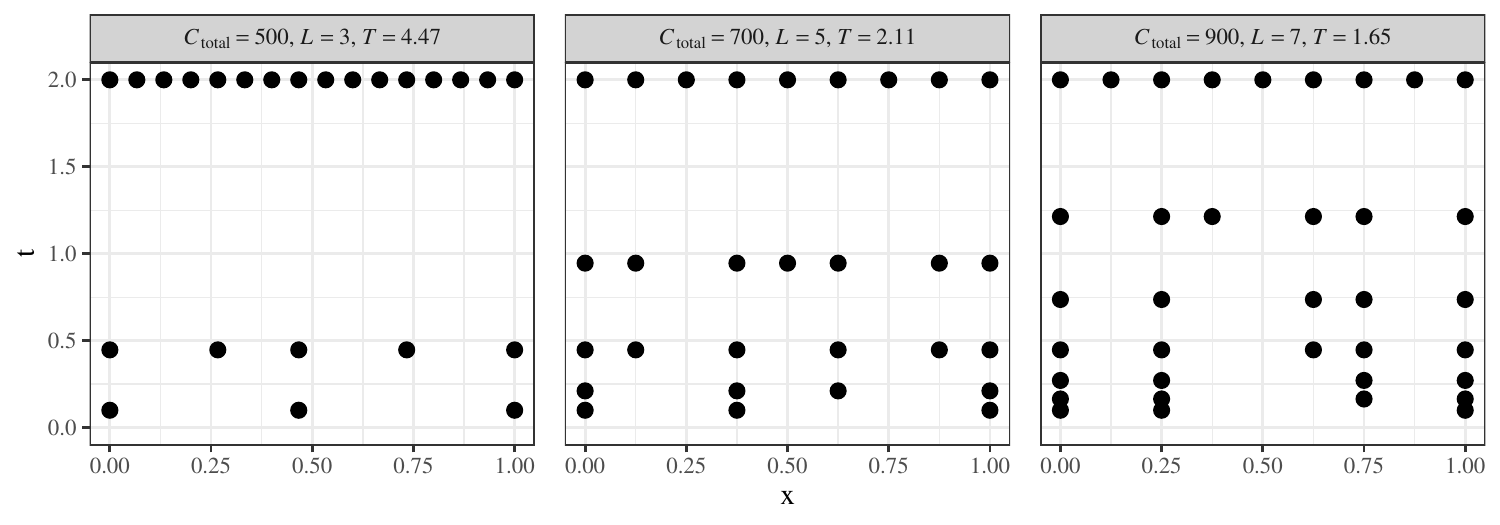}
    \caption{Experimental designs under increasing computational budgets with $n_{\min}=3$. The $x$-axis corresponds to the input $x \in [0,1]$, while the $y$-axis represents the tuning parameter $t$ controlling fidelity. The title of each panel reports the total cost $C_{\rm total}$, the number of fidelity levels $L$, and the scaling factor $T$.}
    \label{fig:Different_B}
\end{figure}

Figure~\ref{fig:Different_B} illustrates the resulting experimental designs under increasing total computational budget $C_{\rm total}$, with $t_1=2$, $t_L=0.1$, $n_{\min}=3$, $\lambda_{\max}=1,\eta=4,c_1=3.5$ and $\gamma=1.6$. A nested space-filling design is employed across fidelity levels. As $C_{\rm total}$ increases from $500$ to $900$, the procedure selects a larger number of fidelity levels, increasing $L$ from $3$ to $7$. Because $t_1$ and $t_L$ are fixed, accommodating additional levels requires reducing the scaling factor $T$ from $4.47$ to $1.65$. Consequently, the resulting tuning parameters become more densely spaced across fidelity levels. This behavior reflects the trade-off between fidelity resolution and sample allocation, where an increased budget allows for finer discretization of the fidelity space.

\section{Numerical Studies}\label{sec:studies}
In this section, we evaluate the proposed model on four synthetic functions, each designed to test a different form of multi-fidelity dependence. The functions are defined as follows:
\begin{align*}
\begin{cases}
 &f(t,x) = \exp{(-1.4x)}\cos{(3.5\pi x)}+t^2\sin(40x)/10, \quad x\in [0,1], \\
 &f(t,x) = \sin\left(\frac{10\pi x}{5+t}\right)+0.2 \sin(8\pi x), \quad x\in [0,1], \\
 &f(t,\mathbf{x}) = \left[\exp(-4t)-\exp\left(-\frac{1}{2x_2}\right)\right] \frac{2300x_1^3+1900x_1^2+2092x_1+60}{100x_1^3+500x_1^2+4x_1+20}, \quad \mathbf{x}=(x_1,x_2)\in [0,1]^2. \\
 &f(t,\mathbf{x}) = \frac{(2\pi-t^2) T_u (H_u-H_l)}{\log(r/r_w)\left(1+t^2+\frac{2LT_u}{\log(r/r_w)r_w^2K_w}+\frac{T_u}{T_l}\right)}, 
\end{cases}  
\end{align*}  
where $r_w\in[0.05, 0.15]$, $r\in[100, 50000]$, $T_u\in[63070, 115600]$, $H_u\in[990, 1110]$, $T_l\in[63.1, 116]$, $H_l\in[700, 820]$, $L\in[1120, 1680]$, and $K_w\in[9855, 12045]$.
The first one-dimensional function is from \cite{tuo2014surrogate}, which exhibits an additive structure (referred to as the additive function) and yields the exact solution $f(0,x)=\exp{(-1.4x)}\cos{(3.5\pi x)}$. The second function is adapted from \cite{higdon2002space} (referred to as the non-additive function), where $t$ is in the denominator, showing a nonlinear relationship between different tuning parameters. The third synthetic example is the multi-fidelity Currin function \citep{currin1988bayesian} from \cite{sung2022stacking}. The last synthetic example is the multi-fidelity Borehole function adapted from \cite{xiong2013sequential}. 

The tuning parameter $t_l$ is chosen using a geometric refinement $t_l=T^{-1}t_{l-1}$ with $T>1$ and $t_1=2.5$. 
In the present numerical studies, we set $T=\frac{10}{7}$. The sample sizes are chosen following the optimal allocation strategy in Section~\ref{sec:theoretical}, using the recursive relation $n_{l-1}=\lceil T^{\gamma} n_l \rceil$ with $\gamma=2$. Additional results under different choices of $T$ and $\gamma$ are provided in Supplementary Material~\ref{supp:figures}. 
The input locations are generated from the nested space-filling design introduced by \cite{le2014recursive}, with the largest tuning parameter $t_1$ given in the last column of Table \ref{tab:numericaltable}, which also summarizes the sample sizes and input dimensions for each function.

\begin{table}[ht!]
\caption{Sample sizes $n_l$ and input dimension $d$ for each synthetic example. \label{tab:numericaltable}}
    \begin{tabular*}{\columnwidth}{@{\extracolsep\fill}llllllll@{\extracolsep\fill}}
\toprule
         &  $n_1$ &  $n_2$&  $n_3$&  $n_4$&  $n_5$&  $n_6$&  $d$\\
\midrule
         Additive &17& 8& 4& 2&1& &1\\
         Non-additive &17& 8& 4& 2&1& &1\\
         Currin &71&35&17&8&4&2&2 \\
         Borehole &71&35&17&8&4&2&8\\
         \bottomrule
    \end{tabular*}
\end{table}

We compare the predictive performance of the proposed model (labeled \texttt{DNAmf}) with four existing methods: two that aim to emulate the highest fidelity output $f(t_L,\mathbf{x})$, including the Recursive Non-Additive emulator (\texttt{RNAmf}) by \cite{heo2023active} and the auto-regressive model (\texttt{CoKriging}) by \cite{le2014recursive}, and the nonlinear auto-regressive GP model (\texttt{NARGP}) by \cite{perdikaris2017nonlinear}. The remaining two methods target the exact solution $f(0,\mathbf{x})$, namely the nonstationary GP model by \cite{tuo2014surrogate} using Brownian motion kernel function (\texttt{BM}), and its extension using fractional Brownian motion kernel function (\texttt{FBM}) proposed by \cite{boutelet2025active}. Due to structural limitations, the \texttt{RNAmf}, \texttt{CoKriging}, and \texttt{NARGP} models are restricted to using only three fidelity levels ($l=1,2,3$ for the additive, non-additive, and Borehole functions, and $l=3,4,5$ for the Currin function). 

The model performance is evaluated using the root-mean-square error (RMSE) and continuous rank probability score (CRPS) \citep{gneiting2007strictly}, which accounts for the predictive distribution (including the mean and variance), based on 100 test points uniformly sampled from the same input space. Lower RMSE and CRPS values indicate better model accuracy.

\begin{figure}[ht]
    \centering
    \includegraphics[width=\linewidth]{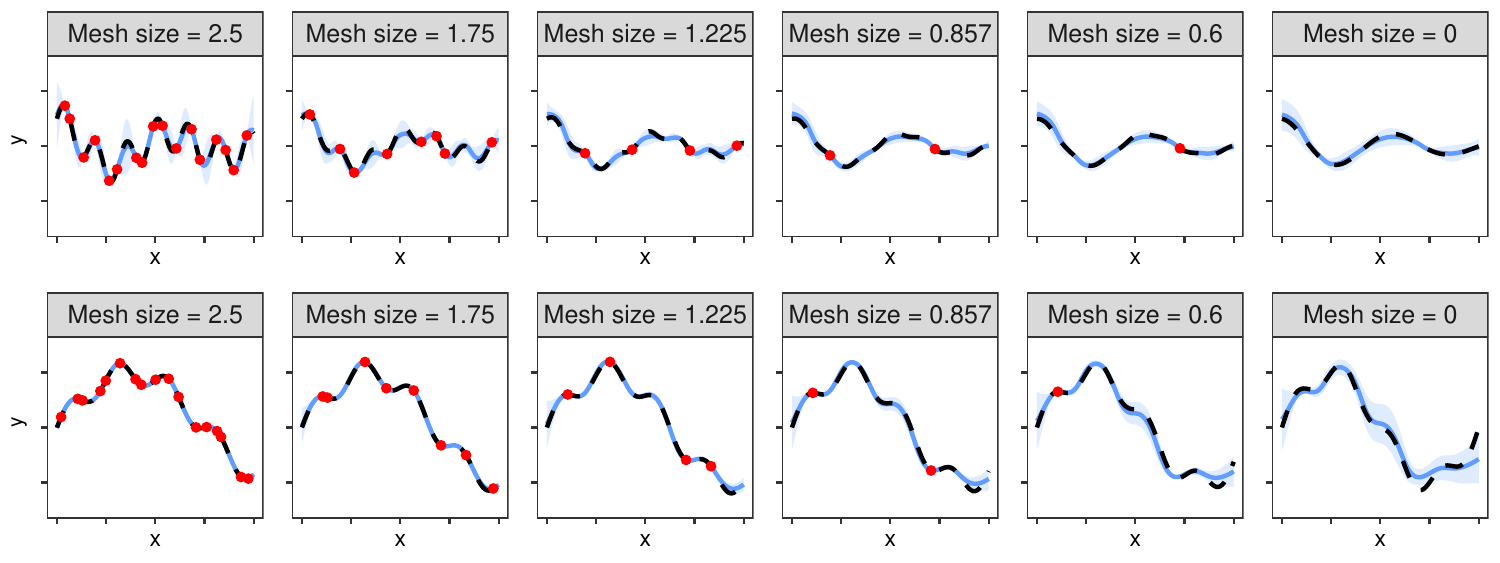}
    \caption{Illustrations of the additive function (top row) and the non-additive function (bottom row). Each row consists of subplots with tuning parameter values decreasing from large (left) to zero (right). In each subplot, the black dashed line represents the true function, red dots denote the design points, the blue line indicates the predicted function, and the shaded region depicts the 99\% confidence interval.}
    \label{fig:tuoillustration}
\end{figure}

Figure \ref{fig:tuoillustration} illustrates model predictions of the DNA model  along with confidence intervals for the additive and non-additive functions across different values of the tuning parameter $t$. By assuming that each fidelity level depends only on the immediately preceding level, the model effectively captures the evolving relationship as $t$ decreases to $0$. Notably, the model accurately predicts the function at $t=0$ despite the absence of design points at this value. The confidence intervals mostly contain the true function, demonstrating the effectiveness of the uncertainty quantification. This extrapolation capability demonstrates the model’s practical utility in situations where data at very small values of $t$ are unavailable, which reduces the need for costly data collection while maintaining predictive accuracy.

\begin{figure}[ht!]
    \centering
\includegraphics[width=\linewidth]{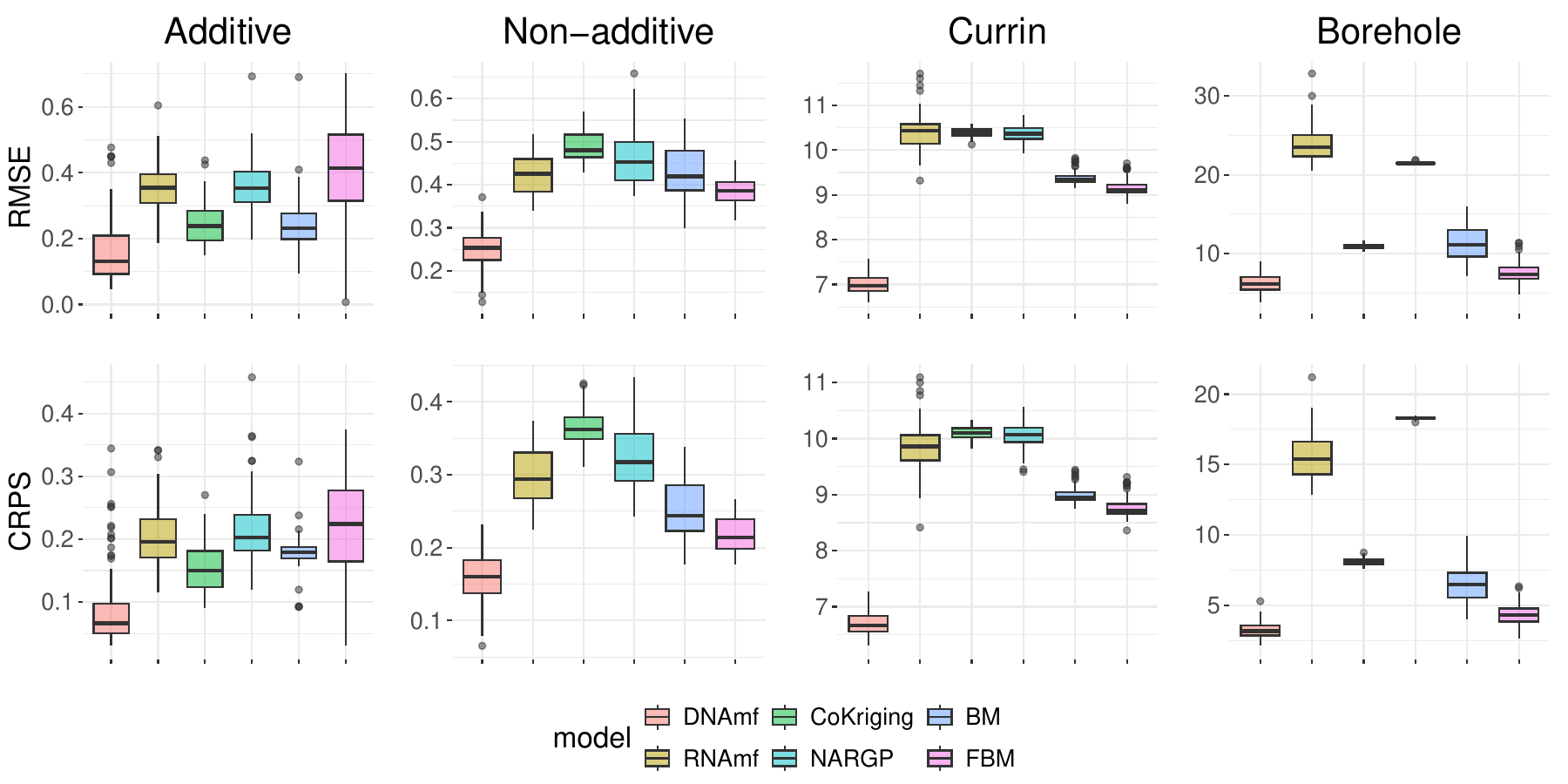} 
\caption{RMSEs and CRPSs of four synthetic examples across 100 repetitions.}
\label{fig:numerical_comparison}
\end{figure}

Figure \ref{fig:numerical_comparison} shows RMSE and CRPS results across 100 independent repetitions, each using randomly selected training input locations. The proposed \texttt{DNAmf} method consistently outperforms all four competing methods in both metrics. Even for the additive function, \texttt{DNAmf} still achieves superior performance. For the non-additive and Currin functions, where the tuning parameter $t$ influences the output nonlinearly, the model's flexible structure leads to substantially greater improvements. For the Borehole function, \texttt{DNAmf} and \texttt{FBM} achieve the best performance among all methods, demonstrating the effectiveness of \texttt{DNAmf} in high-dimensional settings.

Notably, \texttt{RNAmf}, \texttt{CoKriging}, and \texttt{NARGP} show relatively suboptimal performance. This is because they fit separate GPs at each fidelity level, relying only on the limited design points available at that level. As a result, they cannot utilize the highest fidelity level in the additive and non-additive functions, where only a single output is available. Furthermore, these methods focus solely on emulating the highest fidelity output $f(t_L, \mathbf{x})$, without extrapolating to $f(0, \mathbf{x})$. In contrast, \texttt{DNAmf}, \texttt{BM}, and \texttt{FBM} leverage all available data across fidelity levels to build a unified emulator capable of extrapolation. This ability to pool information leads to better predictive performance, particularly when each level has a limited number of design points.

The proposed model also offers interpretability through the parameter $\beta$, which captures the strength of interaction between $\mathbf{x}$ and $t$. For the non-additive function, a stronger coupling is observed, reflected in a higher average estimated interaction parameter $\hat{\beta}=0.007$, in contrast to $\hat{\beta}=0$ for the additive, Currin, and Borehole function. 

\section{Real Applications}\label{sec:real}
In this section, we present three case studies to evaluate the predictive performance of the proposed approach in scenarios where obtaining an exact solution within a finite time or cost is infeasible or difficult. Simulations are conducted at five fidelity levels, corresponding to five different mesh resolutions. As in Section~\ref{sec:studies}, the designs are generated using the nested space-filling design of \cite{le2014recursive}. 

The three case studies are introduced below:
\begin{itemize}
    \item \textbf{Poisson's Equation}: Elliptic PDEs commonly appear in the modeling of many physical phenomena. Specifically, we examine Poisson's equation \citep{evans2010partial, tuo2014surrogate, sung2022stacking} on the square domain $\Omega = [0,1]^2$, defined as follows:
\begin{align*}
    \Delta u = \left( x^2 - 2\pi^2 \right) e^{x s_1} \sin{ \left( \pi s_1 \right) } \sin{ \left( \pi s_2 \right) } + 2x\pi e^{x s_1} \cos{ \left( \pi s_1 \right) } \sin{ \left( \pi s_2 \right) }, \quad \mathbf{s} = \left( s_1, s_2 \right) \in \Omega,
\end{align*}
where $x \in \left[ -1, 1 \right]$ is the one-dimensional input variable, $\Delta = \frac{\partial^2}{\partial x^2} + \frac{\partial^2}{\partial y^2}$ is the Laplace operator, and $\Omega$ represents a square membrane. The Dirichlet boundary condition $u=0$ is imposed on the boundary $\partial \Omega$. Our quantity of interest is the maximum of $u$ over $\Omega$, which captures the highest stress or displacement of the membrane. Poisson's equation admits the known analytical solution at any given $x \in [-1,1]$ and point $\left( s_1, s_2 \right) \in \Omega$ as $u(s_1, s_2; x)=e^{x s_1} \sin{ \left( \pi s_1 \right) } \sin{ \left( \pi s_2 \right) }$, which allows us to evaluate the performance of the model in approximating the exact solution $f(0,x)$. For each combination of $x$ and the mesh size $t$ 
, we solve the PDE via finite element simulations using the \textsf{Partial Differential Equation Toolbox} of \textsf{MATLAB} \citep{MATLAB:R2021b}, as illustrated in Figure \ref{fig:Poisson_visualization}. 
    \item \textbf{Vibration of Square Plate}: The second case study focuses on the fourth natural frequencies in hertz (Hz) of a square elastic plate of size $10 \times 10 \times 1$ \citep{li2020multi}. Three material properties serve as input variables: Young's modulus ($x_1 \in [1 \times 10^{11}, 4 \times 10^{11}]$), Poisson's ratio ($x_2 \in [0.2, 0.4]$), and mass density ($x_3 \in [6 \times 10^3, 9 \times 10^3]$). Fidelity is again controlled by the finite-element mesh size $t$
    . Because an analytical solution for the plate's frequencies at infinite resolution is unavailable and high-resolution simulations are computationally expensive, we treat $t=0.3$ as the finest mesh for evaluating model performance. Simulations at this resolution are conducted at 100 uniform test input locations, with each run taking up to 8 minutes. The frequencies are computed via FEM using \textsf{MATLAB}.
    \item \textbf{Heat Equation}: In the multi-fidelity framework, a time-dependent parabolic PDE can be viewed as a hierarchy of models indexed by their temporal proximity to the target time $S$. We consider a one-dimensional heat equation, a canonical instance of the diffusion equation:
\[ \frac{\partial u(x,s)}{\partial s} = D \frac{\partial^2 u(x,s)}{\partial x^2},\]
where $u(x,s)$ denotes the temperature at spatial coordinate $x \in [0,L]$ and time $s \in [0,S]$, and $D$ denotes the thermal diffusivity coefficient. As $s$ approaches the target time $S$, the solution generally exhibits increasing complexity both in its spatial gradients and in the coupling between space and time. Consequently, direct integration from $s=0$ to $s=S$ on a fine time grid can become prohibitive. 
To address this, we introduce the tuning parameter $t=S-s$ that measures the temporal distance from the simulation time $s$ to the target time $S$. Early time-step solutions (i.e., larger $t$, smaller $s$) provide inexpensive, low-fidelity information, while those obtained closer to the target time (i.e., smaller $t$, larger  $s$) yield high-fidelity results at greater cost. 
To evaluate model performance, we compute the exact solution at $t=0$ at 100 uniform test input locations. 
\end{itemize}

As in Section \ref{sec:studies}, model performance is evaluated using the RMSE and CRPS on 100 test points with 100 repetitions. The sample sizes, input dimensions, the largest value of the tuning parameters, and constant ratio $c$ for each case study are summarized in Table \ref{tab:realtable}. These mesh sizes and sample sizes are chosen in a similar fashion as in Section \ref{sec:studies}.

\begin{table}[ht!]
    \caption{Sample sizes $n_l$, input dimension $d$, the largest tuning parameter $t_1$, and the constant $T$ that determines the other tuning parameters, $t_l=T^{-1}t_{l-1}$, for each case study. \label{tab:realtable}}
    \begin{tabular*}{\columnwidth}{@{\extracolsep\fill}lllllllll@{\extracolsep\fill}}
\toprule
        & $n_1$ & $n_2$& $n_3$& $n_4$& $n_5$& $d$ &$t_1$ & $T^{-1}$ \\
\midrule
         Poisson &17& 11& 7& 5&3&1& 0.1&0.65\\
         Plate &28& 21& 15& 11&8&3& 0.55&0.9\\
         Heat &25&15&9&5&3&1&0.5&0.7\\
\bottomrule
    \end{tabular*}
\end{table}

\begin{figure}[ht!]
    \centering
\includegraphics[width=\linewidth]{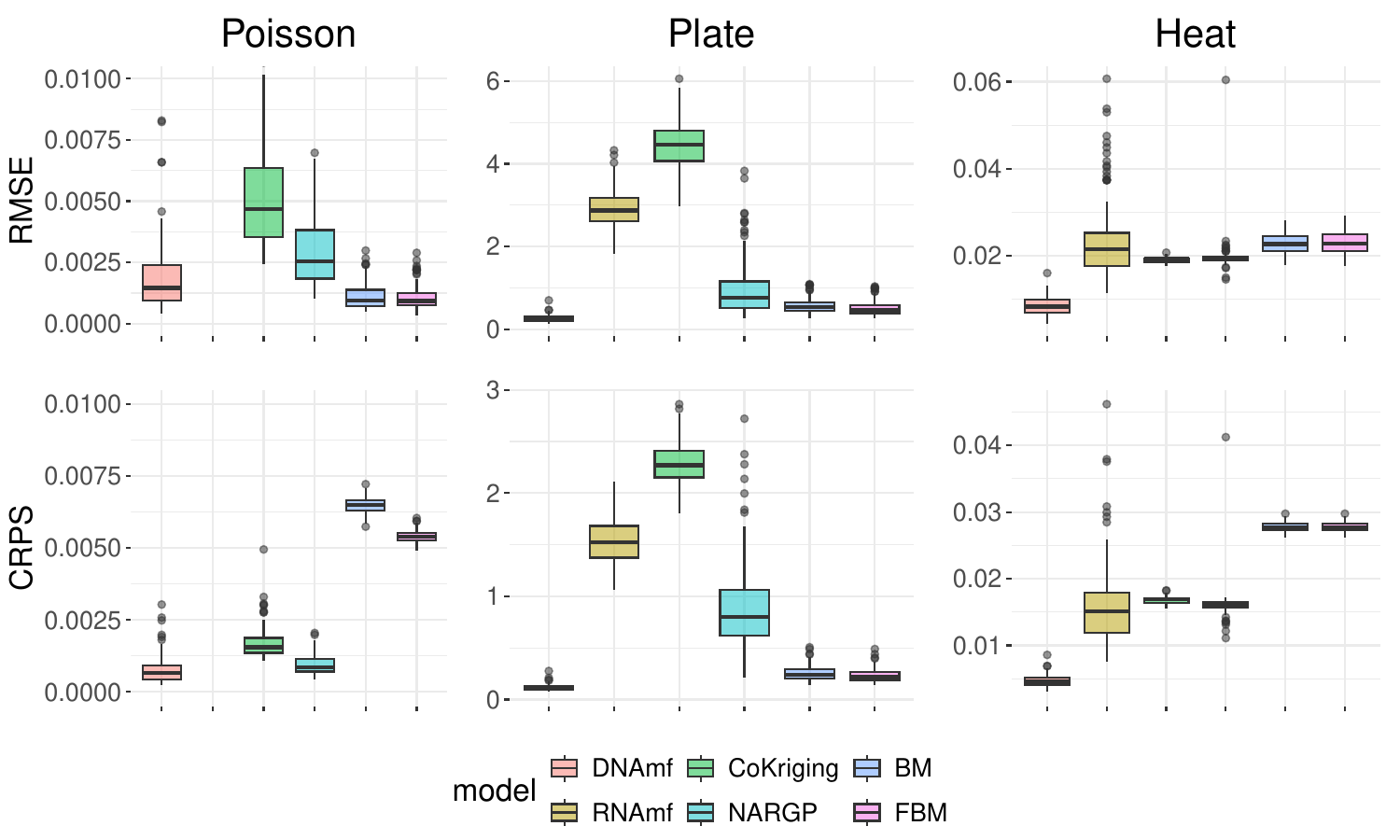} 
\caption{RMSEs and CRPSs of three real case studies across 100 repetitions.}
\label{fig:real_comparison}
\end{figure}
As shown in Figure \ref{fig:real_comparison}, \texttt{DNAmf} generally outperforms all other competitors across all case studies, with the exception of RMSE in the Poisson equation, where it is slightly less accurate than both \texttt{FBM} and \texttt{BM}. However, \texttt{DNAmf} achieves better CRPS in that case, indicating improved uncertainty quantification. Notably, \texttt{RNAmf} performs considerably worse for the Poisson equation, with both RMSE and CRPS values exceeding the displayed range in Figure~\ref{fig:real_comparison}. This degraded performance likely stems from its structural limitation of fitting separate GPs at each fidelity level using relatively few design points, which restricts its ability to effectively capture cross-level dependencies. The estimated interaction parameter $\hat{\beta}$ and the visualizations in Figure~\ref{fig:realillustration} 
together highlight differences in coupling behavior. For both the Poisson and Plate problems, $\hat{\beta} \approx 0$, suggesting that the kernel becomes effectively separable. This indicates that the influence of the input variables remains nearly constant across different values of the tuning parameter $t$, which is consistent with the top row of Figure~\ref{fig:realillustration},
where model predictions and confidence intervals show little variation over $t$. In contrast, for the Heat equation, $\hat{\beta} \approx 1$, revealing the pronounced interaction between inputs and the fidelity level. This behavior is also evident in the bottom row of Figure~\ref{fig:realillustration}, 
where the predictive results vary substantially with $t$, clearly reflecting the strong interaction between the input variables and tuning parameter captured by the nonseparable kernel.

\section{DNA Model with Non-Nested Design}\label{sec:nonnested}

While the nested structure in \eqref{eq:nested} enables efficient implementation of the DNA model, it is not always available in practice. We extend the framework to accommodate non-nested designs while preserving computational efficiency, building on the developments in Sections~\ref{sec:DNA} and \ref{sec:DNAposterior}.

To this end, we adopt the stochastic imputation approach proposed by \cite{suppming2023}. Specifically, we impute \textit{pseudo} outputs at selected \textit{pseudo} inputs to construct an artificial nested design, which allows us to retain the computational advantages of the DNA model. Due to space limitations, detailed derivations and algorithmic steps are provided in Supplementary Material~\ref{supp:non-nested}.

Figure \ref{fig:nonnested} illustrates the prediction results of the DNA model under a non-nested design for the non-additive example. Compared to the nested design (see Figure \ref{fig:tuoillustration}), the non-nested approach generally exhibits increased predictive uncertainty, which is expected due to the additional variability introduced by imputing pseudo-outputs rather than observing true outputs. Overall, the model still performs well: the posterior mean closely aligns with the truth function, and the confidence intervals successfully cover the true function.

\begin{figure}[ht!]
    \centering
    \includegraphics[width=\linewidth]{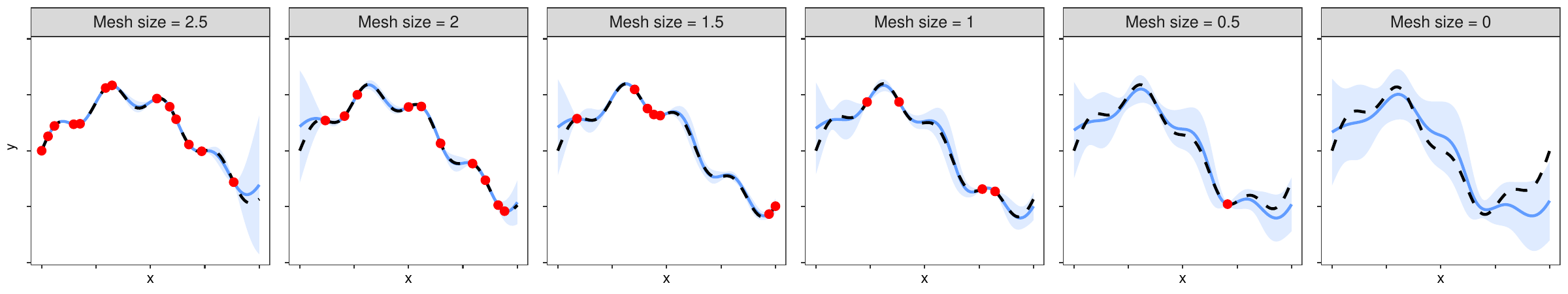}
    \caption{Illustration of the non-additive function in Section \ref{sec:studies} under a non-nested design.}
    \label{fig:nonnested}
\end{figure}


\section{Conclusion and Discussion}\label{sec:conclusion}


We introduce the Diffusion Non-Additive (DNA) model for multi-fidelity simulations with tunable fidelity parameters, offering a flexible framework for modeling complex dependencies across fidelities. By relaxing the additive structure and allowing for a coupling between the fidelity-controlling parameter and the input space, the DNA model can adapt to a wide range of simulation behaviors. Moreover, we establish rigorous convergence error bounds for the DNA model. These theoretical guarantees, together with explicit cost–accuracy complexity results, directly enable an experimental design strategy that optimally allocates finite computational resources across different fidelity levels. Through comprehensive numerical studies, including synthetic functions and PDE-based simulators, we demonstrate that the DNA model achieves strong predictive performance and reliable uncertainty quantification. 

A promising direction for future research is to extend the DNA model to accommodate high-dimensional outputs, which are common in finite element simulations that return spatial fields or outputs over many nodes. In such settings, direct emulation of each output dimension is often computationally infeasible and statistically inefficient. While various approaches have been proposed for high-dimensional outputs, such as principal component decomposition \citep{higdon2008computer}, wavelet-based approach \citep{bayarri2007computer}, P-splines \citep{williamson2012fast}, and optimal basis selection \citep{salter2019uncertainty}, their extensions to the multi-fidelity framework remain scarce. Integrating these approaches within the DNA framework offers a valuable opportunity to achieve scalable emulation of high-dimensional outputs in a multi-fidelity setting.
\\


\noindent
\textbf{Funding}: The authors gratefully acknowledge funding from NSF DMS 2338018.

\noindent
\textbf{Supplementary Materials}:
Additional supporting materials can be found in the Supplementary Materials, including the theoretical proofs, the detailed algorithm, and the supporting figures. 

\noindent
\textbf{Data Availability Statement}:
No real-world observational data were used in this study. All numerical experiments are based on synthetic data generated from computer simulations. The proposed method is implemented in the R package \texttt{DNAmf}  on CRAN. The code used to generate the numerical results is publicly available at https://github.com/heojunoh/DNAmf-Reproducibility.

\bibliography{bib}

\newpage
\setcounter{page}{1}
\bigskip
\bigskip
\bigskip
\begin{center}
{\Large\bf Supplementary Materials for ``Diffusion Non-Additive Model for Multi-Fidelity Simulations with Tunable Precision''}\\
\if1\blind\vspace{5mm}
\setcounter{footnote}{0}
 \renewcommand{\thefootnote}{\fnsymbol{footnote}}
  \centering{Junoh Heo$^{1}$, Romain Boutelet$^{1}$, Wenjia Wang$^{2}$, and Chih-Li Sung$^{1,}$\footnote{Corresponding author. Address for correspondence: Chih-Li Sung, Department of Statistics and Probability, Michigan State University, East Lansing, MI 48824, USA. Email: \texttt{sungchih@msu.edu}}\vspace{0in}\\
        $^{1}$Department of Statistics and Probability, Michigan State University\\
        $^{2}$Department of Industrial Systems Engineering and Management, National University of Singapore\\
        }\fi
\end{center}

\setcounter{section}{0}
\setcounter{equation}{0}
\def\theequation{S\arabic{section}.\arabic{equation}}
\def\thesection{S\arabic{section}}
\def\thefigure{S\arabic{figure}}
\def\thetable{S\arabic{table}}

\section{Proof of Remark~\ref{thm:conditionnumber}
}\label{proofcondition}

Let $\mathbf{K}_{\mathcal{X}} \in \mathbb{R}^{n \times n}$ denote the kernel matrix over $\mathcal{X}$ with entries 
\[
[\mathbf{K}_{\mathcal{X}}]_{ij} = \exp\left(-\sum_{k=1}^d \frac{(x_{ik} - x_{jk})^2}{\theta_j}\right).
\] 
Suppose that the input locations in $\mathcal{X}$ are distinct,  then $\mathbf{K}_{\mathcal{X}}$ is strictly positive definite and thus full rank. 
The full kernel matrix across fidelity levels $t_2, \ldots, t_L$ can be expressed as
\[
\mathbf{K}_\delta = \mathbf{G}_\delta \otimes \mathbf{K}_{\mathcal{X}},
\]
where $\otimes$ denotes the Kronecker product, and $\mathbf{G}_\delta \in \mathbb{R}^{{L-1} \times {L-1}}$ has entries
\[
[\mathbf{G}_\delta]_{ij} = \left( \frac{(t_i - t_j)^2}{\theta_t} + 1 \right)^{-\delta}, \quad \text{for } 2 \le i, j \le L.
\]

When $\delta = 0$, each entry of $\mathbf{G}_0$ equals to 1, i.e., $\mathbf{G}_0= \mathbf{1}_{L-1} \mathbf{1}_{L-1}^\top$, which is a rank-one positive semi-definite matrix. Consequently,
\[
\mathbf{K}_0 = \mathbf{G}_0 \otimes \mathbf{K}_{\mathcal{X}} = (\mathbf{1}_{L-1} \mathbf{1}_{L-1}^\top) \otimes \mathbf{K}_{\mathcal{X}},
\]
is positive semi-definite but has rank only $1 \times n = n$, despite being of size $n(L-1) \times n(L-1)$. Since $\mathbf{K}_{\mathcal{X}} \succ 0$ and $\mathbf{G}_0 \succeq 0$, we conclude that $\mathbf{K}_0$ is positive semi-definite but singular, with $\lambda_{\min}(\mathbf{K}_0) = 0$, and condition number $\kappa(\mathbf{K}_0) = \infty$.

When $\delta > 0$, the function $f_\delta(r) = (r+1)^{-\delta}$ is completely monotonic on $[0,\infty)$. By Schoenberg’s theorem \citep{schoenberg1938metric}, the matrix $\mathbf{G}_\delta$ with entries $f_\delta((t_i - t_j)^2/\theta_t)$ is strictly positive definite, since the tuning parameters $t_2, \ldots, t_L$ are distinct.  Given that  $\mathbf{K}_{\mathcal{X}} \succ 0$, the Kronecker product $\mathbf{K}_\delta = \mathbf{G}_\delta \otimes \mathbf{K}_{\mathcal{X}}$ is also strictly positive definite. Consequently, $\lambda_{\min}(\mathbf{K}_\delta) > 0$. Moreover, since $\mathbf{K}_\delta$ is symmetric and finite-dimensional with size $n(L-1) \times n(L-1)$, all its eigenvalues are real and bounded, implying $\lambda_{\max}(\mathbf{K}_\delta) < \infty$. It follows that $\kappa(\mathbf{K}_\delta) < \infty$.

Thus, for any $\delta > 0$, we conclude that $\kappa(\mathbf{K}_\delta) < \kappa(\mathbf{K}_0)$.

\section{The gradient of the log-likelihood}\label{supp:gradient}

The gradient of the log-likelihood function \eqref{eq:likelihood}
with respect to any parameter $\eta \in (\left\{\theta_j\right\}_{j=1}^d,\theta_y, \theta_t, \beta,\delta)$ is given by
\begin{align*}
-\frac{1}{2} \frac{\partial \log(\det(\mathbf{K}))}{\partial \eta}+ \frac{N_{-1}}{2}\frac{(Y_{-1} -  \hat{\alpha}\mathbf{1}_{N_{-1}})^T \mathbf{K}^{-1} \frac{\partial \mathbf{K}^{-1}}{\partial \eta} \mathbf{K}^{-1} (Y_{-1} - \hat{\alpha} \mathbf{1}_{N_{-1}})}{(Y_{-1} -  \hat{\alpha}\mathbf{1}_{N_{-1}})^T \mathbf{K}^{-1} (Y_{-1} - \hat{\alpha} \mathbf{1}_{N_{-1}})}.
\end{align*}
To compute the partial derivatives, we use
$$\frac{\partial \mathbf{K}^{-1}}{\partial \eta} = -\mathbf{K}^{-1} \frac{\partial \mathbf{K}}{\partial \eta} \mathbf{K}^{-1},\frac{\partial \log(\det(\mathbf{K}))}{\partial \eta}=\text{tr}\left(\mathbf{K}^{-1} \frac{\partial \mathbf{K}}{\partial \eta}\right).$$ Hence, it remains to compute $\frac{\partial \mathbf{K}}{\partial \eta}$.  For the nonseparable squared exponential kernel in \eqref{eq:Gneiting4},  we have the following expressions:
\begin{align*} 
\frac{\partial \mathbf{K}}{\partial \theta_j} &= -\frac{||x_j - x_j'||^2}{\theta_j^2} \frac{\partial \mathbf{K}}{\partial v} , \quad
\frac{\partial \mathbf{K}}{\partial \theta_y} = -\frac{||y - y'||^2}{\theta_y^2} \frac{\partial \mathbf{K}}{\partial v} ,\\   
\frac{\partial \mathbf{K}}{\partial \theta_t} &= \frac{(t-t')^2}{\theta_t^2} u^2 \frac{\partial \mathbf{K}}{\partial u}, \\   
\frac{\partial \mathbf{K}}{\partial \beta} &= \mathbf{K} \log u \left( \frac{d+1}{2} - u^{\beta}v \right), \\   
\frac{\partial \mathbf{K}}{\partial \delta} &= \mathbf{K} \log u , 
\end{align*}
where $u=\left( 1+\frac{(t-t')^2}{\theta_t} \right)^{-1}$, $v=\frac{||y - y'||^2}{\theta_y}+\sum_{j=1}^d\frac{||x_j - x_j'||^2}{\theta_j}$, 
and 
\begin{align*}
\frac{\partial \mathbf{K}}{\partial u} &= \left(\frac{d+1}{2}\beta + \delta -\beta u^{\beta}v \right) u^{\frac{d+1}{2}\beta + \delta -1} \exp(-u^\beta v),\\  
\frac{\partial \mathbf{K}}{\partial v} &= -u^{\frac{d+3}{2}\beta + \delta} \exp(-u^\beta v). 
\end{align*}

For the nonseparable Mat\'ern kernel (introduced in Section~\ref{app:matern}) with $\nu=1.5$, we have
\begin{align*}
\frac{\partial \mathbf{K}}{\partial \theta_j} &= -\frac{v_j}{\theta_j} \frac{\partial \mathbf{K}}{\partial v_j} , \quad \frac{\partial \mathbf{K}}{\partial \theta_y} = -\frac{v_y}{\theta_y} \frac{\partial \mathbf{K}}{\partial v_y}\\
\frac{\partial \mathbf{K}}{\partial \theta_t} &= \frac{(t-t')^2}{\theta_t^2} u^2 \frac{\partial \mathbf{K}}{\partial u}, \\   
\frac{\partial \mathbf{K}}{\partial \beta} &= \mathbf{K} \frac{\log u}{2} \left((d+1)-\left(\sum_{j=1}^d \frac{(u^{\frac{\beta}{2}}v_j)^2}{1+u^{\frac{\beta}{2}}v_j}+\frac{(u^{\frac{\beta}{2}}v_y)^2}{1+u^{\frac{\beta}{2}}v_y}\right) \right), \\   
\frac{\partial \mathbf{K}}{\partial \delta} &= \mathbf{K} \log u, 
\end{align*}
where $u=\left( 1+\frac{(t-t')^2}{\theta_t} \right)^{-1}$, $v_j=\frac{\sqrt{3}||x_j - x_j'||}{\theta_j}$, $v_y=\frac{\sqrt{3}||y - y'||}{\theta_y}$, and
\begin{align*}
\frac{\partial \mathbf{K}}{\partial u} &= \mathbf{K} \left( \frac{\frac{(d+1)}{2}\beta+\delta}{u} - \frac{\beta}{2}u^{-1}\left(\sum_{j=1}^d\frac{(u^{\frac{\beta}{2}}v_j)^2}{1+u^{\frac{\beta}{2}}v_j}+\frac{(u^{\frac{\beta}{2}}v_y)^2}{1+u^{\frac{\beta}{2}}v_y}\right) \right) ,\\  
\frac{\partial \mathbf{K}}{\partial v_j} &= -\mathbf{K} \frac{u^\beta v_j}{1+u^{\frac{\beta}{2}}v_j} , \quad \frac{\partial \mathbf{K}}{\partial v_y} = -\mathbf{K} \frac{u^\beta v_y}{1+u^{\frac{\beta}{2}}v_y}.
\end{align*}

For the nonseparable Mat\'ern kernel with $\nu=2.5$, we have 
\begin{align*}
\frac{\partial \mathbf{K}}{\partial \theta_j} &= -\frac{v_j}{\theta_j} \frac{\partial \mathbf{K}}{\partial v_j} , \quad
\frac{\partial \mathbf{K}}{\partial \theta_y} = -\frac{v_y}{\theta_y} \frac{\partial \mathbf{K}}{\partial v_y} ,\\    
\frac{\partial \mathbf{K}}{\partial \theta_t} &= \frac{(t-t')^2}{\theta_t^2} u^2 \frac{\partial \mathbf{K}}{\partial u}, \\   
\frac{\partial \mathbf{K}}{\partial \beta} &= \mathbf{K} \frac{\log u}{2} \left((d+1)-\left(\sum_{j=1}^d \frac{(u^{\frac{\beta}{2}}v_j)^2(1+u^{\frac{\beta}{2}}v_j)}{3(1+u^{\frac{\beta}{2}}v_j+\frac{(u^{\frac{\beta}{2}}v_j)^2}{3})}+\frac{(u^{\frac{\beta}{2}}v_y)^2(1+u^{\frac{\beta}{2}}v_y)}{3(1+u^{\frac{\beta}{2}}v_y+\frac{(u^{\frac{\beta}{2}}v_y)^2}{3})} \right)\right), \\   
\frac{\partial \mathbf{K}}{\partial \delta} &= \mathbf{K} \log u, 
\end{align*}
where $u=\left( 1+\frac{(t-t')^2}{\theta_t} \right)^{-1}$, $v_j=\frac{\sqrt{5}||x_j - x_j'||}{\theta_j}$, $v_y=\frac{\sqrt{5}||y - y'||}{\theta_y}$, and 
\begin{align*}
\frac{\partial \mathbf{K}}{\partial u} &= \mathbf{K} \left( \frac{\frac{(d+1)}{2}\beta+\delta}{u} - \frac{\beta}{2}u^{-1}\left(\sum_{j=1}^d\frac{(u^{\frac{\beta}{2}}v_j)^2(1+u^{\frac{\beta}{2}}v_j)}{3(1+u^{\frac{\beta}{2}}v_j+\frac{(u^{\frac{\beta}{2}}v_j)^2}{3})}+ \frac{(u^{\frac{\beta}{2}}v_y)^2(1+u^{\frac{\beta}{2}}v_y)}{3(1+u^{\frac{\beta}{2}}v_y+\frac{(u^{\frac{\beta}{2}}v_y)^2}{3})} \right)\right) ,\\  
\frac{\partial \mathbf{K}}{\partial v_j} &= -\mathbf{K} \frac{u^{\frac{\beta}{2}}v_j(1+u^{\frac{\beta}{2}}v_j)}{3(1+u^{\frac{\beta}{2}}v_j+\frac{(u^{\frac{\beta}{2}}v_j)^2}{3})}u^{\frac{\beta}{2}} , \quad
\frac{\partial \mathbf{K}}{\partial v_y} = -\mathbf{K} \frac{u^{\frac{\beta}{2}}v_y(1+u^{\frac{\beta}{2}}v_y)}{3(1+u^{\frac{\beta}{2}}v_y+\frac{(u^{\frac{\beta}{2}}v_y)^2}{3})}u^{\frac{\beta}{2}}.
\end{align*}

\section{Posterior mean and variance of DNA model}\label{supp:sqexposterior}

The posterior mean and variance at the input $\mathbf{x}$ can be derived as follows,
\begin{align*}
\mu_l^*(\mathbf{x})&=\mathbb{E}[f_l(\mathbf{x})|Y_N]= \alpha + \mathbb{E} [ \mathbf{k}(t, \mathbf{x}, f_{l-1}(\mathbf{x}))^T  |Y_N] \mathbf{K}^{-1} (Y_{-1} - \alpha \mathbf{1}_{N_{-1}}) \\
&= \alpha + \sum^{N_{-1}}_{i=1} r_i c_i^{\frac{d+1}{2}+\frac{\delta}{\beta}} \prod_{j=1}^d \exp\left( -c_i\frac{(x_{j}-(\mathbf{X}_{-1})_{ij})^2}{\theta_{j}} \right)  \mathbb{E} \left[ \exp{\left\{ -c_i\frac{((Y_{-L})_i-f_{l-1}(\mathbf{x}))^2}{\theta_{y}} \right\} } \Big|Y_N \right] \\
&= \alpha + \sum^{N_{-1}}_{i=1} r_i c_i^{\frac{d+1}{2}+\frac{\delta}{\beta}} \prod_{j=1}^d \exp\left( -c_i\frac{(x_{j}-(\mathbf{X}_{-1})_{ij})^2}{\theta_{j}} \right) \sqrt{\frac{\theta_{y}}{\theta_{y}+2 \sigma^{*2}_{l-1}(\mathbf{x}) }}  \exp{\left( -c_i\frac{((Y_{-L})_i-\mu^*_{l-1}(\mathbf{x}))^2}{\theta_{y}+2\sigma^{*2}_{l-1}(\mathbf{x})} \right)} \\
&=\alpha + \sqrt{\frac{\theta_{y}}{\theta_{y}+2
\sigma^{*2}_{l-1}(\mathbf{x}) }}\sum^{N_{-1}}_{i=1} r_i c_i^{\frac{d+1}{2}+\frac{\delta}{\beta}} \exp\left( -c_i \left\{\sum_{j=1}^d\frac{(x_{j}-(\mathbf{X}_{-1})_{ij})^2}{\theta_{j}} + \frac{((Y_{-L})_i-\mu^*_{l-1}(\mathbf{x}))^2}{\theta_{y}+2\sigma^{*2}_{l-1}(\mathbf{x})} \right\}\right),
\end{align*}
and
\begin{align*}
\sigma^{*2}_l(\mathbf{x})&=\mathbb{V}[f_l(\mathbf{x})|Y_N]= \mathbb{V} \left[ \mathbb{E} [f_l(\mathbf{x}) | f_{l-1}(\mathbf{x}), Y_N ] \right] + \mathbb{E} \left[\mathbb{V} [f_l(\mathbf{x}) | f_{l-1}(\mathbf{x}), Y_N ] \right] \\
&= \mathbb{E} \left[ \left\{ \mathbb{E} \left[f_l(\mathbf{x}) | f_{l-1}(\mathbf{x}), Y_N \right] \right\}^2 \right] - \mu^{*}_l(\mathbf{x})^2 + \mathbb{E} \left[\mathbb{V} \left[f_l(\mathbf{x}) | f_{l-1}(\mathbf{x}), Y_N \right] \right] \\
&= \mathbb{E}\left[ \left\{\alpha + \mathbf{k}(t_l, \mathbf{x}, f_{l-1}(\mathbf{x}))^T   \mathbf{K}^{-1}  (Y_{-1} - \alpha \mathbf{1}_{N_{-1}}) \right\}^2 |Y_N\right] -\mu^{*}_l(\mathbf{x})^2  \\
&\quad\quad\quad\quad\quad\quad\quad+\mathbb{E} \left[ \tau^2 \left\{ 1 - \mathbf{k}(t_l, \mathbf{x}, f_{l-1}(\mathbf{x}))^T  \mathbf{K}^{-1}  \mathbf{k}(t_l, \mathbf{x}, f_{l-1}(\mathbf{x})) \right\} |Y_N \right]\\
&= \alpha^2 + 2\alpha  (\mu^*_l(\mathbf{x}) - \alpha) + \mathbb{E}\left[ \left\{ \mathbf{k}(t_l, \mathbf{x}, f_{l-1}(\mathbf{x}))^T  \mathbf{K}^{-1} (Y_{-1} - \alpha \mathbf{1}_{N_{-1}}) \right\}^2 |Y_N\right] \\
&\quad\quad\quad\quad\quad\quad\quad-\mu^{*}_l(\mathbf{x})^2 + \tau^2 - \tau^2 \mathbb{E} \left[ \left\{ \mathbf{k}(t_l, \mathbf{x}, f_{l-1}(\mathbf{x}))^T  \mathbf{K}^{-1}  \mathbf{k}(t_l, \mathbf{x}, f_{l-1}(\mathbf{x}))  \right\}  |Y_N\right]
\end{align*}
\begin{align}
&= \tau^2 - (\mu^*_l(\mathbf{x})-\alpha)^2 + \nonumber\\ &+\mathbb{E}\left[  \mathbf{k}(t_l, \mathbf{x}, f_{l-1}(\mathbf{x}))^T  \left\{ \mathbf{K}^{-1} (Y_{-1} - \alpha \mathbf{1}_{N_{-1}})(Y_{-1} - \alpha \mathbf{1}_{N_{-1}})^T \mathbf{K}^{-1} - \tau^2 \mathbf{K}^{-1} \right\}\mathbf{k}(t_l, \mathbf{x}, f_{l-1}(\mathbf{x})) |Y_N\right]\nonumber \\
&= \tau^2 - (\mu^*_l(\mathbf{x})-\alpha)^2 + \left( \sum_{i,k=1}^{N_{-1}} \left(r_i r_k - \tau^2 (\mathbf{K}^{-1})_{ik} \right) (c_ic_k)^{\frac{d+1}{2}+\frac{\delta}{\beta}} \right. \nonumber\\
&\quad\quad\quad\times \left. \exp{ \left(-\sum_{j=1}^d\frac{c_i(x_{j}-(\mathbf{X}_{-1})_{ij})^2+c_k(x_{j}-(\mathbf{X}_{-1})_{kj})^2}{\theta_{j}}  \right)} \right) \nonumber\\ 
&\quad\quad\quad\times \mathbb{E}\left[ \exp{ \left(-\frac{c_i(f_{l-1}(\mathbf{x})-(Y_{-L})_{i})^2+c_k(f_{l-1}(\mathbf{x})-(Y_{-L})_{k})^2}{\theta_{y}}  \right)} \Big|Y_N\right]  \label{eq:eq1} \\ 
&=\tau^2 - (\mu^*_l(\mathbf{x})-\alpha)^2 + \left( \sum_{i,k=1}^{N_{-1}} \zeta_{ik}(\mu^*_{l-1}(\mathbf{x}),\sigma^{*2}_{l-1}(\mathbf{x}))\left(r_i r_k - \tau^2 (\mathbf{K}^{-1})_{ik} \right) (c_ic_k)^{\frac{d+1}{2}+\frac{\delta}{\beta}} \right.\nonumber \\
&\quad\quad\quad\times \left. \exp{ \left(-\sum_{j=1}^d\frac{c_i(x_{j}-(\mathbf{X}_{-1})_{ij})^2+c_k(x_{j}-(\mathbf{X}_{-1})_{kj})^2}{\theta_{j}}  \right)} \right),  \nonumber
\end{align}
where $r_i = (\mathbf{K}^{-1} (Y_{-1}-\alpha \mathbf{1}_{N_{-1}}))_i$, $c_i=c(t,\mathbf{t}_i)=(\frac{(t-\mathbf{t}_i)^2}{\theta_t} + 1)^{-\beta}$ and 
\begin{align*}
\zeta_{ik}(\mu,\sigma^2) &= \sqrt{\frac{\theta_{y}}{\theta_{y}+2(c_i+c_k)\sigma^2}} \\
&\times \exp{\left( - \frac{c_i\left((Y_{-L})_i-\mu\right)^2+c_k\left((Y_{-L})_k-\mu\right)^2+\frac{2}{\theta_{y}}c_ic_k\sigma^2\left((Y_{-L})_i-(Y_{-L})_k\right)^2}{\theta_{y}+2(c_i+c_k)\sigma^2} \right)}.
\end{align*}

For a random variable $X\sim N(\mu, \sigma^2)$, we obtain the expression in \eqref{eq:eq1} as follows:
\begin{align*}
\frac{1}{\sqrt{2\pi\sigma^2}} &\int \exp{\left( -Ax^2 + Bx - C \right)} dx = \frac{1}{\sqrt{2\pi\sigma^2}} \int \exp{\left( -A\left(x-\frac{B}{2A}\right)^2 +\frac{B^2}{4A} - C \right)} dx \\
&= \frac{1}{\sqrt{2\pi\sigma^2}} \exp{\left(\frac{B^2}{4A} - C \right)} \int \exp{\left( -A\left(x-\frac{B}{2A}\right)^2 \right)} dx \\
&= \frac{1}{\sqrt{2\pi\sigma^2}} \exp{\left(\frac{B^2}{4A} - C \right)} \sqrt{\frac{\pi}{A}}\\
&= \frac{1}{\sqrt{2\sigma^2A}} \exp{\left(\frac{B^2}{4A} - C \right)}.
\end{align*}
Thus,
\begin{align*}
\mathbb{E}&\left[ \exp{ \left(-c_1(X-y_{i})^2-c_2(X-y_{k})^2  \right)} \Big|Y_N\right] = \int \frac{1}{\sqrt{2\pi\sigma^2}} e^{ \left(-c_1(x-y_{i})^2-c_2(x-y_{k})^2   \right)} e^{ \left(-\frac{(x-\mu)^2}{2\sigma^2} \right)} dx \\
&= \frac{1}{\sqrt{2\pi\sigma^2}} \int \exp{\left( -\left(c_1+c_2+\frac{1}{2\sigma^2} \right)x^2 + \left(2c_1y_i+2c_2y_k+\frac{2\mu}{2\sigma^2} \right)x - \left(c_1y_i^2+c_2y_k^2+\frac{\mu^2}{2\sigma^2} \right) \right)} dx\\
&= \frac{1}{\sqrt{2\sigma^2\left(c_1+c_2+\frac{1}{2\sigma^2} \right)}} \exp{\left(\frac{\left(2c_1y_i+2c_2y_k+\frac{2\mu}{2\sigma^2} \right)^2}{4\left(c_1+c_2+\frac{1}{2\sigma^2} \right)} - \left(c_1y_i^2+c_2y_k^2+\frac{\mu^2}{2\sigma^2} \right) \right)}\\
&= \frac{1}{\sqrt{1+2\left(c_1+c_2\right)\sigma^2}} \exp{\left(\frac{\left(2\left(c_1y_i+c_2y_k\right)\sigma^2+\mu \right)^2}{2\sigma^2\left(1+2\left(c_1+c_2 \right)\sigma^2\right)} - \frac{2\left(c_1y_i^2+c_2y_k^2\right)\sigma^2+\mu^2}{2\sigma^2}  \right)}\\
&= \frac{1}{\sqrt{1+2\left(c_1+c_2\right)\sigma^2}} \exp{\left(-\frac{c_1\left(y_i-\mu\right)^2+c_2\left(y_k-\mu\right)^2+2c_1c_2\sigma^2\left(y_i-y_k\right)^2}{\left(1+2\left(c_1+c_2 \right)\sigma^2\right)}  \right)},
\end{align*}
where $\mu=\mu^{*}_{l-1}(\mathbf{x})$, $\sigma^2=\sigma^{*2}_{l-1}(\mathbf{x})$, $c_1=\frac{c_i}{\theta_y}$, $c_2=\frac{c_k}{\theta_y}$, $y_i=(Y_{-L})_i$, and $y_k=(Y_{-L})_k$. Therefore, 
\begin{align*}
\zeta_{ik}(\mu,\sigma^2) &= \sqrt{\frac{\theta_{y}}{\theta_{y}+2(c_i+c_k)\sigma^2}} \\
&\times \exp{\left( - \frac{c_i\left((Y_{-L})_i-\mu\right)^2+c_k\left((Y_{-L})_k-\mu\right)^2+\frac{2}{\theta_{y}}c_ic_k\sigma^2\left((Y_{-L})_i-(Y_{-L})_k\right)^2}{\theta_{y}+2(c_i+c_k)\sigma^2} \right)}.
\end{align*}

\section{Nonseparable Mat\'ern kernel functions}\label{app:matern}
The section introduces the nonseparable Mat\'ern kernels with smoothness parameters of $\nu=1.5$ and $\nu=2.5$. Denote $$\psi(x,x';\theta, \nu=1.5) =\left( 1+\frac{1}{\left(\frac{(t-t')^2}{\theta_t} + 1\right)^{ \frac{\beta}{2} }}\frac{\sqrt{3}|x- x'|}{\theta} \right) \exp \left( -\frac{1}{\left(\frac{(t-t')^2}{\theta_t} + 1\right)^{ \frac{\beta}{2} }}\frac{\sqrt{3}|x- x'|}{\theta} \right),$$ 
\begin{align*}
\psi(x,x';\theta, \nu=2.5) &=\left( 1+\frac{1}{\left(\frac{(t-t')^2}{\theta_t} + 1\right)^{ \frac{\beta}{2} }}\frac{\sqrt{5}|x- x'|}{\theta}+\frac{1}{3}\left(\frac{1}{\left(\frac{(t-t')^2}{\theta_t} + 1\right)^{ \frac{\beta}{2} }}\frac{\sqrt{5}|x- x'|}{\theta} \right)^2 \right) \\
& \times \exp \left( -\frac{1}{\left(\frac{(t-t')^2}{\theta_t} + 1\right)^{ \frac{\beta}{2} }}\frac{\sqrt{5}|x- x'|}{\theta} \right).
\end{align*}
The nonseparable Mat\'ern kernel function is denoted by 
\begin{align*}
K((t, \mathbf{x}, y), (t', \mathbf{x}', y') ; \nu)= \left(\frac{(t-t')^2}{\theta_t} + 1\right)^{ - \left(\frac{\beta(d+1)}{2}+\delta \right) } \psi(y,y';\theta_y, \nu) \prod^d_{j=1} \psi(x_j,x_j';\theta_j, \nu).
\end{align*}

\section{Posterior mean and variance under Mat\'ern kernel}\label{supp:maternposterior}

This section is developed along the line of \cite{suppming2021}. The posterior mean and variance at the input $\mathbf{x}$ can be derived as follows,
\begin{align*}
\mu^*_l (\mathbf{x})=&\mathbb{E}[f_l(\mathbf{x})|Y_N] \\
=& \alpha + \mathbb{E} [ \mathbf{k} (t_l,\mathbf{x},f_{l-1}(\mathbf{x}))^T |Y_N] \mathbf{K}^{-1} (Y_{-1} - \alpha \mathbf{1}_{N_{-1}}) \\
=& \alpha + \sum^{N_{-1}}_{i=1} r_i c_i(t_l)^{(d+1)+\frac{2\delta}{\beta}}  \xi_i \prod^d_{j=1}\psi(x_j,(\mathbf{X}_{-1})_{ij};\theta_{lj}, \nu)
, \\
\sigma_l^{*2}(\mathbf{x})=&\mathbb{V}[f_l(\mathbf{x})|Y_N] \\
=& \alpha^2 + 2\alpha  (\mu^*_l(\mathbf{x}) - \alpha) + \mathbb{E}\left[ \left\{ \mathbf{k} (t_l,\mathbf{x},f_{l-1}(\mathbf{x}))^T  \mathbf{K}^{-1} (Y_{-1} - \alpha \mathbf{1}_{N_{-1}}) \right\}^2 |Y_N\right] \\
&-\mu^{*}_l(\mathbf{x})^2 + \tau^2 - \tau^2 \mathbb{E} \left[ \left\{ \mathbf{k} (t_l,\mathbf{x},f_{l-1}(\mathbf{x}))^T  \mathbf{K}^{-1}  \mathbf{k}(t_l,\mathbf{x},f_{l-1}(\mathbf{x}))  \right\}  |Y_N\right]\\
=& \tau^2 - (\mu^*_l(\mathbf{x})-\alpha)^2 
\end{align*}
\begin{align*}
&+\left( \sum_{i,k=1}^{N_{-1}} \zeta_{ik} \left(r_i r_k - \tau^2 (\mathbf{K}^{-1})_{ik} \right) (c_i(t_l)c_k(t_l))^{(d+1)+\frac{2\delta}{\beta}} \prod_{j=1}^d \psi(x_j,(\mathbf{X}_{-1})_{ij};\theta_{j}, \nu) \psi(x_j,(\mathbf{X}_{-1})_{kj};\theta_{j}, \nu) \right)
\end{align*}
where $\psi$ is defined in Section \ref{app:matern}, 
$r_i = (\mathbf{K}^{-1} (Y_{-1}-\alpha \mathbf{1}_{N_{-1}}))_i$,$c_i(t):=c(t,\mathbf{t}_i)=\left(\frac{(t-\mathbf{t}_i)^2}{\theta_t} + 1\right)^{-\frac{\beta}{2}}$,$\xi_i = \mathbb{E} \left[ \psi(f_{l-1}(\mathbf{x}), (Y_{-L})_i ; \theta_{y}, \nu) \Big|Y_N \right]$, and $\zeta_{ik} = \mathbb{E} \left[ \psi(f_{l-1}(\mathbf{x}), (Y_{-L})_i ; \theta_{y}, \nu) \psi(f_{l-1}(\mathbf{x}), (Y_{-L})_k ; \theta_{y}, \nu) \Big|Y_N \right]$. The closed-form expressions of $\xi_i$ and $\zeta_{ik}$ are provided in the following subsections.

\subsection{Mat\'ern-1.5 kernel}

For Mat\'ern kernel with the smoothness parameter of 1.5, $\xi_i$ and $\zeta_{ik}$ are provided as follows,
\begin{align*}
\xi_i &= \exp \left( \frac{3c_i^2(t) \sigma^{*2}_{l-1}(\mathbf{x}) + 2\sqrt{3}c_i(t) \theta_{y} ((Y_{-L})_i - \mu^*_{l-1}(\mathbf{x}) ) }{2 \theta_{y}^2} \right) \\
& \times \left[ E_1' \Lambda_{11} \Phi \left( \frac{\mu^*_{l-1}(\mathbf{x}) - (Y_{-L})_i - \frac{\sqrt{3}c_i(t)\sigma^{*2}_{l-1}(\mathbf{x})}{\theta_{y}}}{\sigma^{*}_{l-1}(\mathbf{x})}  \right) \right. \\
&+ \left. E_1' \Lambda_{12} \frac{\sigma^{*}_{l-1}(\mathbf{x})}{\sqrt{2\pi}} \exp \left( -\frac{ \left((Y_{-L})_i - \mu^*_{l-1}(\mathbf{x}) + \frac{\sqrt{3}c_i(t)\sigma^{*2}_{l-1}(\mathbf{x})}{\theta_{y}} \right)^2}{2\sigma^{*2}_{l-1}(\mathbf{x})} \right) \right] \\
&+ \exp \left( \frac{3c_i^2(t) \sigma^{*2}_{l-1}(\mathbf{x}) - 2\sqrt{3}c_i(t) \theta_{y} ((Y_{-L})_i - \mu^*_{l-1}(\mathbf{x}) ) }{2 \theta_{y}^2} \right) \\
& \times \left[ E_2' \Lambda_{21}  \Phi \left( \frac{-\mu^*_{l-1}(\mathbf{x}) + (Y_{-L})_i - \frac{\sqrt{3}c_i(t)\sigma^{*2}_{l-1}(\mathbf{x})}{\theta_{y}}}{\sigma^{*}_{l-1}(\mathbf{x})}  \right) \right. \\
&+ \left. E_2' \Lambda_{12}  \cdot \frac{\sigma^{*}_{l-1}(\mathbf{x})}{\sqrt{2\pi}} \exp \left( -\frac{ \left( (Y_{-L})_i - \mu^*_{l-1}(\mathbf{x}) - \frac{\sqrt{3}c_i(t)\sigma^{*2}_{l-1}(\mathbf{x})}{\theta_{y}} \right)^2 }{2\sigma^{*2}_{l-1}(\mathbf{x})} \right) \right], \\
\zeta_{ik} &= \exp \left\{ \frac{3\sigma^{*2}_{l-1}(\mathbf{x})(c_i(t)+c_k(t))^2 + 2\sqrt{3} \theta_{y} \left( c_i(t)((Y_{-L})_i-\mu^*_{l-1}(\mathbf{x}))+c_k(t)((Y_{-L})_k -\mu^*_{l-1}(\mathbf{x})) \right)}{2\theta_{y}^2}  \right\} \\
&\times \left[ E_{3}' \Lambda_{31}
\Phi \left\{ \frac{ \left( \mu^*_{l-1}(\mathbf{x}) -(Y_{-L})_k - \sqrt{3}(c_i(t)+c_k(t)) \frac{\sigma^{*2}_{l-1}(\mathbf{x})}{\theta_{y} }  \right) }{\sigma^{*}_{l-1}(\mathbf{x})}  \right. \right\} \\
&+ E_{3}' \Lambda_{32} \left. \frac{\sigma^{*}_{l-1}(\mathbf{x})}{\sqrt{2\pi}} \exp \left(-\frac{\left((Y_{-L})_k-\mu^*_{l-1}(\mathbf{x})+ \sqrt{3}(c_i(t)+c_k(t)) \frac{\sigma^{*2}_{l-1}(\mathbf{x})}{\theta_{y} }  \right)^2}{2\sigma^{*2}_{l-1}(\mathbf{x})}  \right)
\right]\\
&+ \exp \left\{\frac{3\sigma^{*2}_{l-1}(\mathbf{x})(c_i(t)-c_k(t))^2 + 2\sqrt{3} \theta_{y} \left( c_i(t)((Y_{-L})_i-\mu^*_{l-1}(\mathbf{x}))-c_k(t)((Y_{-L})_k -\mu^*_{l-1}(\mathbf{x})) \right) }{2\theta_{y}^2}  \right\} \\
&\times \left[ E_{4}' \Lambda_{41} \left( \Phi \left\{ \frac{ (Y_{-L})_k - \mu^*_{l-1}(\mathbf{x}) }{\sigma^{*}_{l-1}(\mathbf{x})}  \right\} -  \Phi \left\{ \frac{ (Y_{-L})_i - \mu^*_{l-1}(\mathbf{x}) }{\sigma^{*}_{l-1}(\mathbf{x})}  \right\} \right) \right. \\
&+ E_{4}' \Lambda_{42} \frac{\sigma^{*}_{l-1}(\mathbf{x})}{\sqrt{2\pi}} \exp \left(-\frac{\left((Y_{-L})_i-\mu^*_{l-1}(\mathbf{x}) \right)^2}{2\sigma^{*2}_{l-1}(\mathbf{x})} \right)   \left. - E_{4}' \Lambda_{43} \frac{\sigma^{*}_{l-1}(\mathbf{x})}{\sqrt{2\pi}} \exp \left(-\frac{\left((Y_{-L})_k-\mu^*_{l-1}(\mathbf{x}) \right)^2}{2\sigma^{*2}_{l-1}(\mathbf{x})} \right) \right] \\
&+ \exp \left\{ \frac{3\sigma^{*2}_{l-1}(\mathbf{x})(c_i(t)+c_k(t))^2 - 2\sqrt{3} \theta_{y} \left( c_i(t)((Y_{-L})_i-\mu^*_{l-1}(\mathbf{x}))+c_k(t)((Y_{-L})_k -\mu^*_{l-1}(\mathbf{x})) \right) }{2\theta_{y}^2} \right\} 
\end{align*}
\begin{align*}
&\times \left[ E_{5}' \Lambda_{51}
\Phi \left\{ \frac{ \left( -\mu^*_{l-1}(\mathbf{x}) + (Y_{-L})_i - \sqrt{3}(c_i(t)+c_k(t)) \frac{\sigma^{*2}_{l-1}(\mathbf{x})}{\theta_{y} }  \right) }{\sigma^{*}_{l-1}(\mathbf{x})}  \right\} \right. \\
&+ E_{5}' \Lambda_{52} \left. \frac{\sigma^{*}_{l-1}(\mathbf{x})}{\sqrt{2\pi}} \exp \left(-\frac{\left((Y_{-L})_i-\mu^*_{l-1}(\mathbf{x})- \sqrt{3}(c_i(t)+c_k(t)) \frac{\sigma^{*2}_{l-1}(\mathbf{x})}{\theta_{y} }  \right)^2}{2\sigma^{*2}_{l-1}(\mathbf{x})}  \right)
\right],\\
\Lambda_{11} &= 
\begin{pmatrix} 
1 \\ 
\mu^*_{l-1}(\mathbf{x}) - \frac{\sqrt{3}c_i(t)\sigma^{*2}_{l-1}(\mathbf{x})}{\theta_{y}} 
\end{pmatrix},
\Lambda_{12} =
\begin{pmatrix} 
0 \\ 
1 
\end{pmatrix},
\Lambda_{21} = 
\begin{pmatrix} 
1 \\ 
- \mu^*_{l-1}(\mathbf{x}) - \frac{\sqrt{3}c_i(t)\sigma^{*2}_{l-1}(\mathbf{x})}{\theta_{y}} 
\end{pmatrix},
\\
\Lambda_{31} &= 
\begin{pmatrix} 
1 \\ 
\mu^*_{l-1}(\mathbf{x}) - \frac{\sqrt{3}(c_i(t)+c_k(t))\sigma^{*2}_{l-1}(\mathbf{x})}{\theta_{y}} \\
\left( \mu^*_{l-1}(\mathbf{x}) - \frac{\sqrt{3}(c_i(t)+c_k(t))\sigma^{*2}_{l-1}(\mathbf{x})}{\theta_{y}} \right)^2 + \sigma^{*2}_{l-1}(\mathbf{x})
\end{pmatrix},
\Lambda_{32} =
\begin{pmatrix} 
0 \\ 
1 \\
\mu^*_{l-1}(\mathbf{x}) - \frac{\sqrt{3}(c_i(t)+c_k(t))\sigma^{*2}_{l-1}(\mathbf{x})}{\theta_{y}} + (Y_{-L})_k
\end{pmatrix}, \\
\Lambda_{41} &= 
\begin{pmatrix} 
1 \\ 
\mu^*_{l-1}(\mathbf{x}) \\
\left( \mu^*_{l-1}(\mathbf{x}) \right)^2 + \sigma^{*2}_{l-1}(\mathbf{x})
\end{pmatrix},
\Lambda_{42} =
\begin{pmatrix} 
0 \\ 
1 \\
\mu^*_{l-1}(\mathbf{x}) + (Y_{-L})_i
\end{pmatrix},\\
\Lambda_{43} &=
\begin{pmatrix} 
0 \\ 
1 \\
\mu^*_{l-1}(\mathbf{x}) + (Y_{-L})_k
\end{pmatrix},
\Lambda_{51} = 
\begin{pmatrix} 
1 \\ 
-\mu^*_{l-1}(\mathbf{x}) - \frac{\sqrt{3}(c_i(t)+c_k(t))\sigma^{*2}_{l-1}(\mathbf{x})}{\theta_{y}} \\
\left( \mu^*_{l-1}(\mathbf{x}) + \frac{\sqrt{3}(c_i(t)+c_k(t))\sigma^{*2}_{l-1}(\mathbf{x})}{\theta_{y}} \right)^2 + \sigma^{*2}_{l-1}(\mathbf{x})
\end{pmatrix},\\
\Lambda_{52} &=
\begin{pmatrix} 
0 \\ 
1 \\
-\mu^*_{l-1}(\mathbf{x}) - \frac{\sqrt{3}(c_i(t)+c_k(t))\sigma^{*2}_{l-1}(\mathbf{x})}{\theta_{y}} - (Y_{-L})_i
\end{pmatrix},
E_{1} = \frac{1}{\theta_{y}}
\begin{pmatrix} 
\theta_{y} - \sqrt{3}c_i(t)(Y_{-L})_i \\ 
\sqrt{3}c_i(t)
\end{pmatrix},\\
E_{2} &= \frac{1}{\theta_{y}}
\begin{pmatrix} 
\theta_{y} + \sqrt{3}c_i(t)(Y_{-L})_i \\ 
\sqrt{3}c_i(t)
\end{pmatrix},\\
E_{3} &= \frac{1}{\theta_{y}^2}
\begin{pmatrix} 
\theta_{y}^2 + 3c_i(t)c_k(t)(Y_{-L})_i(Y_{-L})_k - \sqrt{3}\theta_{y}\left(c_i(t)(Y_{-L})_i+c_k(t)(Y_{-L})_k\right) \\ 
\sqrt{3} (c_i(t)+ c_k(t))\theta_{y} - 3c_i(t)c_k(t)\left((Y_{-L})_i+(Y_{-L})_k\right)\\
3c_i(t)c_k(t)
\end{pmatrix},
\\
E_{4} &= \frac{1}{\theta_{y}^2}
\begin{pmatrix} 
\theta_{y}^2 - 3c_i(t)c_k(t)(Y_{-L})_i(Y_{-L})_k + \sqrt{3}\theta_{y}\left(c_k(t)(Y_{-L})_k-c_i(t)(Y_{-L})_i\right) \\ 
\sqrt{3}(c_i(t)-c_k(t))\theta_{y}+3c_i(t)c_k(t)\left((Y_{-L})_i+(Y_{-L})_k\right)\\
-3c_i(t)c_k(t)
\end{pmatrix},
\end{align*}
\begin{align*}
E_{5} &= \frac{1}{\theta_{y}^2}
\begin{pmatrix} 
\theta_{y}^2 + 3c_i(t)c_k(t)(Y_{-L})_i(Y_{-L})_k + \sqrt{3}\theta_{y}\left(c_i(t)(Y_{-L})_i+c_k(t)(Y_{-L})_k\right) \\ 
\sqrt{3} (c_i(t)+ c_k(t))\theta_{y} + 3c_i(t)c_k(t)\left((Y_{-L})_i+(Y_{-L})_k\right)\\
3c_i(t)c_k(t)
\end{pmatrix},
\end{align*}
for $ (Y_{-L})_i \leq (Y_{-L})_k$. If $ (Y_{-L})_i > (Y_{-L})_k$, interchange $ (Y_{-L})_i$ and $ (Y_{-L})_k$. $\Phi$ is the cumulative distribution function of a standard normal distribution.

\subsection{Mat\'ern-2.5 kernel}
For Mat\'ern kernel with the smoothness parameter of 2.5, $\xi_i$ and $\zeta_{ik}$ are provided as follows, 
\begin{align*} 
\xi_i &= \exp \left( \frac{5c_i^2(t) \sigma^{*2}_{l-1}(\mathbf{x}) + 2\sqrt{5}c_i(t) \theta_{y} ((Y_{-L})_i - \mu^*_{l-1}(\mathbf{x}) ) }{2 \theta_{y}^2} \right) \\
&\times \left[ E_1' \Lambda_{11} \Phi \left( \frac{\mu^*_{l-1}(\mathbf{x}) - (Y_{-L})_i - \frac{\sqrt{5}c_i(t)\sigma^{*2}_{l-1}(\mathbf{x})}{\theta_{y}}}{\sigma^{*}_{l-1}(\mathbf{x})}  \right) \right. \\
&+ \left. E_1' \Lambda_{12} \frac{\sigma^{*}_{l-1}(\mathbf{x})}{\sqrt{2\pi}} \exp \left( -\frac{ \left( (Y_{-L})_i - \mu^*_{l-1}(\mathbf{x}) + \frac{\sqrt{5}c_i(t)\sigma^{*2}_{l-1}(\mathbf{x})}{\theta_{y}} \right)^2}{2\sigma^{*2}_{l-1}(\mathbf{x})} \right) \right] \\
&+ \exp \left( \frac{5 c_i^2(t)\sigma^{*2}_{l-1}(\mathbf{x}) - 2\sqrt{5}c_i(t) \theta_{y} ((Y_{-L})_i - \mu^*_{l-1}(\mathbf{x}) ) }{2 \theta_{y}^2} \right) \\
&\times \left[ E_2' \Lambda_{21}  \Phi \left( \frac{-\mu^*_{l-1}(\mathbf{x}) + (Y_{-L})_i - \frac{\sqrt{5}c_i(t)\sigma^{*2}_{l-1}(\mathbf{x})}{\theta_{y}}}{\sigma^{*}_{l-1}(\mathbf{x})}  \right) \right. \\
&+ \left. E_2' \Lambda_{22}  \cdot \frac{\sigma^{*}_{l-1}(\mathbf{x})}{\sqrt{2\pi}} \exp \left( -\frac{ \left( (Y_{-L})_i - \mu^*_{l-1}(\mathbf{x}) - \frac{\sqrt{5}c_i(t)\sigma^{*2}_{l-1}(\mathbf{x})}{\theta_{y}} \right)^2 }{2\sigma^{*2}_{l-1}(\mathbf{x})} \right) \right], 
\end{align*}
\begin{align*}
\zeta_{ik} &= \exp \left\{ \frac{5\sigma^{*2}_{l-1}(\mathbf{x})(c_i(t)+c_k(t))^2 + 2\sqrt{5} \theta_{y} \left(c_i(t) ( (Y_{-L})_i-\mu^*_{l-1}(\mathbf{x}) )+c_k(t)((Y_{-L})_k -\mu^*_{l-1}(\mathbf{x})) \right)}{2\theta_{y}^2}  \right\} \\
&\times \left[ E_{3}' \Lambda_{31}
\Phi \left\{ \frac{ \left( \mu^*_{l-1}(\mathbf{x}) - (Y_{-L})_k - \sqrt{5}(c_i(t)+c_k(t)) \frac{\sigma^{*2}_{l-1}(\mathbf{x})}{\theta_{y} }  \right) }{\sigma^{*}_{l-1}(\mathbf{x})}  \right. \right\} \\
&+ E_{3}' \Lambda_{32} \left. \frac{\sigma^{*}_{l-1}(\mathbf{x})}{\sqrt{2\pi}} \exp \left(-\frac{\left( (Y_{-L})_k-\mu^*_{l-1}(\mathbf{x})  + \sqrt{5}(c_i(t)+c_k(t)) \frac{\sigma^{*2}_{l-1}(\mathbf{x})}{\theta_{y} }  \right)^2}{2\sigma^{*2}_{l-1}(\mathbf{x})}  \right)
\right] \\
&+ \exp \left\{\frac{5\sigma^{*2}_{l-1}(\mathbf{x})(c_i(t)-c_k(t))^2 + 2\sqrt{5} \theta_{y} \left( c_i(t)((Y_{-L})_i-\mu^*_{l-1}(\mathbf{x}))-c_k(t)((Y_{-L})_k -\mu^*_{l-1}(\mathbf{x})) \right) }{2\theta_{y}^2}  \right\} \\
&\times \left[ E_{4}' \Lambda_{41} \left( \Phi \left\{ \frac{ (Y_{-L})_k - \mu^*_{l-1}(\mathbf{x}) }{\sigma^{*}_{l-1}(\mathbf{x})}  \right\} -  \Phi \left\{ \frac{ (Y_{-L})_i - \mu^*_{l-1}(\mathbf{x}) }{\sigma^{*}_{l-1}(\mathbf{x})}  \right\} \right) \right. \\
&+ E_{4}' \Lambda_{42} \frac{\sigma^{*}_{l-1}(\mathbf{x})}{\sqrt{2\pi}} \exp \left(-\frac{\left((Y_{-L})_i-\mu^*_{l-1}(\mathbf{x}) \right)^2}{2\sigma^{*2}_{l-1}(\mathbf{x})} \right)  \left. - E_{4}' \Lambda_{43} \frac{\sigma^{*}_{l-1}(\mathbf{x})}{\sqrt{2\pi}} \exp \left(-\frac{\left((Y_{-L})_k-\mu^*_{t-1}(\mathbf{x}) \right)^2}{2\sigma^{*2}_{l-1}(\mathbf{x})} \right) \right] \\
&+ \exp \left\{ \frac{5\sigma^{*2}_{l-1}(\mathbf{x})(c_i(t)+c_k(t))^2 - 2\sqrt{5} \theta_{y} \left( c_i(t)((Y_{-L})_i-\mu^*_{l-1}(\mathbf{x}))+c_k(t)((Y_{-L})_k -\mu^*_{l-1}(\mathbf{x})) \right) }{2\theta_{y}^2} \right\} \\
& \times \left[ E_{5}' \Lambda_{51}
\Phi \left\{ \frac{ \left( -\mu^*_{l-1}(\mathbf{x}) + (Y_{-L})_i - \sqrt{5}(c_i(t)+c_k(t)) \frac{\sigma^{*2}_{l-1}(\mathbf{x})}{\theta_{y} }  \right) }{\sigma^{*}_{l-1}(\mathbf{x})}  \right\} \right. \\
&+ E_{5}' \Lambda_{52} \left. \frac{\sigma^{*}_{l-1}(\mathbf{x})}{\sqrt{2\pi}} \exp \left(-\frac{ \left((Y_{-L})_i -\mu^*_{l-1}(\mathbf{x}) - \sqrt{5} (c_i(t)+c_k(t)) \frac{\sigma^{*2}_{l-1}(\mathbf{x})}{\theta_{y} }  \right)^2}{2\sigma^{*2}_{l-1}(\mathbf{x})}  \right)
\right],\\
\Lambda_{11} &= 
\begin{pmatrix} 
1 \\ 
\mu^*_{t-1}(\mathbf{x}) - \frac{\sqrt{5}c_i(t)\sigma^{*2}_{l-1}(\mathbf{x})}{\theta_{y}} \\
\left( \mu^*_{l-1}(\mathbf{x}) - \frac{\sqrt{5}c_i(t)\sigma^{*2}_{l-1}(\mathbf{x})}{\theta_{y}} \right)^2 + \sigma^{*2}_{l-1}(\mathbf{x})
\end{pmatrix},
\Lambda_{12} =
\begin{pmatrix} 
0 \\ 
1 \\
\mu^*_{l-1}(\mathbf{x}) - \frac{\sqrt{5}c_i(t)\sigma^{*2}_{l-1}(\mathbf{x})}{\theta_{y}} + (Y_{-L})_i
\end{pmatrix},\\
\Lambda_{21} &= 
\begin{pmatrix} 
1 \\ 
- \mu^*_{l-1}(\mathbf{x}) - \frac{\sqrt{5}c_i(t)\sigma^{*2}_{l-1}(\mathbf{x})}{\theta_{y}} \\
\left( \mu^*_{l-1}(\mathbf{x}) + \frac{\sqrt{5}c_i(t)\sigma^{*2}_{l-1}(\mathbf{x})}{\theta_{y}} \right)^2 + \sigma^{*2}_{l-1}(\mathbf{x})
\end{pmatrix},
\Lambda_{22} =
\begin{pmatrix} 
0 \\ 
1 \\
-\mu^*_{l-1}(\mathbf{x}) - \frac{\sqrt{5}c_i(t)\sigma^{*2}_{l-1}(\mathbf{x})}{\theta_{y}} - (Y_{-L})_i
\end{pmatrix}, \\
\Lambda_{31} &= 
\begin{pmatrix} 
1 \\ 
\mu_c\\
\mu_c^2+\sigma^{*2}_{l-1}(\mathbf{x})\\
\mu_c \left( \mu_c^2+3\sigma^{*2}_{l-1}(\mathbf{x}) \right) \\
\mu_c^4 + 6\mu_c^2\sigma^{*2}_{l-1}(\mathbf{x}) + 3\sigma^{*4}_{l-1}(\mathbf{x})
\end{pmatrix}, \mu_c=\mu^*_{l-1}(\mathbf{x}) - \sqrt{5}(c_i(t)+c_k(t))\frac{\sigma^{*2}_{l-1}(\mathbf{x})}{\theta_{y}},
\end{align*}
\begin{align*}
\Lambda_{32} &=
\begin{pmatrix} 
0 \\ 
1 \\
\mu_c+(Y_{-L})_k\\
\mu_c^2 + 2\sigma^{*2}_{l-1}(\mathbf{x}) + \left((Y_{-L})_k\right)^2 + \mu_c (Y_{-L})_k\\
\mu_c^3 + \left((Y_{-L})_k\right)^3 + (Y_{-L})_k \mu_c \left( \mu_c +(Y_{-L})_k \right) + \sigma^{*2}_{l-1}(\mathbf{x}) \left( 5\mu_c + 3(Y_{-L})_k \right)
\end{pmatrix}, \\
\Lambda_{41} &= 
\begin{pmatrix} 
1 \\ 
\mu^*\\
\mu^{*2}+\sigma^{*2}_{l-1}(\mathbf{x})\\
\mu^* \left( \mu^{*2}+3\sigma^{*2}_{l-1}(\mathbf{x}) \right) \\
\mu^{*4} + 6\mu_c^2\sigma^{*2}_{l-1}(\mathbf{x}) + 3\sigma^{*4}_{l-1}(\mathbf{x})
\end{pmatrix}, \mu^*=\mu^*_{l-1}(\mathbf{x}),\\
\Lambda_{42} &=
\begin{pmatrix} 
0 \\ 
1 \\
\mu^*+(Y_{-L})_i\\
\mu^{*2} + 2\sigma^{*2}_{l-1}(\mathbf{x}) + \left((Y_{-L})_i\right)^2 + \mu^*(Y_{-L})_i\\
\mu^{*3} + \left((Y_{-L})_i\right)^3 + (Y_{-L})_i \mu^* \left( \mu^* + (Y_{-L})_i \right) + \sigma^{*2}_{l-1}(\mathbf{x}) \left( 5\mu^* + 3(Y_{-L})_i \right)
\end{pmatrix}, \\
\Lambda_{43} &=
\begin{pmatrix} 
0 \\ 
1 \\
\mu^*+(Y_{-L})_k\\
\mu^{*2} + 2\sigma^{*2}_{l-1}(\mathbf{x}) + \left((Y_{-L})_k\right)^2 + \mu^*(Y_{-L})_k\\
\mu^{*3} + \left((Y_{-L})_k\right)^3 + (Y_{-L})_k \mu^* \left( \mu^* + (Y_{-L})_k \right) + \sigma^{*2}_{l-1}(\mathbf{x}) \left( 5\mu^* + 3(Y_{-L})_k \right)
\end{pmatrix}, \\
\Lambda_{51} &= 
\begin{pmatrix} 
1 \\ 
-\mu_d\\
\mu_d^2+\sigma^{*2}_{l-1}(\mathbf{x})\\
-\mu_d \left( \mu_d^2+3\sigma^{*2}_{l-1}(\mathbf{x}) \right) \\
\mu_d^4 + 6\mu_d^2\sigma^{*2}_{l-1}(\mathbf{x}) + 3\sigma^{*4}_{l-1}(\mathbf{x})
\end{pmatrix}, \mu_d=\mu^*_{l-1}(\mathbf{x}) + \sqrt{5}(c_i(t)+c_k(t))\frac{\sigma^{*2}_{l-1}(\mathbf{x})}{\theta_{y}},
\end{align*}
\begin{align*}
\Lambda_{52} &=
\begin{pmatrix} 
0 \\ 
1 \\
-\mu_d-(Y_{-L})_i\\
\mu_d^2 + 2\sigma^{*2}_{l-1}(\mathbf{x}) + \left((Y_{-L})_i\right)^2 + \mu_d (Y_{-L})_i\\
-\mu_d^3 - \left((Y_{-L})_i\right)^3 - (Y_{-L})_i \mu_d \left( \mu_d + (Y_{-L})_i \right) - \sigma^{*2}_{l-1}(\mathbf{x}) \left( 5\mu_d + 3(Y_{-L})_i \right)
\end{pmatrix}, \\
E_{1} &= \frac{1}{3\theta_{y}^2}
\begin{pmatrix} 
3\theta_{y}^2 - 3\sqrt{5}c_i(t)\theta_{y}(Y_{-L})_i + 5 c_i^2(t)\left((Y_{-L})_i\right)^2 \\ 
3\sqrt{5}c_i(t)\theta_{y} - 10c_i^2(t)(Y_{-L})_i\\
5c_i^2(t)
\end{pmatrix},\\
E_{2} &= \frac{1}{3\theta_{y}^2}
\begin{pmatrix} 
3\theta_{y}^2 + 3\sqrt{5}c_i(t)\theta_{y}(Y_{-L})_i + 5 c_i^2(t)\left((Y_{-L})_i\right)^2 \\ 
3\sqrt{5}c_i(t)\theta_{y} + 10c_i^2(t)(Y_{-L})_i\\
5c_i^2(t)
\end{pmatrix},\\
E_{3} &= \frac{1}{9\theta_{y}^4}
\begin{pmatrix} 
E_{31} &
E_{32} &
E_{33} &
E_{34} &
E_{35} 
\end{pmatrix} ^ \top,\\
E_{31} &= 9\theta_{y}^4 + 25c_i^2(t)c_k^2(t)\left( (Y_{-L})_i \right)^2 \left( (Y_{-L})_k \right)^2 \\
&-3\sqrt{5}\theta_{y} \left( c_i(t)(Y_{-L})_i + c_k(t)(Y_{-L})_k \right)\left( 3\theta_{y}^2 + 5 c_i(t)c_k(t)(Y_{-L})_i (Y_{-L})_k \right)\\
&+15 \theta_{y}^2 \left( c_i^2(t)\left( (Y_{-L})_i \right)^2 + c_k^2(t)\left( (Y_{-L})_k \right)^2 + 3c_i(t)c_k(t)(Y_{-L})_i(Y_{-L})_k  \right) , \\
E_{32} &= 9\sqrt{5} (c_i(t)+c_k(t))\theta_{y}^3 + 15\sqrt{5} c_i(t)c_k(t)\theta_{y} \left( c_i(t)\left( (Y_{-L})_i \right)^2 + c_k(t)\left( (Y_{-L})_k \right)^2 \right) \\
&-15 \theta_{y}^2 \left(c_i(t) (Y_{-L})_i (2c_i(t)+3c_k(t)) + c_k(t)(Y_{-L})_k(2c_k(t)+3c_i(t)) \right) \\
&-50c_i^2(t)c_k^2(t) (Y_{-L})_i (Y_{-L})_k  \left( (Y_{-L})_i + (Y_{-L})_k \right) + 30\sqrt{5} c_i(t)c_k(t)(c_i(t)+c_k(t))\theta_{y} (Y_{-L})_i (Y_{-L})_k ,\\
E_{33} &= 5 \left\{ 5c_i^2(t)c_k^2(t)(\left( (Y_{-L})_i \right)^2 + \left( (Y_{-L})_k \right)^2 + 4 (Y_{-L})_i  (Y_{-L})_k ) + 3(c_i^2(t)+c_k^2(t)+3c_i(t)c_k(t))\theta_{y}^2 \right.\\
&\left. - 3\sqrt{5}c_i(t)c_k(t) \theta_{y} \left( 2c_i(t)(Y_{-L})_i + 2c_k(t)(Y_{-L})_k+c_k(t)(Y_{-L})_i + c_i(t)(Y_{-L})_k \right) \right\} \\
E_{34} &= 5 \left( 3\sqrt{5} c_i(t)c_k(t)(c_i(t)+c_k(t))\theta_{y} - 10c_i^2(t)c_k^2(t) ((Y_{-L})_i + (Y_{-L})_k) \right) , E_{35} = 25c_i^2(t)c_k^2(t),\\
E_{4} &= \frac{1}{9\theta_{y}^4}
\begin{pmatrix} 
E_{41} &
E_{42} &
E_{43} &
E_{44} &
E_{45} 
\end{pmatrix} ^ \top,\\
E_{41} &= 9\theta_{y}^4 + 25c_i^2(t)c_k^2(t)\left( (Y_{-L})_i \right)^2 \left( (Y_{-L})_k \right)^2 \\
&+3\sqrt{5}\theta_{y} \left( c_k(t)(Y_{-L})_k - c_i(t)(Y_{-L})_i \right)\left( 3\theta_{y}^2 - 5 c_i(t)c_k(t)(Y_{-L})_i (Y_{-L})_k \right)\\
&+15 \theta_{y}^2 \left( c_i^2(t)\left( (Y_{-L})_i \right)^2 + c_k^2(t)\left( (Y_{-L})_k \right)^2 - 3c_i(t)c_k(t)(Y_{-L})_i(Y_{-L})_k  \right) , 
\end{align*}
\begin{align*}
E_{42} &= 9\sqrt{5} (c_i(t)-c_k(t))\theta_{y}^3 + 15\sqrt{5} c_i(t)c_k(t)\theta_{y} \left( c_k(t)\left( (Y_{-L})_k \right)^2 - c_i(t)\left( (Y_{-L})_i \right)^2 \right) \\
&-15 \theta_{y}^2 \left(c_i(t) (Y_{-L})_i (2c_i(t)-3c_k(t)) + c_k(t)(Y_{-L})_k(2c_k(t)-3c_i(t)) \right) \\
&-50c_i^2(t)c_k^2(t) (Y_{-L})_i (Y_{-L})_k  \left( (Y_{-L})_i + (Y_{-L})_k \right) + 30\sqrt{5} c_i(t)c_k(t)(c_k(t)-c_i(t))\theta_{y} (Y_{-L})_i (Y_{-L})_k ,\\
E_{43} &= 5 \left\{ 5c_i^2(t)c_k^2(t)(\left( (Y_{-L})_i \right)^2 + \left( (Y_{-L})_k \right)^2 + 4 (Y_{-L})_i (Y_{-L})_k ) + 3(c_i^2(t)+c_k^2(t)-3c_i(t)c_k(t))\theta_{y}^2 \right.\\
&\left. - 3\sqrt{5}c_i(t)c_k(t) \theta_{y} \left( 2c_k(t)(Y_{-L})_k - 2c_i(t)(Y_{-L})_i+c_k(t)(Y_{-L})_i - c_i(t)(Y_{-L})_k \right) \right\} \\
E_{44} &= 5 \left( 3\sqrt{5} c_i(t)c_k(t)(c_k(t)-c_i(t))\theta_{y} - 10c_i^2(t)c_k^2(t) ((Y_{-L})_i + (Y_{-L})_k) \right) , E_{45} = 25c_i^2(t)c_k^2(t),\\
E_{5} &= \frac{1}{9\theta_{y}^4}
\begin{pmatrix} 
E_{51} &
E_{52} &
E_{53} &
E_{54} &
E_{55} 
\end{pmatrix} ^ \top,\\
E_{51} &= 9\theta_{y}^4 + 25c_i^2(t)c_k^2(t)\left( (Y_{-L})_i \right)^2 \left( (Y_{-L})_k \right)^2 \\
&+3\sqrt{5}\theta_{y} \left( c_i(t)(Y_{-L})_i + c_k(t)(Y_{-L})_k \right)\left( 3\theta_{y}^2 + 5 c_i(t)c_k(t)(Y_{-L})_i (Y_{-L})_k \right)\\
&+15 \theta_{y}^2 \left( c_i^2(t)\left( (Y_{-L})_i \right)^2 + c_k^2(t)\left( (Y_{-L})_k \right)^2 + 3c_i(t)c_k(t)(Y_{-L})_i(Y_{-L})_k  \right) ,\\
E_{52} &= 9\sqrt{5} (c_i(t)+c_k(t))\theta_{y}^3 + 15\sqrt{5} c_i(t)c_k(t)\theta_{y} \left( c_i(t)\left( (Y_{-L})_i \right)^2 + c_k(t)\left( (Y_{-L})_k \right)^2 \right) \\
&+15 \theta_{y}^2 \left(c_i(t) (Y_{-L})_i (2c_i(t)+3c_k(t)) + c_k(t)(Y_{-L})_k(2c_k(t)+3c_i(t)) \right) \\
&+50c_i^2(t)c_k^2(t) (Y_{-L})_i (Y_{-L})_k  \left( (Y_{-L})_i + (Y_{-L})_k \right) + 30\sqrt{5} c_i(t)c_k(t)(c_i(t)+c_k(t))\theta_{y} (Y_{-L})_i (Y_{-L})_k ,\\
E_{53} &= 5 \left\{ 5c_i^2(t)c_k^2(t)(\left( (Y_{-L})_i \right)^2 + \left( (Y_{-L})_k \right)^2 + 4 (Y_{-L})_i (Y_{-L})_k ) + 3(c_i^2(t)+c_k^2(t)+3c_i(t)c_k(t))\theta_{y}^2 \right.\\
&\left. + 3\sqrt{5}c_i(t)c_k(t) \theta_{y} \left( 2c_i(t)(Y_{-L})_i + 2c_k(t)(Y_{-L})_k+c_k(t)(Y_{-L})_i + c_i(t)(Y_{-L})_k \right) \right\} \\
E_{54} &= 5 \left( 3\sqrt{5} c_i(t)c_k(t)(c_i(t)+c_k(t))\theta_{y} + 10c_i^2(t)c_k^2(t) ((Y_{-L})_i + (Y_{-L})_k) \right) , E_{55} = 25c_i^2(t)c_k^2(t),
\end{align*}
for $ (Y_{-L})_i \leq (Y_{-L})_k$. If $ (Y_{-L})_i > (Y_{-L})_k$, interchange $ (Y_{-L})_i$ and $ (Y_{-L})_k$.


\section{Proofs in Section \ref{sec:theoretical}}\label{appendix:proofSec4}

\subsection{Proof of Lemma \ref{graph_fill_x}}
For any $\mathbf{x}\in \Omega$, choose $\mathbf{x}_i^{[l]}$ such that
\[
\|\mathbf{x}-\mathbf{x}_i^{[l]}\|_2 = \min_j \|\mathbf{x}-\mathbf{x}_j^{[l]}\|_2.
\]
Then, we have
\begin{align}\label{eq_l2_xfinLemma}
    \|(\mathbf{x},f_{l-1}(\mathbf{x})) - (\mathbf{x}_i^{[l]},f_{l-1}(\mathbf{x}_i^{[l]})) \|_2^2 = \|\mathbf{x}-\mathbf{x}_i^{[l]}\|_2^2 + |f_{l-1}(\mathbf{x})-f_{l-1}(\mathbf{x}_i^{[l]})|^2.
\end{align}
Using Assumption \ref{assum_Lipsf},
\[
|f_{l-1}(\mathbf{x})-f_{l-1}(\mathbf{x}_i^{[l]})|^2 \le B_{l-1}^2\|\mathbf{x}-\mathbf{x}_i^{[l]}\|_2^2,
\]
which, together with \eqref{eq_l2_xfinLemma}, yields
\[
\|(\mathbf{x},f_{l-1}(\mathbf{x})) - (\mathbf{x}_i^{[l]},f_{l-1}(\mathbf{x}_i^{[l]})) \|_2^2 \le (1+B_{l-1}^2)\|\mathbf{x}-\mathbf{x}_i^{[l]}\|_2^2.
\]
Taking square roots and the supremum over $\mathbf{x}\in \Omega$ gives the result.

\subsection{Proof of Lemma~\ref{lemma:recursive-error}}\label{appendix:prooflemma3}
We prove the inequality by repeated application of the triangle inequality and the Lipschitz property in the \(y\)-argument. Start from
\[
f(t,\mathbf{x})-\hat f(t,\mathbf{x}) = f(t,\mathbf{x})-\mu_{L+1}(t,\mathbf{x},\hat f_L(\mathbf{x})).
\]
Add and subtract \(\mu_{L+1}(t,\mathbf{x},f_L(\mathbf{x}))\) to obtain
\[
f(t,\mathbf{x})-\hat f(t,\mathbf{x}) = \underbrace{f(t,\mathbf{x})-\mu_{L+1}(t,\mathbf{x},f_L(\mathbf{x}))}_{\varepsilon_{L+1}(t,\mathbf{x})} + \underbrace{\mu_{L+1}(t,\mathbf{x},f_L(\mathbf{x}))-\mu_{L+1}(t,\mathbf{x},\hat f_L(\mathbf{x}))}_{\Delta_{L+1}}.
\]
By Assumption~\ref{assum_Lipsfl},
\(|\Delta_{L+1}|\le \Lambda_{L+1}\,|f_L(\mathbf{x})-\hat f_L(\mathbf{x})|\). Hence,
\[
|f(t,\mathbf{x})-\hat f(t,\mathbf{x})| \le \varepsilon_{L+1}(t,\mathbf{x}) + \Lambda_{L+1}\,|f_L(\mathbf{x})-\hat f_L(\mathbf{x})|.
\]

Apply the same decomposition to \(f_L(\mathbf{x})-\hat f_L(\mathbf{x})=f_L(\mathbf{x})-\mu_L(\mathbf{x},\hat f_{L-1}(\mathbf{x}))\):
\[
f_L(\mathbf{x})-\hat f_L(\mathbf{x}) = \varepsilon_L(\mathbf{x}) + \big(\mu_L(\mathbf{x},f_{L-1}(\mathbf{x}))-\mu_L(\mathbf{x},\hat f_{L-1}(\mathbf{x}))\big),
\]
and thus
\[
|f_L(\mathbf{x})-\hat f_L(\mathbf{x})| \le \varepsilon_L(\mathbf{x}) + \Lambda_L\,|f_{L-1}(\mathbf{x})-\hat f_{L-1}(\mathbf{x})|.
\]
Substituting back gives
\[
|f(t,\mathbf{x})-\hat f(t,\mathbf{x})| \le \varepsilon_{L+1}(t,\mathbf{x}) + \Lambda_{L+1}\varepsilon_L(\mathbf{x}) + \Lambda_{L+1}\Lambda_L\,|f_{L-1}(\mathbf{x})-\hat f_{L-1}(\mathbf{x})|.
\]
Continuing this unrolling down to level \(1\) yields the claimed bound:
\[
|f(t,\mathbf{x})-\hat f(t,\mathbf{x})| \le \varepsilon_{L+1}(t,\mathbf{x}) + \sum_{l=2}^{L} \Big(\prod_{s=l}^{L} \Lambda_{s+1}\Big)\varepsilon_l(\mathbf{x}) + \Big(\prod_{s=1}^{L} \Lambda_{s+1}\Big)\varepsilon_1(\mathbf{x}),
\]
which completes the proof.

\subsection{Proof of Corollary \ref{cor:rates}}

For each $l=2,\ldots,L$, Lemma~\ref{graph_fill_x} gives
\[
h_{\mathcal{Z}_l^{xy},M_{l-1}} \le \sqrt{1+B_{l-1}^2}\,h_{\mathcal{X}_l,\Omega}.
\]
By Assumption~\ref{assum_Quasi_uniform},
$h_{\mathcal{X}_l,\Omega}\le c_\Omega n_l^{-1/d}$, hence
\[
h_{\mathcal{Z}_l^{xy},M_{l-1}} \le \sqrt{1+B_{l-1}^2}\,c_\Omega\,n_l^{-1/d}.
\]
Substituting this into Lemma~\ref{lemma_inter_bound} yields the inequality.

\subsection{Proof of Theorem~\ref{thm:graph-error-bound}}\label{appendix:prooftheorem2}
Substituting the convergence rates from Corollary~\ref{cor:rates} for each levelwise residual \(\varepsilon_l(\cdot)\), $l=1,\dots,L$, into the recursive bound of Lemma~\ref{lemma:recursive-error}, it remains only to bound the extrapolation error $\varepsilon_{L+1}(t,\mathbf{x})$ at the target precision space.

By the triangle inequality,
\begin{align*}
    & |f(t,\mathbf{x})-\mu_{L+1}(t,\mathbf{x},f_L(\mathbf{x}))|\nonumber\\
    \le & |f(t,\mathbf{x}) - f(t_L,\mathbf{x})| + |f(t_L,\mathbf{x}) - \mu_{L+1}(t_L,\mathbf{x},f_L(\mathbf{x}))| + |\mu_{L+1}(t_L,\mathbf{x}, f_L(\mathbf{x})) - \mu_{L+1}(t,\mathbf{x},f_L(\mathbf{x}))|.
\end{align*}
By the Lipschitz property of $f(t,\mathbf{x})$ and $\mu_{L+1}(t,\mathbf{x},y)$, 
\begin{align*}
    |f(t,\mathbf{x}) - f(t_L,\mathbf{x})| \le & B_t(t_L-t),\nonumber\\
    |\mu_{L+1}(t_L,\mathbf{x}, f_L(\mathbf{x})) - \mu_{L+1}(t,\mathbf{x}, f_L(\mathbf{x}))| \le & \Lambda_{L+1}\,(t_L-t).
\end{align*}
For the second term, because $\mu_{L+1}$ leverages a joint dataset across fidelities that strictly includes the level $L$ data, its predictive error at the boundary $t_L$ is bounded by the interpolation error of the single-level model $\hat{f}_L$. Applying the bound from Corollary~\ref{cor:rates}, we have:
\[
|f(t_L,\mathbf{x}) - \mu_{L+1}(t_L,\mathbf{x},f_L(\mathbf{x}))| \leq C\big(\sqrt{1+B_{L-1}^2}\,c_\Omega \big)^{\eta} n_L^{-\eta/d}\|g_L\|_{\mathcal N_{K_L}(M_{L-1})}.
\]
Combining these bounds yields the bound for $\varepsilon_{L+1}(t,\mathbf{x})$ and finishes the proof.

\subsection{Proof of Theorem~\ref{thm:complexity}}\label{appendix:proof3}
By Theorem~\ref{thm:graph-error-bound}, we have 
\begin{align}\label{eq:thm2decomp}
&|f(0,\mathbf{x})-\hat f(0,\mathbf{x})| \le\  
\underbrace{(B_t+\Lambda_{L+1})t_L}_{(b.1)} +\nonumber\\
&+\underbrace{\sum_{l=2}^L \Big(\prod_{s=l}^{L} \Lambda_{s+1}\Big)\, C\big(\sqrt{1+B_{l-1}^2}\,c_\Omega \big)^{\eta} n_l^{-\eta/d}\|g_l\|_{\mathcal N_{K_l}(M_{l-1})}+ \Big(\prod_{s=1}^{L} \Lambda_{s+1}\Big)\,C c_\Omega^\eta n_1^{-\eta/d}\|g_1\|_{\mathcal N_{K_1}(\Omega)}}_{(b.2)}.
\end{align}
We start by choosing $L$ to be
\begin{equation*}\label{eq:L}
L=\left\lceil \frac{\log(2(c_B+c_\Lambda)t_0\epsilon^{-1})}{\log T}\right\rceil,
\end{equation*}
which implies
\[
\frac{\log(2(c_B+c_\Lambda)t_0\epsilon^{-1})}{\log T}\le L
<
\frac{\log(2(c_B+c_\Lambda)t_0\epsilon^{-1})}{\log T}+1.
\]
Hence
\begin{equation}\label{eq:condition}
\frac{1}{2}T^{-1}\epsilon
<
(c_B+c_\Lambda)t_L
\le
\frac{1}{2}\epsilon,
\end{equation}
and therefore (b.1) is bounded by 
\begin{equation}\label{eq:discrepancy}
(B_t+\Lambda_{L+1})t_L
\le
(c_B+c_\Lambda)t_L
\le
\frac{\epsilon}{2}.
\end{equation}

By Condition 1, (b.2) is bounded by 
\begin{align*}
\sum_{l=2}^L \Big(\prod_{s=l}^{L} \Lambda_{s+1}\Big)\, C\big(\sqrt{1+B_{l-1}^2}\,c_\Omega \big)^{\eta} n_l^{-\eta/d}\|g_l\|_{\mathcal N_{K_l}(M_{l-1})} + &\Big(\prod_{s=1}^{L} \Lambda_{s+1}\Big)\,C c_\Omega^\eta n_1^{-\eta/d}\|g_1\|_{\mathcal N_{K_1}(\Omega)} \\
&\le
c_3 \sum_{l=1}^L  n_l^{-\eta/d},
\end{align*}
where $c_3=Cc_\Omega^\eta c_g (1+c_B^2)^{\eta/2} $. We set
\[
n_l
=
\left\lceil\left(
2c_3\epsilon^{-1}
t_L^{\frac{-\gamma\eta}{\eta+d}}\left(1-T^{-\frac{\gamma\eta}{\eta+d}}\right)^{-1}\right)^{\frac{d}{\eta}}
t_l^{\frac{\gamma d}{\eta+d}},
\right\rceil
\] 
so that 
\begin{align*}
c_3 \sum_{l=1}^Ln_l^{-\eta/d} \leq \frac{\epsilon}{2} t_L^{\frac{\gamma \eta}{\eta+d}}\left(1-T^{-\frac{\gamma\eta}{\eta+d}}\right)  \sum^L_{l=1}t_l^{-\frac{\gamma\eta}{\eta+d}}.
\end{align*}
Because 
\begin{align}\label{eq:xi3}
        \sum^L_{l=1}t_l^{-\frac{\gamma\eta}{\eta+d}}=t_L^{-\frac{\gamma \eta}{\eta+d}}\sum^L_{l=1}(T^{-\frac{\gamma \eta}{\eta+d}})^{L-l}<t_L^{-\frac{\gamma\eta}{\eta+d}}\left(1-T^{-\frac{\gamma \eta}{\eta+d}}\right)^{-1},
\end{align}
combining \eqref{eq:thm2decomp} and \eqref{eq:discrepancy}, it follows that 
\begin{align*}
|f(0,\mathbf{x})-\hat{f}(0,\mathbf{x})|&\leq \frac{\epsilon}{2}+\frac{\epsilon}{2} t_L^{\frac{\gamma \eta}{\eta+d}}\left(1-T^{-\frac{\gamma\eta}{\eta+d}}\right)  \sum^L_{l=1}t_l^{-\frac{\gamma\eta}{\eta+d}}\\
    &\leq\frac{\epsilon}{2}+\frac{\epsilon}{2}=\epsilon.
\end{align*}

To bound the total computational cost $C_{\rm total}$, the sample size for any given level $l$ can be bounded by
\[
n_l\leq c_4 \epsilon^{-\frac{d}{\eta}} t_L^{-\frac{\gamma d}{\eta+d}} t_l^{\frac{\gamma d}{\eta+d}} + 1,
\]
where $c_4 = \big(2c_3\left(1-T^{-\frac{\gamma \eta}{\eta+d}}\right)^{-1}\big)^{\frac{d}{\eta}}$.
The computational cost is then bounded by
\begin{equation*}\label{eq:ctinequality3}
    C_{\rm total}\leq c_1\sum^L_{l=1}n_lt^{-\gamma}_l \leq c_1 c_4 \epsilon^{-\frac{d}{\eta}} t_L^{-\frac{\gamma d}{\eta+d}} \sum^L_{l=1}t^{-\frac{\gamma \eta}{\eta+d}}_l + c_1\sum^L_{l=1}t^{-\gamma}_l.
\end{equation*}
Applying \eqref{eq:xi3} to the first term yields:
\begin{align*}
    c_1 c_4 \epsilon^{-\frac{d}{\eta}} t_L^{-\frac{\gamma d}{\eta+d}} \left( t_L^{-\frac{\gamma\eta}{\eta+d}}\left(1-T^{-\frac{\gamma \eta}{\eta+d}}\right)^{-1} \right) = c_1 c_4 \left(1-T^{-\frac{\gamma \eta}{\eta+d}}\right)^{-1} \epsilon^{-\frac{d}{\eta}} t_L^{-\gamma}.
\end{align*}
Because \eqref{eq:condition} guarantees $t^{-1}_L < 2T(c_B+c_\Lambda)\epsilon^{-1}$, we have $t_L^{-\gamma} < (2T(c_B+c_\Lambda))^\gamma \epsilon^{-\gamma}$. Thus, the first term is strictly bounded by $c_5 \epsilon^{-\frac{d}{\eta} - \gamma}$, where $c_5 = c_1 c_4 \left(1-T^{-\frac{\gamma \eta}{\eta+d}}\right)^{-1} (2T(c_B+c_\Lambda))^\gamma$.

Similarly, the second term in \eqref{eq:ctinequality3} is a geometric series bounded by
\begin{align*}
    c_1 \sum^L_{l=1}t_l^{-\gamma} < c_1 t_L^{-\gamma}(1-T^{-\gamma})^{-1} < c_1 \frac{(2T(c_B+c_\Lambda))^{\gamma}}{1-T^{-\gamma}}\epsilon^{-\gamma}.
\end{align*}
Because $\epsilon < e^{-1} < 1$, the $\epsilon^{-\frac{d}{\eta} - \gamma}$ term dominates the $\epsilon^{-\gamma}$ term. Thus, we arrive at the final cost bound:
\[
C_{\rm total}\leq c_2\epsilon^{-\frac{d}{\eta}-\gamma},
\]
where $c_2 = c_5 + c_1 \frac{(2T(c_B+c_\Lambda))^{\gamma}}{1-T^{-\gamma}}$.

\section{DNA Model with Non-Nested Design}\label{supp:non-nested}

Let $\mathcal{X}^*_L = \mathcal{X}_L$ and $\mathcal{X}^*_l = \mathcal{X}_l \cup \mathcal{X}^*_{l+1}$ for $l = 1, \dots, L-1$. Assume that $\mathcal{X}_l \cap \mathcal{X}_{l+1} = \varnothing$. Under this construction, the resulting design satisfies the nested structure:
\[
\mathcal{X}^*_L \subseteq \mathcal{X}^*_{L-1} \subseteq \cdots \subseteq \mathcal{X}^*_1 \subseteq \Omega.
\]
Denote the \textit{pseudo} inputs as $\widetilde{\mathcal{X}}_l := \mathcal{X}^*_l \setminus \mathcal{X}_l$, their corresponding \textit{pseudo} outputs are given by $\widetilde{\mathbf{y}}_l := f_l(\widetilde{\mathcal{X}}_l)$. Furthermore, we define $\mathbf{y}^*_l$ as the combined output, incorporating both the original and \textit{pseudo} outputs: $\mathbf{y}^*_l = f_l(\mathcal{X}^*_l).$

\subsection{Estimation}

We first focus on estimating the unknown parameters $\boldsymbol{\varphi}=(\alpha_1,\alpha,\tau^2_1,\tau^2,\beta, \delta,  \boldsymbol{\theta})$ where $\boldsymbol{\theta}=(\left\{\theta_{j}\right\}_{j=1}^d,\theta_y,\theta_{t})$, for which we employ the stochastic expectation-maximization (SEM) method \citep{celeux1985sem}. The procedure is detailed in Section \ref{appendix:stochasticEM}. 

In the initialization step, we fit independent GPs at each fidelity level and generate initial \textit{pseudo} outputs $\{\widetilde{\mathbf{y}}^{(0)}_l\}^{L-1}_{l=1}$ using their respective GP posterior mean. 
The initial parameter estimates are obtained using the \textit{pseudo}-complete dataset $\{\{\mathbf{y}^{*(0)}_l\}_{l=1}^{L}, \{\mathcal{X}^*_l\}_{l=1}^{L}, \{t_l\}_{l=1}^{L}\}$.

In the E-step (imputation), we sample the pseudo outputs $\widetilde{\mathbf{y}}^{(m)}_1$ from the posterior distribution $p(\widetilde{\mathbf{y}}_1 | \mathbf{y}_1)$ and sample $\{\widetilde{\mathbf{y}}^{(m)}_l\}_{l=2}^{L-1}$ from the conditional normal distribution
$p(\{\widetilde{\mathbf{y}}_l\}_{l=2}^{L-1} | Y_{-1})$, given fixed parameter estimates $\boldsymbol{\varphi}^{(m-1)}$ and previous-step outputs $Y_{-L}^{*(m-1)}:=\{\mathbf{y}^{*(m-1)}_l\}^{L-1}_{l=1}$. 

With the pseudo-complete dataset $\{\{\mathbf{y}^{*(m)}_l\}_{l=1}^{L}, \{\mathcal{X}^*_l\}_{l=1}^{L}, \{t_l\}_{l=1}^{L}\}$, which follows a nested structure, we update the parameter estimates $\boldsymbol{\varphi}^{(m)}$ by maximizing the likelihood function as developed in Section~\ref{eq:estimation}. The iteration continues until reaching a prespecified number of iterations $M$, where we set $M = 100$ in the example presented later.

By alternating between a stochastic E-step and a deterministic M-step, the SEM algorithm constructs a Markov chain that does not converge to a single value but instead fluctuates around estimates that maximize the complete-data likelihood. To obtain a final estimate, we follow the approach of \citet{suppming2021}, taking  the average of the chain after discarding an initial burn-in period $B$:
$$\hat{\alpha}=\frac{1}{M-B}\sum^M_{m=B+1}\hat{\alpha}^{(m)}.$$
The same approach is used to estimate other parameters. Following \citet{suppming2023}, we recommend setting $B = \lceil 0.75 M \rceil$ (i.e., the smallest integer greater than or equal to $0.75 M$).

\subsection{Prediction}

With the estimated parameters $\hat{\boldsymbol{\varphi}}$, we generate $M$ pseudo-complete outputs, $\{Y^{*(m)}\}^M_{m=1}$, where each $Y^{*(m)} := \{\mathbf{y}^{*(m)}_l\}_{l=1}^{L}$, following the same procedure as in the E-step of the previous subsection. Using these generated samples, we compute the posterior mean and variance as follows: 
$$\mu^{(m)}(t,\mathbf{x})=\mathbb{E}[f(t,\mathbf{x})|Y^{*(m)}]\quad\text{and}\quad\sigma^{2(m)}(t,\mathbf{x})=\mathbb{V}[f(t,\mathbf{x})|Y^{*(m)}]$$
for $l=2,\ldots,L$, and $\mu^{(m)}_l(\mathbf{x})=\mathbb{E}[f_l(\mathbf{x})|Y^{*(m)}]$ and $\sigma^{2(m)}_l(\mathbf{x})=\mathbb{V}[f_l(\mathbf{x})|Y^{*(m)}]$. 
These posterior means and variances follow the same closed-form expressions as $h_1$ and $h_2$ in Proposition~\ref{prop:closedform}, since the data has a nested structure with pseudo inputs and outputs. The only modification is the replacement of $X_{-1}$ and $Y_{-L}$ by $X^*_{-1}$ and $Y^{*(m)}_{-L}$. The final predictive mean and variance are then approximated as follows:
\begin{align*}
\mu^*&(t,\mathbf{x})\approx \frac{1}{M}\sum^M_{m=1}\mu^{(m)}(t,\mathbf{x}),\\
\sigma^{2*}&(t,\mathbf{x})\approx \frac{1}{M}\sum^M_{m=1}(\mu^{(m)}(t,\mathbf{x})^2+\sigma^{2(m)}(t,\mathbf{x})) - \mu^*(t,\mathbf{x})^2.
\end{align*}

\subsection{Stochastic EM for non-nested DNA model}\label{appendix:stochasticEM}

Let $Y_{-L}^{*(m)} = (\mathbf{y}_1[1:n_2]^T, \widetilde{\mathbf{y}}^{(m)}_1[1:\widetilde{n}_2]^T, \mathbf{y}_2[1:n_3]^T, \widetilde{\mathbf{y}}^{(m)}_2[1:\widetilde{n}_3]^T, \ldots, \mathbf{y}_{L-1}[1:n_L]^T)^T$, and denote $\mathbf{X}^*_{-1} = ((\mathcal{X}^*_2)^T, \ldots, (\mathcal{X}^*_L)^T)^T$ as the combined inputs with corresponding tuning parameters $\mathbf{t}^*_{-1} = (t_2\mathbf{1}^T_{n^*_2}, \ldots, t_L\mathbf{1}^T_{n^*_L})^T$. Here, $n^*_l$ and $\widetilde{n}_l$ represents the sample sizes of the combined outputs $\mathbf{y}^{*(m)}_l$ and the pseudo outputs $\widetilde{\mathbf{y}}^{*(m)}_l$, respectively. Across levels $2,\ldots,L$, we further define the total number of the combined observations as $N_{-1}^*=\sum^L_{l=2}n^*_l$ and the number of imputed pseudo observations as $\widetilde{N}_{-1}=N_{-1}^{*}-N_{-1}$. 

\begin{itemize}
    \item \textbf{Initialization}:
    \begin{itemize}
        \item For each $l = 1, \dots, L-1$, fit independent GPs to $(\mathcal{X}_l, \mathbf{y}_l)$ and use the posterior mean to initialize $\widetilde{\mathbf{y}}_l^{(0)}$. Update the combined outputs as $\mathbf{y}^{*(0)}_l := \mathbf{y}_l \cup \widetilde{\mathbf{y}}^{(0)}_l.$
        \item Estimate parameters $\hat{\boldsymbol{\varphi}}^{(0)}$ using MLE with the pseudo-complete dataset $\{\{\mathbf{y}^{*(0)}_l\}_{l=1}^{L}, \allowbreak \{\mathcal{X}^*_l\}_{l=1}^{L}, \{t_l\}_{l=1}^{L}\}$.
    \end{itemize}
    \item \textbf{For $m = 1, \dots, M$}:
    \begin{itemize}
        \item \textbf{Imputation Step}:
        \begin{itemize}
            \item Sample $\widetilde{\mathbf{y}}_1$ from the posterior distribution
            \[
            p(\widetilde{\mathbf{y}}_1 | \mathbf{y}_1; \hat{\alpha}_1, \hat{\tau}^2_1,\hat{\boldsymbol{\theta}}_1) \sim \mathcal{N} \left(\mu_1(\widetilde{\mathcal{X}}_1), \hat{\tau}^2_1 ( K_1(\widetilde{\mathcal{X}}_1, \widetilde{\mathcal{X}}_1) - \mathbf{k}_1(\widetilde{\mathcal{X}}_1)^T \mathbf{K}_1^{-1} \mathbf{k}_1(\widetilde{\mathcal{X}}_1) )\right),
            \]
            where 
            $$
            \mu_1(\widetilde{\mathcal{X}}_1)=\hat{\alpha}_1\mathbf{1}_{\widetilde{n}_1} +\mathbf{k}_1(\widetilde{\mathcal{X}}_1)^T \mathbf{K}_1^{-1} (\mathbf{y}_1-\hat{\alpha}_1 \mathbf{1}_{n_1}),
            $$
            \( \mathbf{k}_1(\widetilde{\mathcal{X}}_1) \) is an \( \widetilde{n}_1 \times n_1 \) matrix with elements \( (\mathbf{k}_1(\widetilde{\mathcal{X}}_1))_{i,j} = K_1((\widetilde{\mathcal{X}}_1)_i, \mathbf{x}^{[1]}_j) \), and $K_1(\widetilde{\mathcal{X}}_1, \widetilde{\mathcal{X}}_1)$ is an \( \widetilde{n}_1 \times \widetilde{n}_1 \) matrix with elements \( (K_1(\widetilde{\mathcal{X}}_1,\widetilde{\mathcal{X}}_1))_{i,j} = K_1((\widetilde{\mathcal{X}}_1)_i, (\widetilde{\mathcal{X}}_1)_j) \).
            \item Sample $\{\widetilde{\mathbf{y}}^{(m)}_l\}_{l=2}^{L-1}$ from
            $p(\{\widetilde{\mathbf{y}}_l\}_{l=2}^{L-1} | Y_{-1}; \hat{\boldsymbol{\varphi}}^{(m-1)}, Y_{-L}^{*(m-1)})$, which follows a normal distribution with with the mean
$$
\hat{\alpha}^{(m-1)}\mathbf{1}_{\widetilde{N}_{-1}}-\mathbf{B}^{(m-1)}(\mathbf{C}^{(m-1)})^{-1}(Y_{-1}-\hat{\alpha}^{(m-1)}\mathbf{1}_{N_{-1}})
$$
and the covariance 
$$
\mathbf{A}^{(m-1)}-\mathbf{B}^{(m-1)}(\mathbf{C}^{(m-1)})^{-1}(\mathbf{B}^{(m-1)})^T.
$$
Here, $\mathbf{A} \in \mathbb{R}^{\widetilde{N}_{-1} \times \widetilde{N}_{-1}}$ is defined as
\begin{align*}
    \mathbf{A}^{(m-1)}_{ij} = K(&((\mathbf{t}^*_{-1}\setminus\mathbf{t}_{-1})_i, (\mathbf{X}^*_{-1}
\setminus\mathbf{X}_{-1})_i, (Y^{*(m-1)}_{-L}\setminus Y^{(m-1)}_{-L})_i), \\&((\mathbf{t}^*_{-1}\setminus\mathbf{t}_{-1})_j, (\mathbf{X}^*_{-1}
\setminus\mathbf{X}_{-1})_j, (Y^{*(m-1)}_{-L}\setminus Y^{(m-1)}_{-L})_j)), 
\end{align*}
where $\mathbf{t}_{-1}$ and $Y^{(m-1)}_{-L}$ represent the subset of $\mathbf{t}^*_{-1}$ and $Y^{*(m-1)}_{-L}$ corresponding to the indices of $\mathbf{X}_{-1}$. The matrix $\mathbf{B} \in \mathbb{R}^{\widetilde{N}_{-1} \times N_{-1}}$ is given by
\[
\mathbf{B}^{(m-1)}_{ij} = K\left(((\mathbf{t}^*_{-1}\setminus\mathbf{t}_{-1})_i, (\mathbf{X}^*_{-1}
\setminus\mathbf{X}_{-1})_i, (Y^{*(m-1)}_{-L}\setminus Y^{(m-1)}_{-L})_i), ((\mathbf{t}_{-1})_j, (\mathbf{X}_{-1})_j, (Y^{(m-1)}_{-L})_j)\right). 
\]
Similarly, the matrix $\mathbf{C}^{(m-1)}\in\mathbb{R}^{N_{-1}\times N_{-1}}$ is defined as
\[
\mathbf{C}^{(m-1)}_{ij} = K\left(((\mathbf{t}_{-1})_i, (\mathbf{X}_{-1})_i, (Y^{(m-1)}_{-L})_i), ((\mathbf{t}_{-1})_j, (\mathbf{X}_{-1})_j, (Y^{(m-1)}_{-L})_j)\right). 
\]
The hyperparameters of the kernel function $K$, namely $\boldsymbol{\theta},\beta$ and $\delta$, are plugged in by their estimates from the previous iteration, $\hat{\boldsymbol{\theta}}^{(m-1)},\hat{\beta}^{(m-1)}$ and $\hat{\delta}^{(m-1)}$.
        \end{itemize}
\item \textbf{Maximization Step}: Given the pseudo-complete data $\{\{\mathbf{y}^{*(m)}_l\}_{l=1}^{L}, \{\mathcal{X}^*_l\}_{l=1}^{L}, \{t_l\}_{l=1}^{L}\}$, update the parameter estimates $\hat{\boldsymbol{\varphi}}^{(m)}$ by maximizing the likelihood function, as described in Section~\ref{eq:estimation}.
    \end{itemize}
\end{itemize}

\section{Supporting figures in Sections~\ref{sec:studies} and \ref{sec:real}}\label{supp:figures}
\begin{figure}[ht!]
    \centering
\includegraphics[width=\linewidth]{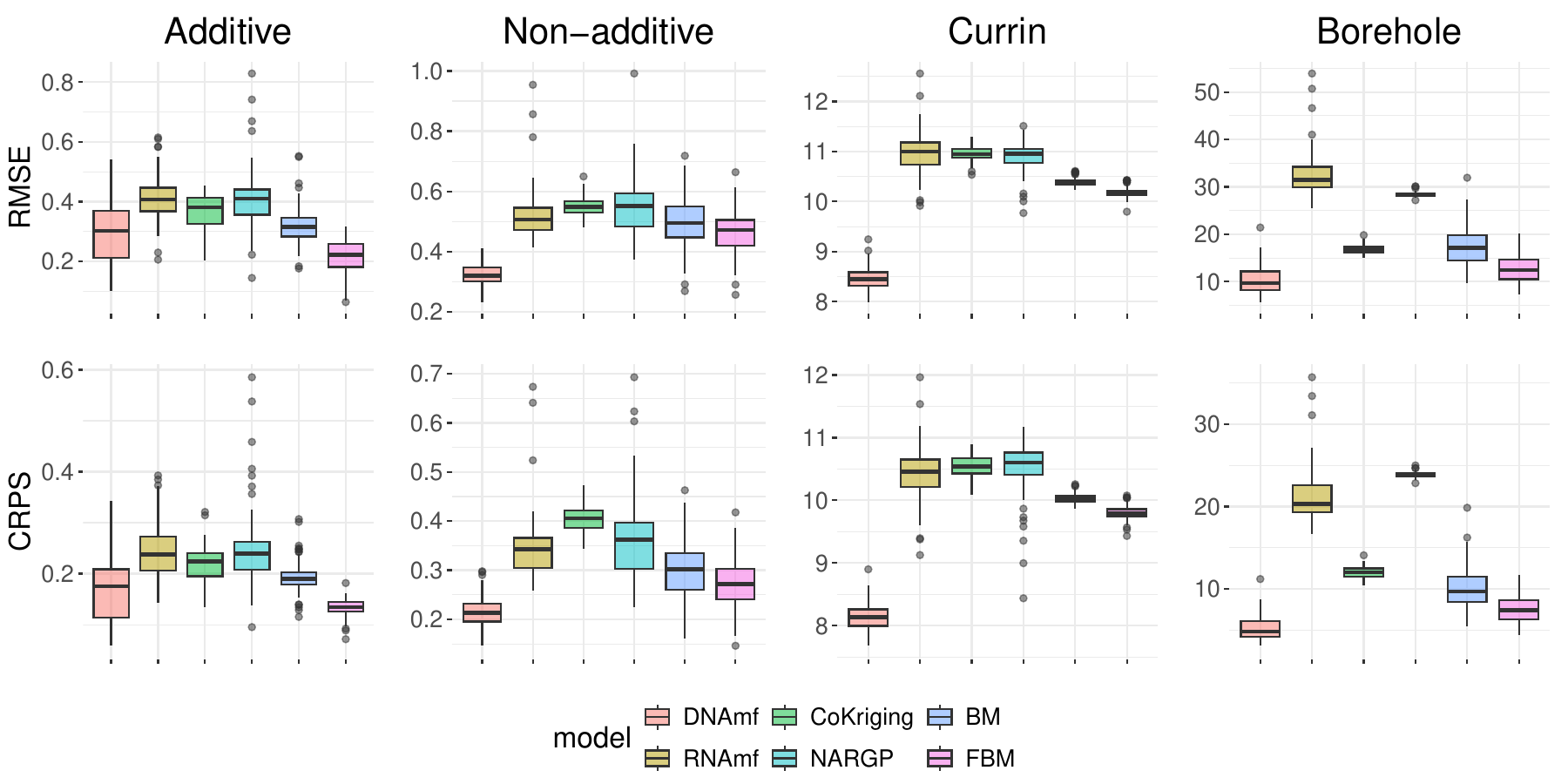} 
\caption{RMSEs and CRPSs of four synthetic examples across 100 repetitions with $T=4/3$ and $\gamma=2$.}
\label{fig:numerical_comparison_0.75_2}
\end{figure}

\begin{figure}[ht!]
    \centering
\includegraphics[width=\linewidth]{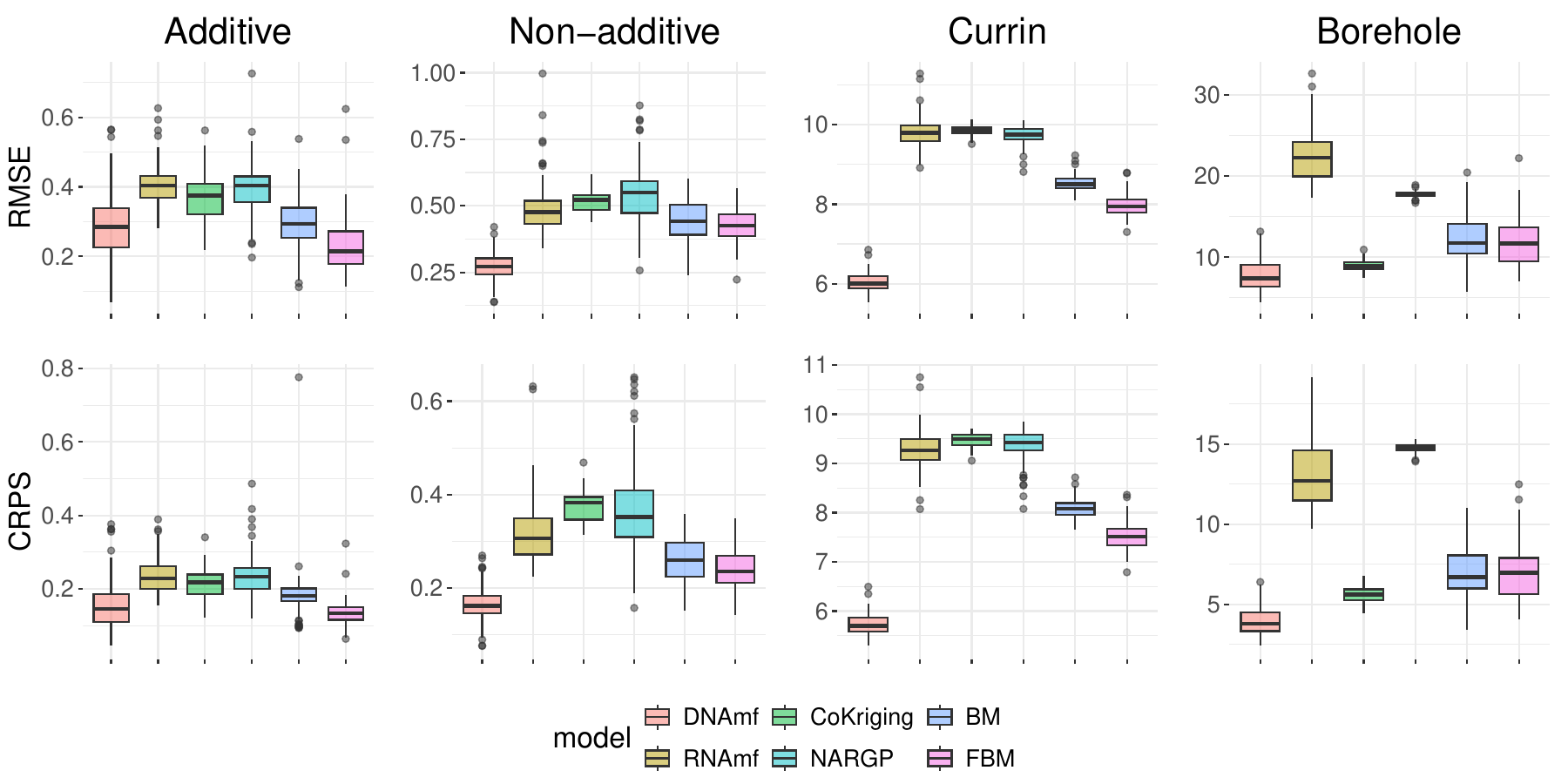} 
\caption{RMSEs and CRPSs of four synthetic examples across 100 repetitions with $T=3/2$ and $\gamma=1.5$.}
\label{fig:numerical_comparison_0.66_1.5}
\end{figure}

\begin{figure}[ht!]
    \centering
\includegraphics[width=\linewidth]{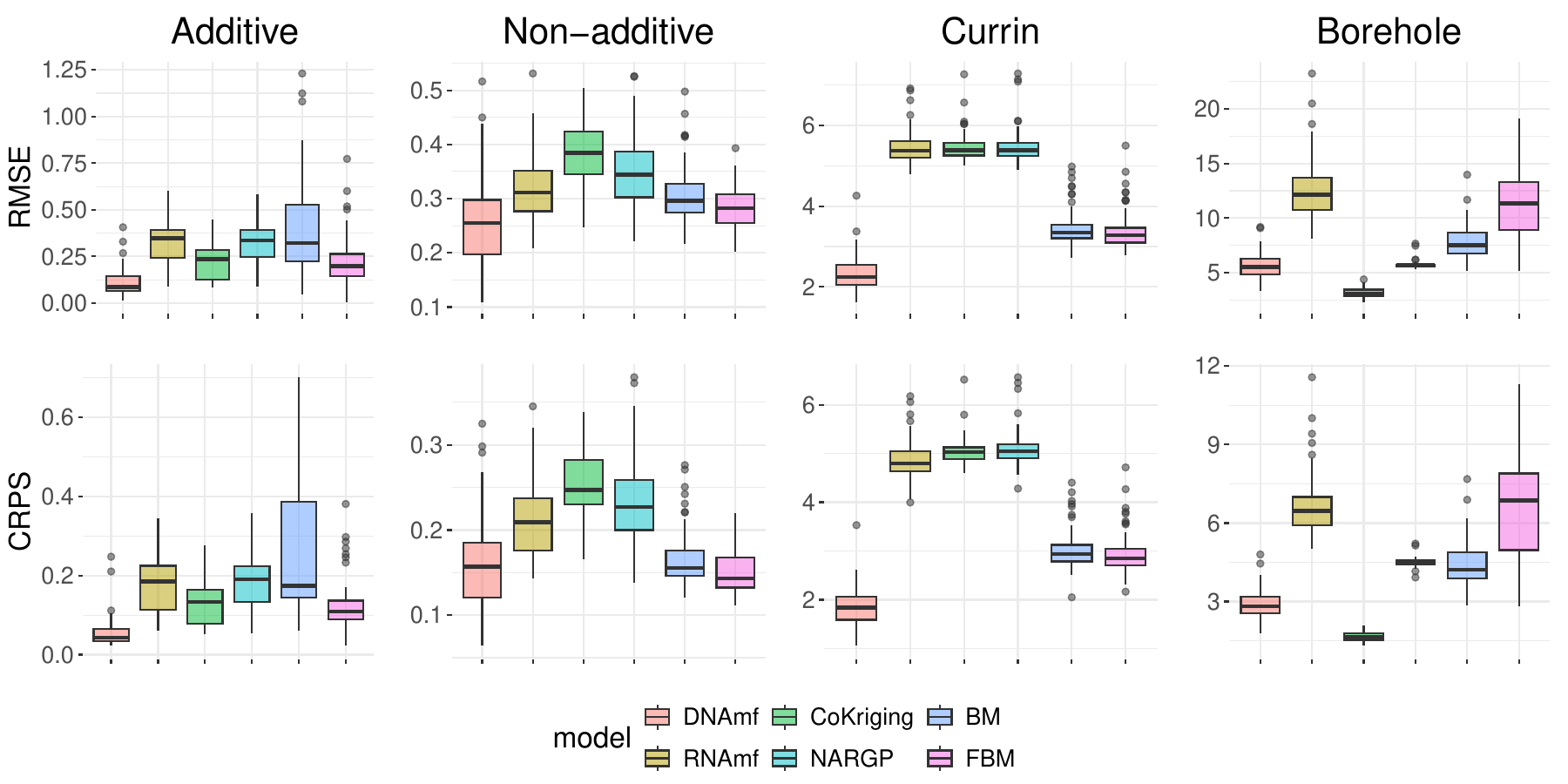} 
\caption{RMSEs and CRPSs of four synthetic examples across 100 repetitions with $T=2$ and $\gamma=1$.}
\label{fig:numerical_comparison_0.5_1}
\end{figure}

\begin{figure}[ht!]
    \centering
\includegraphics[width=\linewidth]{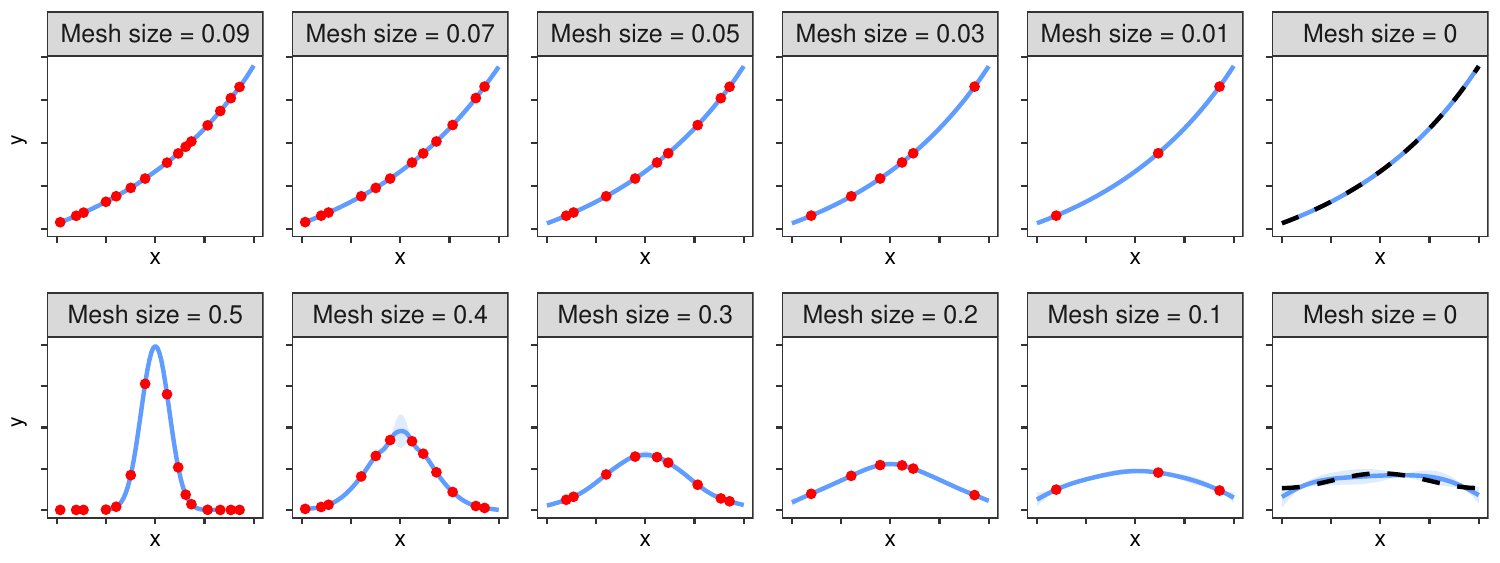} 
\caption{Illustrations of Poisson's equation (upper panel) and the heat equation (bottom panel) case studies. Each subplot represents subplots with tuning parameter values decreasing from large (left) to zero (right). In each subplot, the black dashed line represents the true function, red dots denote the design points, the blue line indicates the predicted function, and the shaded region depicts the 99\% confidence interval.}
\label{fig:realillustration}
\end{figure}

This section provides the comparison results in Section~\ref{sec:studies} (Figure \ref{fig:numerical_comparison_0.75_2}, \ref{fig:numerical_comparison_0.66_1.5},  and \ref{fig:numerical_comparison_0.5_1}), and the demonstration of the model predictions with confidence intervals for Poisson's equation and the heat equation in Section~\ref{sec:real} (Figure \ref{fig:realillustration}).

\end{document}